\documentclass[aps,twocolumn,prd,showpacs]{revtex4}

\usepackage{graphicx}

\usepackage{amsfonts}

\usepackage{amssymb}

\usepackage{array}
\usepackage{graphicx}
\usepackage{pstricks}
\usepackage{color}
\usepackage{xcolor}

\usepackage{subfigure}
\usepackage{amsmath}

\usepackage{latexsym}

\begin{document}
%

%
\title{Frieden wave function representations via an EPR-Bohm experiment}\author{J. Syska  }\email[E-mail: ]{jacek.syska@us.edu.pl}
\affiliation{Department of Field Theory and Particle Physics,
\\Institute of Physics, University of Silesia,  Uniwersytecka 4,
40-007 Katowice, Poland}


\date{\today}


\begin{abstract}
The appearance of the spin-$\frac{1}{2}$ and spin-1 representations
in the Frieden-Soffer extreme physical information (EPI) statistical approach
to the Einstein-Podolsky-Rosen-Bohm (EPR-Bohm) experiment is shown.
In order to obtain the EPR-Bohm result, in addition to the observed
structural and variational information principles of the EPI method,
the condition of the regularity of the probability distribution is used.
The observed structural information principle is obtained
from the analyticity of the logarithm of the likelihood function.
It is suggested that, due to the self-consistent
analysis of both information principles,
quantum mechanics is covered by the statistical
information theory.
The estimation of the angle between the analyzers in the EPR-Bohm experiment
is discussed.
\end{abstract}


\maketitle


\section{Introduction}

\label{Introduction}

\vspace{-4mm}

In the
1990s the generalization of
the maximum likelihood method (MLM) and Fisher information ($I_{F}$)
analysis in the statistical inference \cite{Fisher} was proposed by
Frieden and Soffer \cite{Frieden-Soffer} (more information is given at the end of this section).
It concerns the nonparametric estimation that permits the
selection of the equation of motions (or generating equations) of
various field theory or statistical physics models. They  called it the
extreme physical information (EPI) method. Although the EPI method is mainly
used to describe physical phenomena
\cite{Frieden-Soffer,Frieden-book}, other applications \cite{Frieden-Soffer,Frieden-book,Dziekuje za
skrypt,Dziekuje za produkcyjnosc} have also been considered. This
paper is devoted to the application of EPI for the description of
the Einstein-Podolsky-Rosen-Bohm (EPR-Bohm) experiment proposed by Frieden  \cite{Frieden-book}.
The basis for this kind of statistical analysis of the
phenomena was introduced in the
1920s by Fisher \cite{Fisher}.
His statistical model selection method, which is particularly useful for
a small sample, was constructed independently from the contemporary development of the
physical models, especially quantum mechanics.
It appeared that the EPI method leads to the estimation of the equations
of physical field theories for which a small size of the sample
is also a characteristic feature of the physical models
\cite{Frieden-Soffer,Frieden-book}.
For example, the size of the sample of the EPI model of the EPR-Bohm
problem is equal to $N=1$ \cite{Frieden-book}.
Previously, the EPR-Bohm type problems have mainly been understood as
manifestations of the quantum-mechanical reality.

The central quantity in an EPI analysis is the information channel
capacity  $I$, which is the trace of the expectation value of the
Fisher information matrix. The basic tools of the EPI method of
estimation are two information principles.

In \cite{Dziekuje informacja_2} the derivations, first of the observed and,
second, of the expected {\it structural information principle}
from basic principles, was given. It was based first on the analyticity
of the logarithm of the likelihood function, which allows for its
Taylor expansion in the neighborhood of the true value of the
vector parameter $\Theta$ \cite{Dziekuje informacja_2}, and second on the
symmetric and assumed positive definite form of the Fisher information
matrix.
The Fisher information enters into the EPI formalism
as the second order coefficient of this
expansion  \cite{Dziekuje informacja_2}.
%
%
%
The output of the solution of the information principles comprises the
equations of motion of basic field theories
or a distribution generating equation.
For example, in \cite{Frieden-book} the Maxwell equations
for $N=4$ were obtained, while
in \cite{Dziekuje za channel,Dziekuje za skrypt} the
construction of the information channel capacity for the vector
position parameter in the Minkowskian space-time was completed.
This has laid the statistical
foundations of the kinematical term of the Lagrangian of the
physical action for many field theory models that have been derived by
the EPI method of Frieden and Soffer \cite{Frieden-Soffer},
where the metricity of the statistical space ${\cal S}$ of
the system is also used.
%
%
%
Additionally, the fact that the formalism of the information principles is used
for the derivation of the distribution generating equation
signifies that the microcanonical description of the thermodynamic
properties of a compound system has to meet the analyticity and
metricity assumptions as well (in agreement with the Jaynes'
principle \cite{Jaynes}).

Although the analysis of the EPR-Bohm problem presented  below
originates in the Frieden  approach \cite{Frieden-book}, the paper
gives both a mathematical background \cite{Dziekuje
informacja_2,Dziekuje za skrypt} and a modified  physical
interpretation. First, the EPI analysis follows from a
meaning different than in \cite{Frieden-Soffer,Frieden-book}
of the structural information principle \cite{Dziekuje informacja_2}, which
originates in the above-mentioned analyticity
condition of (in this case) the one-dimensional statistical
space ${\cal S}$ (Sec.~\ref{Warunki brzegowe}) \cite{Dziekuje
informacja_2}.
Second, the
joint space of events (see Sec.~\ref{Warunki brzegowe})
for the construction of the Fisher information for the
EPR-Bohm problem is different than in \cite{Frieden-book}.
The information channel capacity that is necessary
to solve the EPR-Bohm problem in EPI method is presented
in Sec.~\ref{Pojemnosc informacyjna zagadnienia EPR}.
The new meaning of the structural information principle is also connected
with a more physical \cite{Dziekuje informacja_1,Dziekuje
informacja_2} and less informational \cite{Frieden-Soffer,Frieden-book}
oriented interpretation of the physical information $K$.
The variational information principle is connected with the
extremization of $K$ \cite{Frieden-Soffer,Frieden-book,Dziekuje informacja_1,Dziekuje
informacja_2}.
Third, the probability distribution that is characteristic for the
EPR-Bohm problem [see Eq.(\ref{wynikEPR})
in Sec.~\ref{analysis with Rao-Fisher metric}] is obtained
as the solution of two information
principles:
the observed structural principle, which is the differential one,
and the variational principle
(Sec.~\ref{Informacja strukturalna EPR}).
The condition of the regularity of the probability distribution
on the
one-dimensional statistical space ${\cal S}$, equipped with the
Rao-Fisher metric \cite{Rao,Efron,Dawid,Amari,Amari Nagaoka book},
is also used
(Sec.~\ref{restrictions from the regularity condition}).
This metric appears to be constant on ${\cal S}$.
Thus, the result is not obtained using the quantum mechanical
precondition of the basis wave functions orthogonality, as was
done in \cite{Frieden-book}. Also, a minute aspect of the
analysis is that the boundary conditions are put in order
\cite{Mroziakiewicz}.  \\
The experimental settings (Sec.~\ref{physical settings}) for
spin-$\frac{1}{2}$  particles (electrons) are the same as in
\cite{Frieden-book}; however, for spin-1 (massless photons), they are
different than in \cite{Frieden-book}. They are arranged in
such a way that the joint space of the events of the pair of spin
projections in the analyzers (see Sec.~\ref{Warunki brzegowe})
is the same in both cases. This makes the EPI results
in both cases basically of the same form and, therefore, they are
more easily interpreted as unveiling the origin of the rotation
group representation to which a particular particle
in the EPI-Bohm experiment belongs.
Next, it is suggested that the Fermi-Dirac (for
electrons) and the Bose-Einstein (for photons) statistics used in
the quantum mechanical description of the EPR-Bohm problem \cite{Manoukian} seem
to be a reminiscence of two consecutive steps.
In the case of fermions that are ruled by the Dirac equation,
the first step  consists of
the appearance of the generalized
Einstein-Brillouin-Keller quantization conditions \cite{Keppeler}.
In the case of the free electromagnetic
field that fulfills the Maxwell equations, the first step consists of
the existence of the conserved generalized helicity \cite{helicity of photon}. For both cases
the relevant equations of motion are obtained in advance
by the EPI method \cite{Frieden-book}.
The second step consists of the appearance of
the statistical information principles,
which, as shown in
Secs.~\ref{Pojemnosc informacyjna zagadnienia EPR}--\ref{Informacja strukturalna EPR},
follow the analyticity of the log-likelihood function
of the statistical space of the model.
%
%
%
Thus, in the case of fermions,
the Pauli exclusion principle has a statistical information
theory background.

Next, for the measurement performed by the experimentalist, in
Sec.~\ref{Niepewnosc wyznaczenia kata} the statistical
estimation of the angle $\vartheta$ between the analyzers in the
EPR-Bohm experiment (see Sec.~\ref{physical settings}) is
presented. In this context, both the difference between the inner
accuracy of the estimation \cite{Frieden-book,Dziekuje za channel} of
the angle $\vartheta$ in the EPI analy\-sis of the inner
$N=1$-dimensional sample collected by the system alone
\cite{Frieden-Soffer,Frieden-book} and the accuracy of the estimation of
$\vartheta$ in the experiment performed by the experimentalist
are discussed.
In the latter case the asymptotic local unbiasedness of
the $\vartheta$ estimator is analyzed. \\
Finally, throughout the analysis the priority
of the {\it analytical form} of the observed Fisher information is kept
and the {\it metric form} of the observed Fisher information is absolutely
secondary; i.e., on the observed level, it is consequently the full analytical model
that is solved (see Sec.~\ref{Informacja strukturalna EPR}).

{\it The Frieden-Soffer original form of the physical information
and information principles}.
Frieden and Soffer use two Fisherian information quantities: the intrinsic
information ${\rm J}$ of the source phenomenon and the information
channel capacity $I$, which connects the phenomenon and observer.
Using the information channel capacity $I$ and ``bound''
information ${\rm J}$, the physical information $K \equiv I - {\rm
J}$ was postulated \cite{Frieden-Soffer,Frieden-book}, which (together
with its densities) was used for the construction of the
Frieden-Soffer information principles, at both the observed and the
expected levels.
Although the structural information principle is postulated at the expected level,
in the Frie\-den and Soffer approach it is then reformulated to
the observed one, giving, together with the variational principle,
two coupled differential equations.
Thus, Frieden and Soffer, along with Plastino and Plastino \cite{Frieden-Soffer},
put the solution of the (differential) information principles for various EPI models
into practice.
The above-mentioned analytical background of the structural information principle was derived in \cite{Dziekuje informacja_2} and postulated previously in \cite{Dziekuje
informacja_1} (where the right direction for the transfer of
information during the measurement, i.e., ${\rm J} \rightarrow I$,
for the description of the Frieden interpretation should be used).
In our approach \cite{Dziekuje informacja_2} to the construction of the information
principles (see Sec.~\ref{general form of information principles})
the notion of the (total) {\it physical
information} ($TPI$) $K=Q+I \,$ instead of $K=I - {\rm J}$ introduced in
\cite{Frieden-Soffer} is used, where $Q$ is the {\it structural information}.
This difference does not affect the derivation of
the equation of motion or the generating equation for the problems that have been
analyzed until now \cite{Frieden-Soffer,Frieden-book},
since, identically, $Q = - {\rm J}$. Nevertheless, the above-mentioned analyticity
condition appears to be fruitful for the EPI modeling
(see Sec.~\ref{EPR-Bohm information principles}).
%
%
%

\vspace{-3mm}

\section{The physical settings and boundary conditions}

\label{physical settings and boundary conditions}

\vspace{-2mm}

\subsection{The physical settings and statistical space ${\cal S}$}

\vspace{-2mm}

\label{physical settings}

\begin{figure}[top]
\begin{center}
\includegraphics[width=70mm,height=35mm]{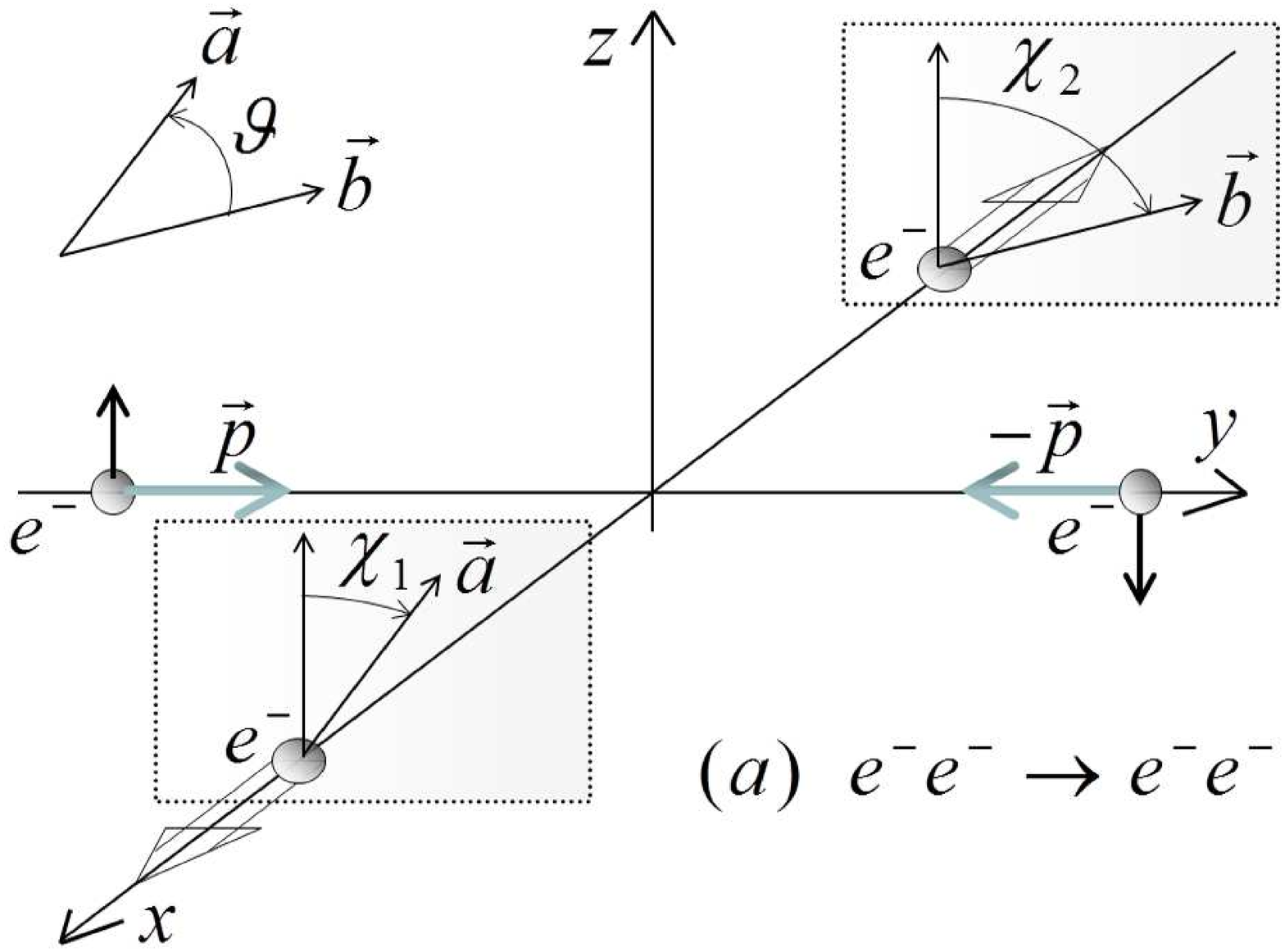}\\
\includegraphics[width=70mm,height=35mm]{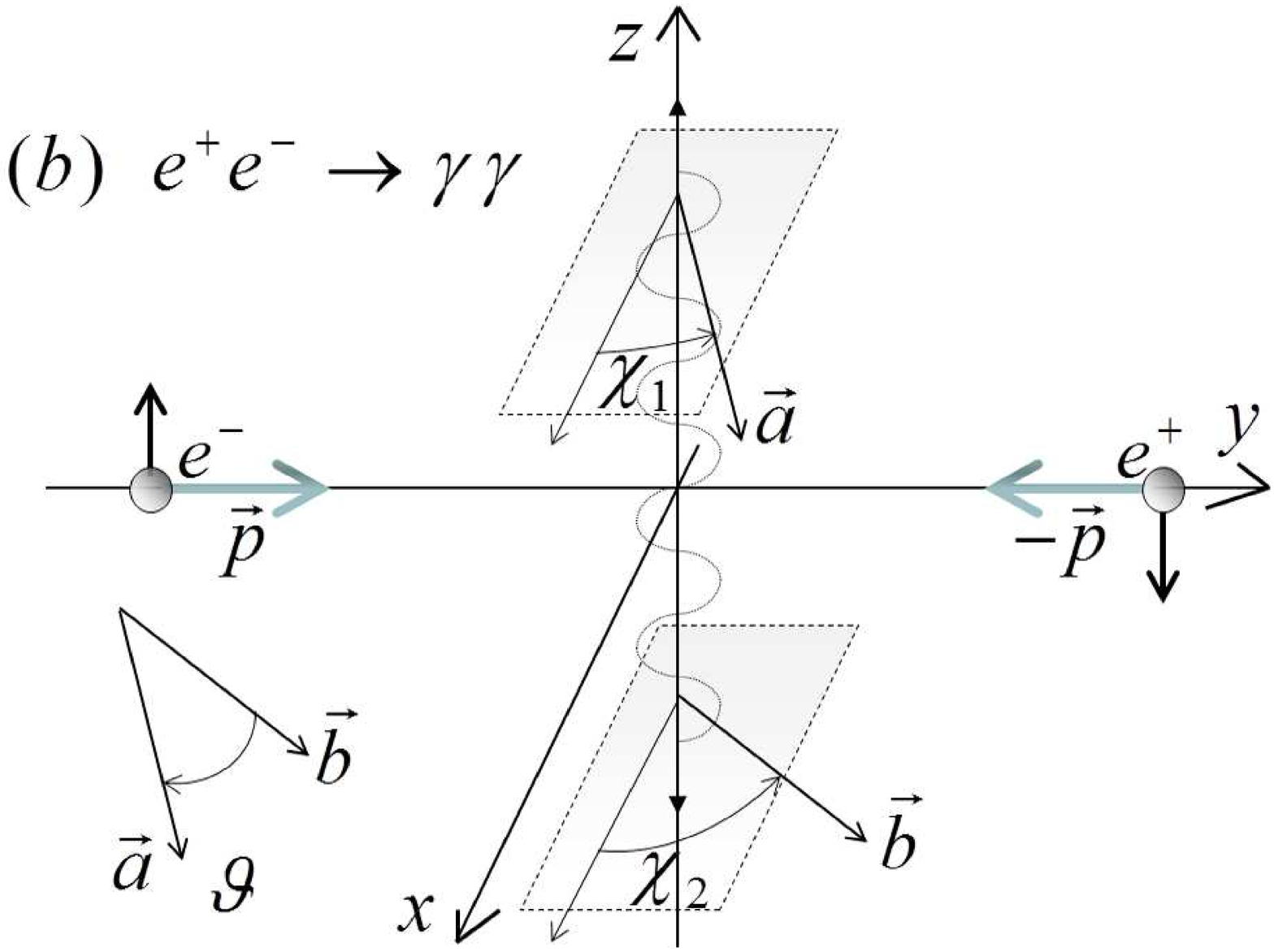}
\end{center}
\vspace{-2mm}
\caption{The EPR-Bohm  experiment.
The possible configurations of processes are presented (see the text).
\newline
({\bf a}) Process
$e^{-}e^{-} \rightarrow e^{-}e^{-}$. One of the initial electrons
has the spin up and the other down along the $z$ axis.
Their momenta along the $y$ axis are
$\vec{p}$ and $-\vec{p}$, respectively. The final electrons of the
bipartite system
move along the $x$ axis with opposite momenta and are
registered in the Stern-Gerlach devices. The angle $\vartheta$
between the directions $\vec{a}$ and $\vec{b}$ of the
Stern-Gerlach devices is signified.
$\chi_{1}$ and $\chi_{2}$ are the angles of the analyzers with the $z$ axis;
$\vartheta = \chi_{2} - \chi_{1}$.
({\bf b}) Process $e^{-}e^{+} \rightarrow \gamma\, \gamma$.
The initial electron and positron have their spins up and down along the
$z$ axis and their momenta  along the $y$-axis are $\vec{p}$ and $-\vec{p}$,
respectively. The created photons ``1'' and ``2''  of the bipartite system move up and down
along the $z$ axis with the opposite momenta, respectively, and their polarizations
are measured by two polarizers, which
make angles $\chi_{1}$ and $\chi_{2}$ with the $x$ axis; $\vartheta = \chi_{2} - \chi_{1}$.
%
%
}
\vspace{-2mm}
\label{fig-1}
\end{figure}

Let us consider two types of the EPR-Bohm experiment: the first
one for the spin-0 charged molecule which decays into bipartite system of
two identical  spin-$\frac{1}{2}$  particles (electrons $e^{-}$ in this paper)
and the second one for the spin-0 neutral molecule which decays
into a bipartite system of two spin-1 (massless) photons.
In both cases the total angular
momentum along the $z$ axis is zero (see Fig.~1).
For the first case, such a bipartite state for the EPR-Bohm experiment can be
effectively prepared, e.g., as the final state in the scattering
process $e^{-}e^{-} \to e^{-}e^{-}$, where the spins of the
initial electrons of the process are arranged to be one up and the
other down along the $z$ axis, whereas their momenta $\vec{p}\,$
and $-\vec{p}$ are along the $y$ axis \cite{Manoukian}  [see
Figure~\ref{fig-1}(a)]. There is the nonzero probability
that two
scattered particles (here final electrons) move along the $x$ axis
with the opposite momenta  \cite{Manoukian-Yongram}.
In the EPR-Bohm experiment, the measurement of the spin projections
of the scattered particles is performed.
%
%

The analyzer ``$a$'', which is the Stern-Gerlach device, measures
the projection  $S_{a}$
of the  spin $\vec{S}_{1}$ of the particle ``1'' along the direction
of the unite vector
$\vec{a}$ and similarly the analyzer  ``$b$'' measures the
projection  $S_{b}$
of the  spin $\vec{S}_{2}$ of the particle ``2'' along the direction
of the unite vector
$\vec{b}$.
The angle between the planes of the vectors $\vec{a}$ and $\vec{b}$ that
include the $x$ axis is equal to
$\vartheta = \chi_{1}-\chi_{2}$, $0\le  \vartheta < 2\pi$.
Similarly,  for the second case, the process  $e^{-}e^{+} \to \gamma
\gamma\,$ is the relevant one, where electron $e^{-}$ and positron
$e^{+}$ are prepared with spins up and down along the $z$ axis and
with momenta $\vec{p}\,$ and $-\vec{p}$ along the $y$ axis,
respectively  [see Fig.~\ref{fig-1}(b)]. The electron and positron
annihilate into two photons. The created photons ``1'' and ``2''
move up and down along the $z$ axis with the opposite momenta,
and there is also a nonzero probability
that such a process occurs
\cite{Manoukian-Yongram}.
The polarizations of photons are measured
by polarizers with polarization vectors making angles $\chi_{1}$
and $\chi_{2}$ with the $x$ axis, respectively \cite{Manoukian}.

Below, in Sec.~\ref{Warunki brzegowe} we derive the  boundary conditions
for the EPR-Bohm problem in the EPI method,
taking care not to appeal to the quantum mechanical vision of
reality.
%
However, we take into account the general statement
that if $P(AB)$ is the joint probability distribution for two
random variables $A$  and $B$ and  ${P(A)}$ and ${P(B)}$ are
their  marginal distributions, then \cite{Frieden-book}:
\begin{eqnarray}
\label{subaddytywnosc informacji} I_{F}\left({P(AB)}\right)  \ge
I_{F} \left({P(A)}\right) + I_{F} \left({P(B)}\right)\equiv
\tilde{C} \, ,
\end{eqnarray}
where the equality holds if the variables are independent. Here $\tilde{C}$
is the information channel capacity of the composite system. Relation
(\ref{subaddytywnosc informacji}) means that if there is any
dependence among the variables,
then, if we know the result of
the experiment for the first variable,
the amount of information needed for the determination of the result of the
experiment for the other one diminishes. So, the existence of the
dependence in the system increases the Fisher information $I_{F}$
on the parameters that characterize the probability distribution
of the system.

{\it Note on the common notion for electron and photon spin projections.}
The notation is common for the measurement of the spin
projection of final particles, whether they are electrons [Fig.~1(a)]
or photons [Fig.~1(b)]. The final electrons [Fig.~1(a)]
are recorded in the Stern-Gerlach devices
and the final photons [Fig.~1(b)] are recorded in the polarizers.
When the spin-$\frac{1}{2}$ particle (e.g., electron 1) is registered then ``$+$''
signifies the observed value of its polarization along the direction $\vec{a}$ and
the spin projection on $\vec{a}$ that is equal to $S_{a}=S_{+}=+\hbar/2$,
whereas ``$-$'' signifies
the observed value of its polarization along the direction $-\vec{a}$ and
the spin projection on $\vec{a}$ equal to $S_{a}=S_{-}=-\hbar/2$.
When the spin-1 photon (e.g., photon 1) is registered then $+$
signifies the observed value of its polarization along the direction $\vec{a}$ and
the spin projection (helicity) {\it on the $z$-axis}
(direction of its propagation) equal to $S_{a}=S_{+}=+\hbar$,
whereas $-$ signifies the observed value of its polarization
along the perpendicular direction $\vec{a}_{\perp}$ and the spin
projection {\it on the $z$-axis} equal to $S_{a}=S_{-}=-\hbar$.
Similar notation is used for the particle 2 (electron or photon)
that is registered in the polarizer $b$.

\vspace{-2mm}

\subsection{The boundary conditions for the EPR-Bohm problem}

\label{Warunki brzegowe}

In order to determine the problem of the differential equations of the
EPI  method, it is necessary to indicate the boundary conditions for the
probabilities. They follow the phenomenological premises,
conservation laws and special symmetries of the system being
examined.

\vspace{-1mm}

\subsubsection{Determination of the spin projection}

\label{determination of the spin projection}


%
The basic EPI method concept from which the derivation of the Dirac equation follows is
the (density of the) information channel capacity $I$,
which for the spinorial field is invariant under the isometry transformations in
$N/2$-dimensional complex space $\mathbb{C}^{N/2}$
of fields $\psi\equiv[\psi_{n}({\bf x})]_{n=1}^{N/2}$
\cite{Dziekuje za models building}.
After decomposing the density of $I$ into the components \cite{Frieden-book},
each of them can be factorized into terms that are the elements
of the Clifford group ${\rm Pin(1,3)}$ \cite{Dziekuje za models building}.
The group ${\rm Pin(1,3)}$ is a subset of the Clifford algebra $C(1,3)$.
As the spinor representations of an orthonormal basis in
$C(1,3)$ are the Dirac $\gamma^{\mu}$ matrices; therefore,
in summary, the Dirac equation appears via the factorization of the density
of the information channel capacity \cite{Frieden-book}
and due to the previously mentioned observed
structural information principle and the variational one,
in which the density of the information channel capacity $i$ forms the kinetic term.
In this way in \cite{Frieden-book} for the $N=8$-dimensional sample
the Dirac equation for the spinorial field [i.e., for the complex field
$\psi\equiv[\psi_{n}({\bf x})]_{n=1}^{4}$ of the rank $N=8$]
was obtained. Then, under the assumption of the conservation of momentum,
it was supplemented  \cite{Frieden-Soffer-de-Broglie}
by the EPI method background of the de~Broglie-Fourier representation
of a particle \cite{Broglie}.
%
%
%
%

Next, in \cite{Keppeler} the semiclassical theory for spinning
particles (ruled by the Dirac equation) was presented. Here, Keppeler
derived a
form of the Einstein-Brillouin-Keller
quantization conditions
generalized for the particles with spin,
in which the spin quantum number equal to $\pm 1/2$ appears. In \cite{Keppeler} the
latitude of the classical spin vector appears as a constant of motion.
This type of
semiclassical quantization also appears in \cite{Dziekuje za neutron}
for the description of the spin of the neutron.
The suitable steps for the photon are as follows. For $N=4$,
the EPI method leads to the Maxwell equations \cite{Frieden-book,Dziekuje za channel}
(with the Lorentz condition additionally imposed).
Then, for the free electromagnetic field seen as the composition
of the circularly polarized right- and left-handed photons \cite{Manoukian,Roychoudhuri},
the existence of the conserved classical generalized helicity was proven
in \cite{helicity of photon}.
%
%
%
%
%
%
Finally,
in \cite{stochastic projection}
the stochastic trial to explain the quantization of
the spin projection was presented.

[Meanwhile, the classical
statistics estimation
relevant for stochastic processes is connected with
the analysis of the (pure)
likelihood function and, hence, it omits
the logarithmic structure of the system
\cite{Dziekuje_Jacek_nova_2,Dziekuje informacja_2,Dziekuje za skrypt} (which is
present in the EPI method)].


\subsubsection{The joint space of events}

\label{joint space of events}

\vspace{-1mm}

Thus, in accordance with the above section,
{\it the boundary condition} connected
with the existence of strictly (up to the noise in the measuring devices; see
Sec.~\ref{zaszumienie pomiaru}) two possible spin projections appears, which
for a particle with spin $\hbar/2$ are on the arbitrary space direction
and for the massless photon with spin $\hbar$ on the direction of the propagation.
In agreement with the above {\it Note on the common notation}
(Sec.~\ref{physical settings}), we
let $+$ signify the observed value of the spin projection
$S_{a}=+\hbar/2$ or $S_{a}=+\hbar$, while $-$ signifies
$S_{a} = - \hbar/2$ or $S_{a} = - \hbar$ for the particle with
spin $\hbar/2$ or a photon with spin $\hbar$, respectively.
%
We introduce the common denotation for {\it the base space}
$\mathbb{S}$ of the random variable $S_{a}$ for both cases:
\begin{eqnarray}
\label{przestrzen dla Sa}
\mathbb{S} = \left\{ S_{-},S_{+}
\right\} \equiv \left\{-,+\right\} \; .
\end{eqnarray}
Similar notation is used for the observed value of the spin
projection $S_{b}$, for which the base space is also $\mathbb{S}$.
%
%
%

{\it The joint space of events} $\Omega_{ab}$ of the pair of spin
projections for particles 1 and 2 is as
follows \cite{Frieden-book}:
\begin{eqnarray}
\label{4 zdarzenia EPR}
\!\!\!\!\!\!\! \quad S_{a}S_{b} \equiv S_{ab} & \equiv & ab \in \Omega_{ab} = \left\{ S_{++},S_{--},S_{+-},S_{-+} \right\}
\nonumber \\
& \equiv &
\left\{++,--,+-,-+\right\}
\, .
\end{eqnarray}
It can be checked that the form of $\Omega_{ab}$ is different for a spin-1 massive particle
or a higher than spin-1 particle.
The analysis of these cases is not included in this paper.

Let us suppose that four joint conditional
probabilities can
be defined $P\left(S_{ab}|\vartheta \right)$,
\begin{eqnarray}
\label{spinprawdop}
\!\!\!\!\! P\left(++|\vartheta\right) \, , \; P\left(--|\vartheta\right) \, , \;
P\left(+-|\vartheta\right) \, , \; P\left(-+|\vartheta\right) \;
,
\end{eqnarray}
where $\vartheta \in \langle 0, \, \pi) \equiv V_{\vartheta}$ and
$V_{\vartheta}$ is the parameter space.

{\it Let us remark} that in the case
of the lack of the joint
probability  space $\Omega_{AB}$ for the random variables, let us say
$A$ and $B$, there is no possibility to define the joint
probability distribution $P(A,B)$ for these two variables, even if
their marginal distributions $P(A)$ and $P(B)$ exist. This means that
they cannot be simultaneously measured.
Also, in general, despite the existence of the joint marginal
probability distribution $P(A,B)$ for the variables $A$ and $B$
and the existence of the joint marginal probability distribution
$P(B,C)$ for the variables $B$ and $C$, the
joint probability distribution $P(A,B,C)$ for variables $A$, $B$
and $C$ does not exist. Let us notice that in the proof of the Bell inequality
\cite{Bell,Khrennikov}, it is taken for granted that the joint
distribution $P(A,B,C)$ does exist \cite{Bell,Khrennikov}. Such a
possibility always exists only if the joint  events space
$\Omega_{ABC}$  for these three random variables exists, which is the
Cartesian product $\Omega_A \times \Omega_B \times \Omega_C$.
On the other hand, Bell-type inequalities are known from
Boole's times as the test, which, if failed,
confirms the impossibility of the construction of the joint probability
distribution. The above consideration could be extended for an
arbitrary number of random variables \cite{Khrennikov,Accardi}. In
the full description of the EPR-Bohm experiment,
not only the random variables of the spin projection
measured in the analysers $a$ and $b$ but two random
angle variables measured for these particles in the moment of
their production should be taken into account.

Because of the mutual exclusion of different number
$\aleph=4$ of events $S_{ab}$, (\ref{4 zdarzenia EPR}), the condition of
the probability
normalization $P\left(S_{a}S_{b}|\vartheta\right)$  in the
EPR-Bohm problem can  be
written as follows:
\begin{eqnarray}
\label{normalizacja P daje wsp w qab}
\!\!\!\!\!\!\!\!
&P&\!\!\!\!\left(\bigcup\limits_{ab}{S_{ab}|\vartheta}\right) =
\sum\limits_{ab}P\left(S_{ab}|\vartheta\right) = \\
&=& P\left(S_{++}|\vartheta\right) + P\left(S_{--}|\vartheta\right)
+ P\left(S_{+-}|\vartheta\right) + P\left(S_{-+}|\vartheta\right) \nonumber \\
& = &   1 \;, \,\;
\quad \quad \quad \quad \quad \quad \quad \quad \quad \quad \quad \quad \quad \quad \quad \quad \forall  \vartheta \in V_{\vartheta} \, .  \nonumber
\end{eqnarray}
These four joint probabilities $P\left(S_{ab}|\vartheta\right)$
form the statistical (sub)space ${\cal S}$,
\begin{eqnarray}
\label{statistical space S}
{\cal S} =
\{
P\left(S_{ab}|\vartheta\right)|\;\;
\vartheta \in \langle 0 \; 2 \,\pi) \equiv V_{\vartheta} \subset
\mathbb{R}^{1} \}
\end{eqnarray}
of the EPR-Bohm problem, which
is the one-dimensional submanifold of the $\aleph-1=3$ dimensional
probability simplex \cite{Bengtsson_Zyczkowski} coordinatized by the parameter $\vartheta$ (see
the Appendix).

Taking into account the {\it Note on the common notation} in Sec.~\ref{physical settings}
and keeping the order of the consecutive summands in
Eq.(\ref{normalizacja P daje wsp w qab}), we obtain the following.
In the case of a spin-$\frac{1}{2}$ particle and using the notation from
Fig.~1(a), we see that Eq.(\ref{normalizacja P daje wsp w qab}) has the form \cite{Manoukian}
\begin{eqnarray}
\label{normalizacja P dla one-half}
\!\!\! & & \sum\limits_{ab}P\left(S_{ab}|\vartheta\right)  = \nonumber \\
\!\!\! &=&
P\left(\vec{a},\vec{b}\right) + P\left(-\vec{a},-\vec{b}\right)
+ P\left(\vec{a},-\vec{b}\right) + P\left(-\vec{a},\vec{b}\right)
=  1 \; , \nonumber
\end{eqnarray}
where for the particle 1  and the polarizer $a$, the event $\vec{a}$ under the denotation of the probability $P$ signifies the particle polarization in the direction $\vec{a}$, i.e., $S_{a}=S_{+}$, and $-\vec{a}$ signifies the event of the particle polarization in the direction $-\vec{a}$, i.e., $S_{a}=S_{-}$ (and similarly for  the particle 2 and the polarizer $b$).
In the case of spin-1 photon and using the notation from Fig.~1(b),  we see that Eq.(\ref{normalizacja P daje wsp w qab}) has the form \cite{Manoukian}
\begin{eqnarray}
\label{normalizacja P dla one}
\!\!\! & & \sum\limits_{ab}P\left(S_{ab}|\vartheta\right)   \nonumber \\
\!\!\! &=&
P\left(\vec{a},\vec{b}\right) + P\left(\vec{a}_{\perp},\vec{b}_{\perp}\right)
+ P\left(\vec{a},\vec{b}_{\perp}\right) + P\left(\vec{a}_{\perp},\vec{b}\right)
= 1 \; , \nonumber
\end{eqnarray}
where for the photon 1  and the polarizer $a$, the event $\vec{a}$ under the denotation of the probability $P$ means the photon polarization in the direction $\vec{a}$, which signifies the spin projection on the $z$ axis equal to $S_{a}=S_{+}$, and $\vec{a}_{\perp}$ means the event of the photon polarization in the direction perpendicular to $\vec{a}$ and therefore $S_{a}=S_{-}$ (and similarly for  the photon 2 and the polarizer $b$).
%
%
%

{\it Note on the method of the estimation of $\vartheta$.}
In Sec.~\ref{Niepewnosc wyznaczenia kata} we see that due to the fact that the frequencies of
the events $S_{ab}$, (\ref{4 zdarzenia EPR}),
observed in the analyzers $a$ and $b$, are locally unbiased
estimators of the probabilities (\ref{spinprawdop}), the value of the angle
$\vartheta$ can be robustly estimated by an experimentalist
by taking a large enough $\textsc{m}$-dimensional (outer)
sample. However, to make this possible, an analytical model which enables
the derivation of the formulas for the probability distribution
(\ref{spinprawdop}) is needed, and it can be obtained using the
EPI method. The dimension of the (inner) sample in the EPI
analysis has to be taken as equal to $N=1$ (Section~\ref{Pojemnosc
informacyjna zagadnienia EPR}). Although the result was originally
obtained by Frieden \cite{Frieden-book}, we regain it without any
reference to the quantum mechanical formalism (i.e., here to
the orthogonality of the quantum mechanical wave functions).

\vspace{-1mm}

\subsection{The formulation of the boundary conditions}

\label{formulation of the boundary conditions}

\vspace{-1mm}

For every event $S_{ab}$, (\ref{4 zdarzenia EPR}),
the probability $P\left(S_{ab}\right)$ of the
appearance of $S_{ab}$, irrespective of the value of $\vartheta$,
is equal to
\begin{eqnarray}
\label{srednia z  PSab theta}
P\left(S_{ab}\right)
= \int\limits_{0}^{2\pi} {P\left(S_{ab},\vartheta\right) d\vartheta}
\; .
\end{eqnarray}
Here, using the definition of the conditional probability,
the probability $P\left(S_{ab},\vartheta\right)$ is defined as
\begin{eqnarray}
\label{P Sab warunek vartheta}
P\left(S_{ab},\vartheta\right) := P\left({S_{ab}\left|{\vartheta}\right.}\right) r\left({\vartheta}\right)
\; ,
\end{eqnarray}
where $r(\vartheta)$ is the so-called {\it lack of
knowledge function}.

(A) Because the events (\ref{4 zdarzenia EPR}) are mutually
exclusive and they span the entire space of events $\Omega_{ab}$, the first
boundary condition is the {\it normalization}
\begin{eqnarray}
\label{normalizacja prawd dla EPR}
\!\!\!\!\!\!\! \!\!\!\!\!\!\! & &  \sum_{ab}{P\left(S_{ab}\right)} = \nonumber \\
\!\!\!\!\!\!\! \!\!\!\!\!\!\! & = & P\left(S_{++}\right) + P\left(S_{--}\right) + P\left(S_{+-}\right) + P\left(S_{-+}\right)
= 1  \, , \,
\end{eqnarray}
which is fulfilled irrespective of the value of $\vartheta$.

Because of the normalization conditions (\ref{normalizacja P daje
wsp w qab}) and (\ref{normalizacja prawd dla EPR}) and using
Eqs.(\ref{srednia z  PSab theta}) and (\ref{P Sab warunek vartheta}),
it can be noticed that the possible form of $r(\vartheta)$
can be chosen as
\begin{eqnarray}
\label{prawdkat}
r(\vartheta)=\frac{1}{2\pi} \; ,\quad\quad
\vartheta \in V_{\vartheta} \; ,
\end{eqnarray}
which means that in the range $\left\langle 0, 2 \pi \right)$ of
the measuring apparatus variability of the angle $\vartheta$ (see
Fig.~1) and due to the lack our knowledge, every value of
$\vartheta \in V_{\vartheta}$ is equally possible.

{\it Remark on the value of the angle $\vartheta$.} The
value of  $\vartheta$ is connected with the arrangement of the
measuring
analyzers $a$ and $b$ (Stern-Gerlach devices or polarizers),
and it is hardly to be (seriously) treated as possessing
the distribution. Essentially,
$\vartheta$ is the parameter which is characteristic for the
experiment being carried out. Nonetheless,
$r(\vartheta)$ is seen as {\it a priori} known ``probability'' by some,
which means that quantum mechanics that is deduced in this way should be
treated as the Bayesian  statistical theory \cite{stany koherentne}.

(B) The consecutive conditions follow from {\it the symmetry of
the system and the law of the total spin conservation}
(assuming that {\it the relative orbital angular momentum
of the particles is equal to zero} in the experiment).

Let us consider a simple case $\vartheta=0$ when both planes in
which the measuring devices (Stern-Gerlach or polarizer) are
set up are oriented in the same way. From the condition of the
total spin conservation it follows that
\begin{eqnarray}
\label{vartheta zero} P\left(++|\vartheta=0\right)=P\left(--|\vartheta=0\right)=0 \; ,
\end{eqnarray}
which means that with this arrangement of the apparatus, we never
see both spins simultaneously directed up or down.
Also, we never see the spins directed one up and the other down in the following situations:
\begin{eqnarray}
\label{vartheta pi}
& {\rm for} &  {\rm spin}-\frac{1}{2} \; {\rm if} \; \vartheta = \pi
\nonumber \\
&\Rightarrow& \;  P\left(+-|\vartheta=\pi\right)=P\left(-+|\vartheta=\pi\right)=0 \; ,
\nonumber\\
\!\!\!\!\!\!\! \!\!\!\!\!\!\! \!\!\!\!\!\!\! {\rm and} & & \\
& {\rm for} & {\rm spin}-1 \;  {\rm if} \; \vartheta = \frac{\pi}{2}
\nonumber \\
&\Rightarrow& \;  P\left(+-|\vartheta= \frac{\pi}{2}\right)=P\left(-+|\vartheta= \frac{\pi}{2}\right)=0 \; ,
 \nonumber
\end{eqnarray}
respectively.
As a result, from the law of the total spin
conservation, we find that if we know that in the case
of spin-$\frac{1}{2}$ particle $\vartheta=0$ or
$\vartheta=\pi$  (or in the case of spin-1 photon $\vartheta=0$ or
$\vartheta=\pi/2$), then the observation of one spin projection gives us
complete knowledge about the second one. In this case, the spin states
are clearly not independent.
This conclusion is incorporated
in their preparation method intuitively (discussed in Sec.~\ref{physical settings}).

Next, as $P\left(S_{b}|\vartheta\right)$ is the marginal
probability of the appearance of the particular value of the spin
projection of the particle 2, it does not depend on
$S_{a}$, i.e., on the orientation of the spin projection of the
particle 1. From the normalization condition for
$P\left(S_{b}|\vartheta\right)$, it follows that
\begin{eqnarray}
\label{C theta const} \!\!\!\!\!\!\!\!\!
P\left(S_{b}=+|\vartheta\right) + P\left(S_{b} = -
|\vartheta\right) = 1 \;  .
\end{eqnarray}
Next, due to the fact that, because of the
symmetry, the definition of the {\it up} vs. {\it down}
spin projection $S_{b}$ is chosen arbitrarily, we see that Eq.(\ref{C theta
const}) leads to
\begin{eqnarray}
\label{Sa Sb const} P\left(S_{b}=+|\vartheta\right) =
P\left(S_{b}=-|\vartheta\right) = \frac{1}{2} \; .
\end{eqnarray}
Thus, we have obtained
\begin{eqnarray}
\label{pol}
P\left(S_{b}|\vartheta\right) = \frac{1}{2} \quad \;\;
{\rm for} \;\;\; S_{b} = +,- \;\, ,
\end{eqnarray}
independent of the angle $\vartheta$ between the vectors
$\vec{a}$ and $\vec{b}$.
It follows that
\begin{eqnarray}
\label{Sb wycalkowane}
\!\!\!\!\!\!\! P\left(S_{b}\right) &=&
\int\limits_{0}^{2\pi} {P\left(S_{b}|\vartheta\right)
r\left(\vartheta\right) d\vartheta}
\nonumber \\
&=& \frac{1}{2} \int\limits_{0}^{2\pi} {r\left(\vartheta\right)
d\vartheta} = \frac{1}{2} \quad \;\; {\rm for} \;\;\; S_{b} = +,-
\;\, ,
\end{eqnarray}
where $r(\vartheta)$ is specified in Eq.(\ref{prawdkat}).
%
%
%

The other important property of the spacial symmetry of the
experiment is {\it the lack of the  preference for upward or downward
orientation of the the spin}; i.e.,
\begin{eqnarray}
\label{symetriaEPR}
& & \!\!\!\!\!\!\! \!\!\!\!\!\!\! P\left(S_{+-}|\vartheta\right) \!
= \! P\left(S_{- +}|\vartheta\right) \; \nonumber \\
& & \!\!\!\!\!\!\! \!\!\!\!\!\!\! \!\!\!\!\!\!\!
\!\!\!\!\!\!\! \!\!\!\!\!\!\! \!\!\!\!\!\!\!
 {\rm and} \;  \\
& & \!\!\!\!\!\!\! \!\!\!\!\!\!\!  P\left(S_{++}|\vartheta\right) \!
= \! P\left(S_{--}|\vartheta\right)  \, . \; \nonumber
\end{eqnarray}
This means that if we observe the experiment for the system rotated around the $x$-axis by the angle $\pi$,
then the
probability would be exactly the same.

From Eq.(\ref{symetriaEPR}) and Eqs.(\ref{srednia z  PSab theta}) and (\ref{P Sab warunek vartheta}),
we obtain
\begin{eqnarray}
\label{++ = - -}
P\left(S_{++}\right) &=& \int\limits_{0}^{2\pi} {P\left(S_{++}|\vartheta\right)r\left(\vartheta\right)d\vartheta} \nonumber \\
& = & \int\limits_{0}^{2\pi}  {P\left(S_{--}|\vartheta\right)r\left(\vartheta\right)d\vartheta}=P\left(S_{--}\right)
\end{eqnarray}
and, analogously,
\begin{eqnarray}
P\left(S_{+-}\right)=P\left(S_{-+}\right)\label{22} \; .
\end{eqnarray}
\begin{widetext}
Additionally, we obtain
\begin{eqnarray}
\label{doda}
P\left(S_{+-}\right) & = & \int_{0}^{2\pi} {P\left(S_{+-}|\vartheta\right) r\left(\vartheta\right)d\vartheta}  \\
& = & \!\!\!\!  \int_{0}^{2\pi} {P\left(S_{+-}|\vartheta+\pi\right) r\left(\vartheta+\pi\right)d\vartheta}
= \int_{0}^{2\pi} {P\left(S_{++}|\vartheta\right) r\left(\vartheta+\pi\right)
d\vartheta}  \; , \;\;\;\;\; \quad \quad\quad\quad   {\rm for}\; {\rm spin} \; \frac{1}{2} \nonumber
\\
& = & \!\!\!\! \int_{0}^{2\pi} {P\left(S_{+-}|\vartheta+\frac{\pi}{2}\right) r\left(\vartheta+\frac{\pi}{2}\right)d\vartheta}
= \int_{0}^{2\pi} {P\left(S_{++}|\vartheta\right) r\left(\vartheta+\frac{\pi}{2}\right)
d\vartheta}  \; , \;\; \quad\quad\quad {\rm for}\; {\rm spin} \; 1 \nonumber \\
&=& P\left(S_{++}\right) \; , \nonumber
\end{eqnarray}
\vspace{-7mm}
\end{widetext}
where in the second and third lines the simple change of the (circle) variables
$\vartheta \to \vartheta + \pi$ or $\vartheta \to \vartheta + \pi/2$ for ${\rm spin}-\frac{1}{2}$
or ${\rm spin}-1$ particle, respectively, has been used (i.e., the starting point of integration around a closed circle is insignificant).
Then, in the second line, for spin-$\frac{1}{2}$ particle we use the equality
\begin{eqnarray}
\label{theta plus pi sym for spin one half}
P\left(S_{++}|\vartheta\right) = P\left(S_{+-}|\vartheta+\pi\right)   \; , \;\;\;\;\; \quad   {\rm for}\; {\rm spin} \; \frac{1}{2}
\end{eqnarray}
and in the third line for spin-1 photon,
\begin{eqnarray}
\label{theta plus pi sym for spin 1}
P\left(S_{++}|\vartheta\right) = P\left(S_{+-}|\vartheta+\frac{\pi}{2}\right)    \; , \;\; \quad \quad {\rm for}\; {\rm spin} \; 1 \; .
\end{eqnarray}
The conditions (\ref{theta plus pi sym for spin one half}) or (\ref{theta plus pi sym for spin 1})
follow from the equivalence of
the events of obtaining the ``$+$'' polarization in the direction $\vec{a}$
of the measuring device and the ``$-$'' polarization in the direction $- \vec{a}$ or $\vec{a}_{\perp}$
in the case of an electron or photon, respectively.
Finally, in the last equality in Eq.(\ref{doda}), we use $r(\vartheta) = r(\vartheta+\pi)=  r(\vartheta+\frac{\pi}{2})=1/{(2\pi)}$.

As a consequence of  Eqs.(\ref{++ = - -})--(\ref{doda}), we
obtain
\begin{eqnarray}
P\left(S_{-+}\right) = P\left(S_{--}\right) \; .
\label{doda2}
\end{eqnarray}
Taking into account Eqs.(\ref{++ = - -})--(\ref{doda2}) and
Eq.(\ref{normalizacja prawd dla EPR}), we obtain
\begin{eqnarray}
P\left(S_{ab}\right) = \frac{1}{4} \;  \; , \quad {\rm for
\; every }\;\; S_{ab} \in \Omega_{ab}\; .
\label{laczneEPR}
\end{eqnarray}
Finally, the Bayes' formula  for the {\it conditional probability}
gives
\begin{eqnarray}
\label{warunkSab}
\!\!\!\! P\left(S_{a}|S_{b}\right) =
\frac{P\left(S_{ab}\right)}{P\left(S_{b}\right)} = \frac{1/4}{1/2}
= \frac{1}{2} \, , \;\; {\rm for \; every } \;\; S_{a}, S_{b} \, .
\end{eqnarray}
By reasoning similar to that which gave Eqs.(\ref{pol}),
(\ref{Sb wycalkowane}), (\ref{laczneEPR}), (\ref{warunkSab}) previously, we
obtain
\begin{eqnarray}
\label{pol dla Sa}
\!\!\!\! P\left(S_{a}|\vartheta\right) = \frac{1}{2} \;
, \;\;\; P\left(S_{a}\right) = \frac{1}{2} \; , \;\;\; {\rm for}
\;\;\; S_{a} = +,- \; ,
\end{eqnarray}
\begin{eqnarray}
\label{laczneEPR ba}
\!\!\!\! P\left(S_{ba}\right) = \frac{1}{4} \; \, ,
\;\;\; P\left(S_{b}|S_{a}\right) = \frac{1}{2} \; , \quad {\rm for
\; every }\;\; S_{a}, S_{b} \; ,
\end{eqnarray}
respectively.

{\it Final remarks on the boundary conditions.} Let us
note that, to this point, we have not used the EPI method
in the derivation of formulas (\ref{vartheta zero})--(\ref{warunkSab}).
This is because these are the boundary conditions for
the equations of the EPI method. These conditions follow from (i)
the initial observation of the existence of exactly two possible
spin projections
for the spin-$\frac{1}{2}$ particle (here the electron) or for the spin-1 (massless) photon
(see the {\it Note on the common notation} in Sec.~\ref{physical settings}),
(ii) the normalization of the probability distribution,
(iii) the law of the total angular momentum conservation, and
(iv) the symmetry of the system.
Finally, in what follows, when a solution of the
equation generating the distribution is chosen, in Sec.~\ref{EPR-Bohm information principles}
we use the geometrical symmetry of the system under the rotation by the angle
$2 \pi$ once more.

\vspace{-1mm}

\section{The information channel capacity for the EPR-Bohm problem}

\label{Pojemnosc informacyjna zagadnienia EPR}

\vspace{-1mm}

Below we derive the EPI method results for the
probabilities (\ref{spinprawdop}).
While they are consistent with the ones obtained in
\cite{Frieden-book}, there are also some
noticeable differences (Sec.~\ref{Pojemnosc informacyjna
zagadnienia EPR}) and extensions of the method (e.g., in
Secs.~\ref{forms of the amplitude} and \ref{Niepewnosc wyznaczenia kata}).

{\it The general form of $I$ for independent data.}
Let an original random variable $Y$ (a discrete or continuous one) take
values ${\bf y} \in {\cal Y}$, where  ${\cal Y}$ is the base
space,  and suppose that parameter $\theta$ of the regular (in
$\theta$) distribution $p({\bf y})$ in which we are interested,
is the scalar one.
Let the data $y = ({\bf y}_{n})_{n=1}^{N}$ be a realization
of the $N$-dimensional sample $\widetilde{Y} = (Y_{1},Y_{2},...,Y_{N})
\equiv( Y_{n})_{n=1}^{N}$ and $p_{n}({\bf y}_{n}|\theta_{n})$ be a point distribution for
the $n$-th observation in the $N$-dimensional sample
\cite{Dziekuje za channel,Dziekuje za skrypt}.
The set of all possible realizations $y$ of the sample $\widetilde{Y}$ forms the {\it sample space}
${\cal B}$ of the system.
We assume that the variables $Y_{n}$ of the sample $\widetilde{Y}$ are independent.
%
%
It is also supposed that $p_{n}({\bf y}_{n}|\theta_{n})$ does not depend on the parameter
$\theta_{m}$ for $m \neq n$.
Thus, the data are generated in agreement with the
point probability distributions, which fulfill the condition $p_{n}({\bf y}_{n}|\Theta) = p_{n}({\bf y}_{n}|{\theta}_{n})\,$ ($n=1,...,N$), where $\Theta \equiv (\theta_{n})_{n=1}^{N}$,
and {\it the likelihood function} $P(y\,|\Theta)$ of the sample $y = ({\bf y}_{n})_{n=1}^{N}$ is the product
\begin{eqnarray}
\label{funkcja wiarygodnosci proby - def}
P(\Theta) \equiv P\left({y|\Theta}\right) = \prod\limits_{n=1}^{N} {p_{n}\left({{\bf y}_{n}|\theta_{n}}\right)} \; .
\end{eqnarray}
The expected
Fisher information matrix on the statistical (sub)space ${\cal S}$ at point $P(\Theta)$
\cite{Amari Nagaoka book} is defined as
\begin{eqnarray}
\label{infoczekiwana}
I_F \left(\Theta\right) \equiv E_{\Theta} \left({\rm i^{F}}(\Theta)\right) = \int_{\cal B} dy \, P(y|\Theta) \, {\rm i^{F}}(\Theta) \; ,
\end{eqnarray}
where ${\cal B}$ is the sample space, the differential element is given by
$dy \equiv d^{N}{\bf y} = d{\bf y}_{1} d{\bf y}_{2} ... d{\bf y}_{N}$,
and ${\rm i^{F}}(\Theta)$ is the {\it observed Fisher information matrix}
\cite{Pawitan,Dziekuje informacja_2,Dziekuje za skrypt,Dziekuje za channel}.
%
The subscript $\Theta$ in the expected value
signifies the true value of the parameter
under which the data $y$
are generated.
The Fisher information matrix defines
on ${\cal S}$ the Riemannian Rao-Fisher metric
\cite{Amari Nagaoka book,Dziekuje za channel}.

The central quantity of EPI analysis is the information channel capacity  $I$, which
is the trace of the (expected) Fisher information matrix. Because under above conditions,
the observed Fisher information matrix is diagonal ${\rm i^{F}}(\Theta)$ $= {\rm diag}({\rm i^{F}}_{\!\!nn}(\Theta))$; hence,
the information channel capacity $I(\Theta)$ is equal to
\begin{eqnarray}
\label{def I dla diag IF}
I(\Theta) = \sum_{n=1}^{N}
\int_{\cal B} dy \, P(y|\Theta) \, {\rm i^{F}}_{\!\!nn}(\Theta)
=
\int_{\cal B} dy \, \textit{i} \; ,
\end{eqnarray}
where $\textit{i} :=  P(\Theta) \; \sum_{n=1}^{N} {\rm i^{F}}_{\!\!nn}(\Theta) $ is
the {\it information channel density} \cite{Dziekuje za skrypt,Dziekuje za channel}.

When expressed by the point probability distributions,
the {\it analytical form} of the information channel capacity $I(\Theta)$
is as follows \cite{Frieden-Soffer,Frieden-book}:
\begin{eqnarray}
\label{postac I analytical point}
I(\Theta) &=& \sum_{n=1}^{N}
I_{Fn}(\theta_{n})
\\
&=& - \sum_{n=1}^{N} {\int_{\cal Y}{d{\bf
y}_{n}\, {p_{n}\left({{\bf
y}_{n}|\theta_{n}}\right)}\, {\frac{{\partial^{2}\ln p_{n}
\left({{\bf
y}_{n}|\theta_{n}}\right)}}{{\partial\theta_{n}^{\,2}}}}}}  \nonumber
\\
&=& -
\sum_{n=1}^{N}  {\int_{\cal Y} {d{\bf y}_{n}\,
{\frac{{\partial^{2} p_{n}\left({{\bf
y}_{n}|\theta_{n}}\right)}}{{\partial\theta_{n}^{2}}}}}}  \nonumber
\\
&+&
\sum_{n=1}^{N}  {\int_{\cal Y} {d{\bf y}_{n}\,
\frac{1}{{p_{n}\left({{\bf y}_{n}|\theta_{n}}\right)}} \,
\left({\frac{{\partial p_{n}\left({{\bf
y}_{n}|\theta_{n}}\right)}}{{\partial\theta_{n}}}}\right)^{2}}} \,
  , \nonumber
\end{eqnarray}
where in the second line both Eq.(\ref{funkcja wiarygodnosci proby - def}) and the normalization
of the point distributions $p_{n}({{\bf y}_{n}|\theta_{n}})$, $n=1,...,N$ were used \cite{Frieden-book}.
Due to the normalization and the condition of regularity of the probability distribution (in every $\theta_{n}$) \cite{Pawitan}, it follows that
\begin{eqnarray}
\label{reg and norm general}
\!\!\!\!\!\!\! \!\!\!\!\!\!\! & & \int_{\cal Y} d{\bf y}_{n}\,
\frac{{\partial^{2} p_{n}\left({{\bf
y}_{n}|\theta_{n}}\right)}}{{\partial\theta_{n}^{2}}}  \nonumber
\\
\!\!\!\!\!\!\! \!\!\!\!\!\!\! &=& \frac{\partial^{2}}{\partial\theta_{n}^{2}}
\int_{\cal Y} d{\bf y}_{n}\, p_{n}\left({{\bf
y}_{n}|\theta_{n}}\right) = \frac{\partial^{2}}{\partial\theta_{n}^{2}} 1
 = 0
\end{eqnarray}
and the information channel capacity (\ref{postac I analytical point}) for the vector parameter
$\Theta \equiv (\theta_{n})_{n=1}^{N}$ can be written
in its {\it metric form} \cite{Frieden-book,Amari Nagaoka book}:
\vspace{-3mm}
\begin{eqnarray}
\label{postac I koncowa w pn}
I(\Theta) &=& \sum_{n=1}^{N}
I_{Fn}(\theta_{n})   \\
&=&
\sum_{n=1}^{N}  {\int_{\cal Y}{d{\bf
y}_{n} \, {p_{n}  \left({{\bf
y}_{n}|\theta_{n}}\right)}\left({\frac{{\partial\ln
p_{n}\left({{\bf
y}_{n}|\theta_{n}}\right)}}{{\partial\theta_{n}}}}\right)^{2}}}
 \; . \nonumber
\end{eqnarray}
The Fisher information $I_{Fn}$ is the measure
of the precision of the estimation of one scalar parameter
$\theta_{n}$ \cite{Dziekuje za channel,Dziekuje za skrypt}.
Below the forms
of the Fisher information will be adopted for the purpose of the
estimation of the angle $\vartheta$ (see Fig.~1).

%
{\it Note}. According to {\it the main assumption of the EPI method}
proposed by Frieden and Soffer, the system alone
samples the the space of the positions that is accessible to it
using its Fisherian, kinematical degrees of freedom
\cite{Frieden-Soffer,Dziekuje za channel}.
%
%
%
%
Thus,  the
bipartite system alone measures \cite{Frieden-book,Dziekuje za channel}
the values of the spin projections of two particles,
i.e., particle 1, which is $S_{a}$, and particle 2, which is $S_{b}$.
%
%

{\it The sample space in the EPR-Bohm problem.} The particular form of the
information channel capacity
takes into account the {\it measurement channel} \cite{Dziekuje za
channel}, i.e., the channel which is
indivisible from an experimental point of view. In the EPR-Bohm
problem the sample appears to be $N=1$-dimensional \cite{Frieden-book},
which (as we will see) gives the required form of the estimated probabilities
and the measurements of $S_{a}$ and $S_{b}$ appear
dependent \cite{Dziekuje za channel}. Thus, the
{\it measurement channel} consists of
one joint measurement (performed by the system
alone) of the pair of spin projections $S_{a}$ and $S_{b}$,
i.e.,
of the random variable $S_{ab}$, which takes the value from the
joint space of events $\Omega_{ab}$ that is given by Eq.(\ref{4 zdarzenia EPR}).
Using the general formula (\ref{postac I koncowa w pn}) the
assignments which pertain to the EPR-Bohm problem are
\begin{eqnarray}
\label{N oraz wartosci y w funkcji wiaryg dla EPR}
\!\!\!\!\!\!\! {\bf y_{1}} & \equiv & S_{ab} \; ,
\;\; {\cal Y} \equiv \Omega_{ab} \; ,
\;\; \theta_{1} \equiv \vartheta \; , \nonumber \\
\int_{\cal Y}{d{\bf y}_{1}} &\equiv&
\sum\limits_{S_{a}=-}^{+} \sum\limits_{S_{b}=-}^{+}
\equiv \sum\limits_{ab} \;, \;\;\; (N = 1) \; ,
\end{eqnarray}
where the summation over $ab$ in $\sum_{ab}$
is performed over the joint space of events $\Omega_{ab}$.
As $N=1$ the sample space ${\cal B}$ is equivalent to the base space
$\Omega_{ab}$. Because, in accordance with
Eq.(\ref{N oraz wartosci y w funkcji wiaryg dla EPR}),
the dimension of the sample is equal to $N=1$, the information channel
capacity $I$ in Eq.(\ref{postac I koncowa w pn}) (let us
call it the ``single-value-$\vartheta$''  information channel
capacity) reduces to the Fisher information
$I_{F\,n=1}(\theta_{1}) = I_{F}(\vartheta)$ for
$\theta_{1}  \equiv \vartheta$.

{\it The likelihood function}. As a bipartite system (of two particles)
performs the measurement of $S_{ab}$ by itself, the likelihood
function for the problem  stated above is
\begin{eqnarray}
\label{finkcja wiaryg dla vartheta}
\!\!\!\!\!\!\!
& & P\left(S_{ab}|\vartheta\right) \; \; \; {\rm is \; the\;likelihood} \;
\nonumber \\
& & {\rm of \; the \; sample \; for} \;\;  \vartheta \;\; {\rm parameter},
\end{eqnarray}
which means that in Eq.(\ref{postac I koncowa w pn}) the  assignment
$p_{1} \equiv P\left(S_{ab}| \vartheta\right)$ also has to be performed.
Now the form of $P\left(S_{ab}| \vartheta\right)$ is searched for using the EPI method.

{\it The probability amplitudes} $q_{ab}$ are defined in the following way \cite{Bengtsson_Zyczkowski,Amari Nagaoka book}:
\begin{eqnarray}
\label{inffishEPR2}
P\left(S_{ab}\left|{\vartheta}\right.\right) = \frac{1}{4} \,
q_{ab}^{2}\left(\vartheta\right) \;\;\;\; {\rm for
\; every }\;\; S_{ab} \in \Omega_{ab} \;  .
\end{eqnarray}
Because the dimension of the sample is equal to $N=1$,  the rank of
the amplitude $q_{ab}\left(\vartheta\right)$ of the field is also equal
to 1 \cite{Frieden-book}.

\vspace{-2mm}

\subsection{The expected Fisher information
and the information capacity of the channel $(\vartheta)$}

\label{expected Fisher information}

\vspace{-2mm}

As $\vartheta$ is the scalar parameter and the dimension of the
sample is equal to $N=1$, thus taking into account the assignments
given by Eq.(\ref{N oraz wartosci y w funkcji wiaryg dla EPR}), the
information channel capacity $I({\vartheta})$, (\ref{postac I koncowa w
pn}), of the {\it measurement channel} $(\vartheta)$ is equal to
{\it the Fisher information
$I_{F}(\vartheta)$ of the parameter} $\vartheta$,
\begin{eqnarray}
\label{IF dla vartheta w EPR}
I({\vartheta}) = I_{F}(\vartheta)
\, ,
\end{eqnarray}
where the {\it analytical form} (\ref{postac I analytical point}) of the (expected)
Fisher information on parameter $\vartheta$ is equal to
\begin{eqnarray}
\label{postac I analytical EPR}
I_{F}(\vartheta) &=&
\sum\limits_{a=-}^{+} \sum\limits_{b=-}^{+}
P\left(S_{ab}\left|{\vartheta}
\right.\right) {\rm i^{F}}_{\!\!ab}(\vartheta)  \nonumber \\
&= &
\sum\limits_{ab}
{P\left(S_{ab}\left|{\vartheta}
\right.\right) \left( - \;\frac{\partial^{2}
\ln P \left(S_{ab}\left|{\vartheta}\right.\right)}{\partial
\vartheta^{2}}\right)}  \nonumber
\\
&\equiv&  \sum\limits_{ab} i_{ab} =
\sum\limits_{ab}
\left( - \, \frac{\partial^{2}P\left(S_{ab}\left|{\vartheta}\right.\right)}{\partial
\vartheta^{2}} \,
+
{\left( q_{ab}^{'} \right)^{2}} \,
\right)  \nonumber \\
&=&
\sum\limits_{ab}
\left( - \, q_{ab} {
q_{ab}^{''} } + \frac{\partial^{2}P\left(S_{ab}\left|{\vartheta}\right.\right)}{\partial
\vartheta^{2}} \,
\right) \, ,
\end{eqnarray}
where Eq.(\ref{inffishEPR2}) and the denotations $\sum\limits_{ab} \equiv \sum\limits_{a=-}^{+} \sum\limits_{b=-}^{+}\,$, $q_{ab}^{'} \equiv
\frac{dq_{ab}(\vartheta)}{d\vartheta}$, and $q_{ab}^{''} \equiv
\frac{d^{2}q_{ab}(\vartheta)}{d\vartheta^{2}}$ have been used.  In the last equality the relation
\begin{eqnarray}
\label{P na q}
\frac{\partial^{2}P\left(S_{ab}\left|{\vartheta}\right.\right)}{\partial
\vartheta^{2}} = \frac{1}{2} \,(q_{ab}^{'})^{2} + \frac{1}{2} \, q_{ab} \,q_{ab}^{''}
\end{eqnarray}
was also applied.
Due to the normalization, (\ref{normalizacja P daje wsp w qab}), and the regularity condition
[see Eq.(\ref{reg and norm general})],
\begin{eqnarray}
\label{regularity condition}
\!\!\!\!\! \sum\limits_{ab}
\frac{\partial^{2}P\left(S_{ab}|\vartheta\right)}{\partial\vartheta^{2}}  = \frac{\partial^{2} }{\partial\vartheta^{2}} \sum\limits_{ab}  P\left(S_{ab}|\vartheta\right)= \frac{\partial^{2} }{\partial\vartheta^{2}} 1 = 0 \; , \;
\end{eqnarray}
the {\it analytical form} (\ref{postac I analytical EPR}) of the Fisher information transforms into the following {\it metric form}:
\begin{eqnarray}
\label{inffishEPR-origin}
\!\!\!\!\!\!\!\! & & \!\!\!\!\!\!\!\!
I_{F}(\vartheta)  =  \sum\limits_{ab}
P\left(S_{ab}\left|{\vartheta}
\right.\right) \widetilde{{\rm i^{F}}}_{\!\!ab}(\vartheta)
 \nonumber \\
\!\!\!\!\!\!\!\! &=&
\sum\limits_{ab}
{\frac{1}{P\left(S_{ab}\left|{\vartheta}
\right.\right)}\left(\frac{\partial P
\left(S_{ab}\left|{\vartheta}\right.\right)}{\partial
\vartheta}\right)^{2}}  =
\sum\limits_{ab} {\left(q_{ab}^{'} \right)^{2}} \; .
\end{eqnarray}
The {\it analytical form} of the observed Fisher information ${\rm i^{F}}(\vartheta)$
in Eq.(\ref{postac I analytical EPR}) differs from its {\it metric form} $\widetilde{{\rm i^{F}}}(\vartheta)$ in
Eq.(\ref{inffishEPR-origin}) by $\frac{-1}{P\left(S_{ab}|\vartheta\right)} \frac{\partial^{2}P\left(S_{ab}|\vartheta\right)}{\partial\vartheta^{2}} \; $.
%
However, also due to Eq.(\ref{regularity condition}) and in accordance with Eq.(\ref{IF dla vartheta w EPR}),  we see that both the {\it EPI method form} of the (expected) Fisher information for the EPR problem and its single-value-$\vartheta$ information channel capacity
for the measurement channel $(\vartheta)$ are equal to
\begin{eqnarray}
\label{inffishEPR}
\;\;\;\;\;\;\;\;
I({\vartheta}) = I_{F}(\vartheta)
= -  \sum\limits_{ab} q_{ab}(\vartheta) q_{ab}^{''}(\vartheta)    \; .
\end{eqnarray}
%
{\it Remark.} Thus, $I_{F}(\vartheta)$ is the information
about the unknown angle $\vartheta$ confined
in the $N=1$-dimensional sample
for the random variable $S_{ab}$ (which is the pair of  spin projections
for particles $1$ and $2$ of the bipartite system).

{\it  The total information capacity $I$ for the parameter
$\vartheta$}.  As the angle  $\vartheta$ is the parameter whose
value can change continuously in the interval $\langle0,2\pi)$,
in accordance with  Eq.(\ref{inffishEPR}), there are an infinite
number of channels
for which the Fisher information on $\vartheta$ can be
calculated.
To handle such a situation, the EPI method uses single,
scalar information \cite{Frieden-Soffer,Frieden-book} (let us denote it simply
$I$) called the (total) information channel capacity. This
quantity is constructed by summing all of the possible
single-value-$\vartheta$ information channel capacities.
%
Thus, first, the information channel capacity
$I \left(\vartheta_{k}\right)$ of one channel $\left(\vartheta_{k}\right)$,
where $\vartheta = \vartheta_{k}$, is given by
Eq.(\ref{inffishEPR}), and second,
the summation runs over all of the values of $\vartheta_{k}$.
As a result of this summing
we obtain the ({\it total}) information
channel capacity $I$ for the parameter  $\vartheta \in V_{\vartheta} = \langle 0, 2 \, \pi)$:
\begin{eqnarray}
\label{pojemnoscEPR}
\!\!\!\!\!\!\! & & \!\!\!\!\!\!\! \!\!\!\!\!\!\!
I  \equiv  \sum\limits_{k} I
\left(\vartheta_{k}\right)  \nonumber \\
\!\!\!\!\!\!\!  & & \!\!\!\!\!\!\!\!\!
\rightarrow I = {\int\limits_{0}^{2\pi}
I \left(\vartheta\right) \, d\vartheta}
=  -  \sum\limits_{ab} {\int\limits_{0}^{2\pi} {q_{ab}(\vartheta) \, q_{ab}^{''}(\vartheta) \, d\vartheta}} \; ,
\end{eqnarray}
where  the integration appears by the reason of the substitution
of the summation over the discrete index $k$ by the integration
over the continuous set of values of the parameter  $\vartheta$.
This means that after
determining what the single $k$ channel connected with $\vartheta$ is,  we perform
the integration, which runs in accord with Eq.(\ref{statistical space S})
from $0$ to $2\pi$ in order to obtain the total information channel capacity.
The single-value-$\vartheta$ information channel capacity
(\ref{inffishEPR}) has been used in the last equality.
{\it The information channel capacity $I$ is the one which
enters into the estimation procedure of
the} EPI {\it method}. The
derivation of the form of the amplitude $q_{ab}$ as the solution of the
information principles was presented in \cite{Frieden-book};
however, the sample space for $I_{F}(\vartheta)$ and the definition of $q_{ab}$ are
different in
\cite{Frieden-book} and \cite{Comment-Frieden_ampl_q}.
\\
%
%
%

\vspace{-1mm}

\section{The information principles and generating equation}

\label{Informacja strukturalna EPR}

\vspace{-1mm}

\subsection{The general form of the information principles}

\label{general form of information principles}

\vspace{-1mm}

In \cite{Mroziakiewicz,Dziekuje informacja_1} the existence of the (total) {\it physical information} $K$,
\begin{eqnarray}
\label{physical K}
K = I + Q  \geq 0 \;
\end{eqnarray}
was postulated (see also
Sec.~\ref{Introduction}
on the Frieden-Soffer original form of the physical information
and information principles).
The choice of the intuitive condition $K \geq 0$ is connected with the
{\it expected structural information principle} of the EPI method,
\begin{eqnarray}
\label{condition from K}
I + \kappa \, Q  = 0 \; ,
\end{eqnarray}
derived for $\kappa=1$ in \cite{Dziekuje informacja_2},
where $\kappa$ is the so-called efficiency coefficient \cite{Frieden-book}.
The general form of $I(\Theta)$ was given previously by Eq.(\ref{def I dla diag IF}).
Here, $Q$ is the {\it structural information}, whose general form  is as
\cite{Dziekuje informacja_2}
\begin{eqnarray}
\label{Q dla niezaleznych Yn w d4y}
Q = \int_{\cal B} dy\, \textit{q} = \sum_{n=1}^{N} \int_{\cal
Y}  d{\bf y}_{n} \, p_{n}({\bf y}_{n}|\theta_{n}) \,
{\rm q^{F}}_{\!\!n}\,( q_{n}({\bf y}_{n}))  \; ,
\end{eqnarray}
where ${\rm q^{F}}_{\!\!n}\,( q_{n}({\bf y}_{n})) \equiv {\rm q^{F}}_{\!\!nn}\,( q_{n}({\bf y}_{n}))$
is the {\it observed structural information} \cite{Dziekuje informacja_2} under the assumption
that the variables $Y_{n}$ of the sample $\widetilde{Y}$ are independent and
$p_{n}({\bf y}_{n}|\theta_{n})$ does not depend on the parameter
$\theta_{m}$ for $m \neq n$ (see the text at the beginning of
Sec.~\ref{Pojemnosc informacyjna zagadnienia EPR}).
Then, the {\it observed structural information matrix} is equal
to ${\rm q^{F}} = {\rm diag}({\rm q^{F}}_{\!\!n}\,( q_{n}({\bf y}_{n})))$.
The quantity $\textit{q} :=  P(\Theta) \; \sum_{n=1}^{N} {\rm q^{F}}_{\!\!n}( q_{n}({\bf y}_{n}))$
in Eq.(\ref{Q dla niezaleznych Yn w d4y})  is the {\it structural information density}  \cite{Dziekuje za skrypt,Dziekuje informacja_2,Dziekuje za channel}.
%
%
%
The form of the information principle, which is more fundamental than
(\ref{condition from K}), is the {\it observed structural information
principle} that has the form $ {\rm q^{F}} + {\rm i^{F}} = 0$
(derived for the EPR-Bohm problem in the following section).
This follows from the analyticity of the logarithm of the
likelihood function, which allows for its Taylor
expansion in the neighborhood of the true value of the vector
parameter $\Theta \equiv (\theta_{n})_{n=1}^{N}$ \cite{Dziekuje
informacja_2} and in information densities it reads \cite{Dziekuje
informacja_2,Dziekuje za skrypt}
\begin{eqnarray}
\label{observed general struct inf princ}
\textit{i} +  \kappa \, \textit{q} = 0 \;  ,
\end{eqnarray}
where $\textit{i}$ and $\textit{q}$
are the information channel density  [see Eq.(\ref{def I dla diag IF})]
and structural information density, respectively.\\
It has to be stressed that it is the
{\it (modified)
observed structural information principle} (and not the expected one),
which is the one that is solved self-consistently together with the {\it variational information
principle} \cite{Frieden-book,Dziekuje za skrypt},
\begin{eqnarray}
\label{variational inf princ general}
\delta(I + Q) = 0 \;  ,
\end{eqnarray}
which for the EPR-Bohm problem is introduced in Eq.(\ref{zskalarnaEPR}).
%

%
%
[{\it The modified observed structural information principle.} The observed structural information principle (\ref{observed general struct inf princ}) and the
modified observed structural information principle
$\widetilde{\textit{i}} + C + \kappa \, \textit{q} = 0$ are connected, and
they are equivalent under the integral [in the sense that both of them lead to
the expected structural information principle (\ref{condition from K})], i.e.,
\begin{eqnarray}
\label{rownowaznosc strukt i zmodyfikowanego strukt}
\int_{\cal B} \! dy \, ( \, \widetilde{\textit{i}}  + C + \kappa \, \textit{q} )
= 0 \; \;\; & \Leftrightarrow & \;\;\; \int_{\cal B} \! dy \,
( \textit{i} + \kappa \, \textit{q} ) = 0  \; \;\; \Rightarrow \nonumber \\
&\Rightarrow& \;\;\; I + \kappa \, Q = 0 \; ,
\end{eqnarray}
where the integration is over the entire sample space ${\cal B}$ and $C$ is a constant.
The transition from the second to the first integral
in Eq.(\ref{rownowaznosc strukt i
zmodyfikowanego strukt}) is due to Eq.(\ref{reg and norm general})
and other transformations, which are equivalent under the integral
\cite{Frieden-book,Dziekuje za skrypt}].

\subsection{The information principles for the EPR-Bohm problem}

\label{EPR-Bohm information principles}

%
%

\subsubsection{The structural information principle}

\label{structural information principle}

\vspace{-1mm}

After the assignments (\ref{N oraz wartosci y w funkcji wiaryg dla EPR})
and integration over all of the possible values of $\vartheta$ [similar to that in
Eq.(\ref{pojemnoscEPR}) for $I$], the general form of the structural information
$Q$ given by Eq.(\ref{Q dla niezaleznych Yn w d4y})
results in the following form of the (total)
structural information in the EPR-Bohm problem
for the system described by the set of amplitudes $q_{ab}$:
\begin{eqnarray}
\label{strukturalnaEPR}
Q  \equiv \frac{1}{4}
\sum\limits_{ab}{\int\limits_{0}^{2\pi}{
q_{ab}^{2}(\vartheta) \, {\rm q^{F}}_{\!\!ab}(q_{ab})\, d\vartheta}} \; .
\end{eqnarray}
Now the physical information $K$, (\ref{physical K}),
in the EPR-Bohm problem is
\cite{Frieden-book,Dziekuje informacja_2}
\begin{eqnarray}
\label{TPI diag}
K = I + Q = \sum\limits_{ab} \int_{0}^{2 \pi} k_{ab}(\vartheta) \, d\vartheta \; ,
\end{eqnarray}
where $I$ is
given by Eq.(\ref{postac I analytical EPR}). \\
%
Let us notice that with this
general understanding of $K$,
the diversity of the equations of the EPI method is a consequence
of the diverse preconditions dictated by physics.
These could be, e.g., the continuity equation (which by itself is
the result of a statistical estimation
\cite{Dziekuje informacja_2,Dziekuje za skrypt}) and  some symmetries
that are characteristic for the phenomena and the normalization conditions
\cite{Frieden-book,PPSV}.

In Eq.(\ref{TPI diag}), $k_{ab}(\vartheta)$ is the {\it density of the
physical information}, which according to Eqs.(\ref{postac I analytical EPR}), (\ref{strukturalnaEPR}) and
(\ref{P na q}) is equal
to \cite{Comment-Frieden-k-dens}
\begin{eqnarray}
\label{k EPR}
k_{ab}(\vartheta) &=&
- \, \frac{1}{2} \, q_{ab} { q_{ab}^{''} } + \frac{1}{2} (q_{ab}^{'})^2
+ \frac{1}{4} \, q_{ab}^{2} \,  {\rm q^{F}}_{\!\!ab}(q_{ab})    \nonumber \\
&=&   - \frac{1}{2} \, q_{ab} {
q_{ab}^{''} }  + \frac{1}{4} \, q_{ab}^{2} \, \widetilde{{\rm q^{F}}}_{\!\!ab}(q_{ab})  \;\;\;
\nonumber \\
& & {\rm for
\; every }\; S_{ab} \in \Omega_{ab}  \; ,
\end{eqnarray}
where the {\it modified observed structural information} $\widetilde{{\rm q^{F}}}_{\!\!ab}$ used in the EPI method has been introduced:
\begin{eqnarray}
\label{qF in EPI}
\;\;\; \widetilde{{\rm q^{F}}}_{\!\!ab}(q_{ab})  := \frac{2}{q_{ab}^{2}(\vartheta)}
(q_{ab}^{'})^{2}  +  {\rm q^{F}}_{\!\!ab}(q_{ab}) \; .
\end{eqnarray}

In what follows we see [compare Eq.(\ref{jEPR})] that
$\widetilde{{\rm q^{F}}}_{\!\!ab}$ is free from the first derivative
of $q_{ab}$ for the EPR-Bohm problem,
which means that $\frac{2}{q_{ab}^{2}(\vartheta)}
(q_{ab}^{'})^{2}$ cancels a
term in ${\rm q^{F}}_{\!\!ab}(q_{ab})$.\\
Let us suppose the analyticity of the log-likelihood function $\ln P\left(S_{ab}\left|{\vartheta}\right.\right)$.
After Taylor expanding $\ln P\left(S_{ab}\left|{\tilde{\vartheta}}\right.\right)$
around the true value of $\vartheta$ we obtain
\begin{eqnarray}
\label{Freiden like equation in EPR-Bohm}
\!\!\!\! & & {\rm q^{F}}_{\!\!ab}(P\left(S_{ab}\left|{\vartheta}\right.\right)) \;(\Delta\vartheta)^{2} \!  \nonumber \\
\!\!\!\! &\equiv& \!
2 \left(\ln \frac{P\left(S_{ab}\left|{\tilde{\vartheta}}\right.\right)}{ P\left(S_{ab}\left|{\vartheta}\right.\right)}  - \frac{\partial \ln P\left(S_{ab}\left|{\tilde{\vartheta}}\right.\right)}{\partial \tilde{\vartheta}}\mid_{\tilde{\vartheta}=\vartheta} \Delta\vartheta - R_{3} \right) \nonumber \\
\!\!\!\! & = &
\frac{\partial^{2} \ln P\left(S_{ab}\left|{\tilde{\vartheta}}\right.\right)}{\partial \tilde{\vartheta}^{2}}\mid_{\tilde{\vartheta}=\vartheta}
(\Delta\vartheta)^{2} \equiv - {\rm i^{F}}_{\!\!ab}(\vartheta) \; (\Delta\vartheta)^{2} \; ,
\nonumber \\
\!\!\!\!
& &   \quad\quad\quad\quad\quad\quad\quad\quad\quad \;\;\;
{\rm for \; every }\;\; S_{ab} \in \Omega_{ab} \; ,
\end{eqnarray}
where $\Delta\vartheta \equiv(\tilde{\vartheta} - \vartheta )$ and $R_{3}$ is the remainder of Taylor series.
The observed structural information ${\rm q^{F}}_{\!\!ab}$ is defined by the left-hand side (LHS) of Eq.(\ref{Freiden like equation in EPR-Bohm}) and  the right-hand side (RHS) is equal to $- {\rm i^{F}}_{\!\!ab}(\vartheta) \; (\Delta\vartheta)^{2}$, where ${\rm i^{F}}_{\!\!ab}(\vartheta)$ is the observed Fisher information on the parameter $\vartheta$ \cite{Dziekuje informacja_2}. Let us note that
after omitting $(\Delta\vartheta)^{2}$ on both sides, the observed structural information principle
${\rm i^{F}}_{\!\!ab} + {\rm q^{F}}_{\!\!ab}=0$  \cite{Dziekuje informacja_2}  for the EPR-Bohm problem is obtained.
This arises purely as a result of the analyticity of the log-likelihood function. \\
After using the denotations
of the kind $\frac{\partial^{2} \ln P(\vartheta)}{\partial \vartheta^{2}} \equiv \frac{\partial^{2} \ln P(\tilde{\vartheta})}{\partial \tilde{\vartheta}^{2}}\mid_{\tilde{\vartheta}=\vartheta}$ and
$q_{ab}^{'}(\vartheta) \equiv \frac{\partial q_{ab}(\tilde{\vartheta})}{\partial \tilde{\vartheta}}\mid_{\tilde{\vartheta}=\vartheta}$, next passing on the RHS of Eq.(\ref{Freiden like equation in EPR-Bohm}) from the derivative of $\ln P(S_{ab}|{\tilde{\vartheta}})$ to the one of $P(S_{ab}|{\tilde{\vartheta}})$ and using the relation (\ref{P na q}) on the RHS of Eq.(\ref{Freiden like equation in EPR-Bohm}),  we can rewrite this equation
as follows:
\begin{eqnarray}
\label{Freiden like equation symbol}
\!\!\!\!\!\!\! & & \!\!\!\!\!\!\! \!\!\!\!\!\!\! {\rm q^{F}}_{\!\!ab} \;(\Delta\vartheta)^{2} = \nonumber \\
\!\!\!\!\!\!\! &=&  \frac{1}{q_{ab}^{2}(\vartheta)} \left( 2 q_{ab}(\vartheta) q_{ab}^{''}(\vartheta)  -  2 (q_{ab}^{'}(\vartheta))^{2} \right) \;(\Delta\vartheta)^{2} \; .
\end{eqnarray}
Omitting $(\Delta\vartheta)^{2}$ on both sides, Eq.(\ref{Freiden like equation symbol}) can be rewritten in the form
\begin{eqnarray}
\label{Freiden like equation with qF modyfikowane}
\;\;\;\; \widetilde{{\rm q^{F}}}_{\!\!ab}(q_{ab}) &\equiv&
\left( {\rm q^{F}}_{\!\!ab}(q_{ab}) +  \frac{1}{q_{ab}^{2}(\vartheta)} \, 2  \, (q_{ab}^{'}(\vartheta))^{2} \right) = \nonumber \\
&=& \frac{1}{q_{ab}^{2}(\vartheta)}  2 \, q_{ab}(\vartheta) q_{ab}^{''}(\vartheta)   \; ,
\end{eqnarray}
where the appearance of $q_{ab}$, (\ref{inffishEPR2}), in the argument of ${\rm q^{F}}_{\!\!ab}$ means that the probability  $P(S_{ab}|{\tilde{\vartheta}})$ (and its derivatives) present in ${\rm q^{F}}_{\!\!ab}$, which is defined by Eq.(\ref{Freiden like equation in EPR-Bohm}), has been replaced with the amplitude $q_{ab}$ (and its derivatives).
Now the forms of the amplitudes $q_{ab}$ that are the solution to the EPR-Bohm problem are sought among the $sin$ and $cos$ trigonometric functions. Thus, because of the form of the RHS of the above equation, terms with the first derivative $q_{ab}^{'}(\vartheta)$ on its LHS also have to cancel each other.

{\it The modified observational structural information principle.}
Let us
rewrite Eq.(\ref{Freiden like equation with qF modyfikowane}) as
\begin{eqnarray}
\label{mikroEPR}
&-&  2 \, q_{ab}(\vartheta) q_{ab}^{''}(\vartheta) + q_{ab}^{2}(\vartheta) \, \widetilde{{\rm q^{F}}}_{\!\!ab}(q_{ab})  = 0  \;  \;\;\;\; \nonumber \\
& & {\rm for \; every }\;\; S_{ab} \in \Omega_{ab} \;  ,
\end{eqnarray}
which, because of moving the term $\frac{1}{2}\, (q_{ab}^{'})^{2}$ in Eqs.(\ref{Freiden like equation symbol}) and (\ref{Freiden like equation with qF modyfikowane}) [and as a consequence in Eq.(\ref{k EPR})] from the Fisher information part to the structural one, will be called the {\it modified observed structural information principle}
of the EPR-Bohm problem.
The LHS of Eq.(\ref{mikroEPR}) is (up to the factor $\frac{1}{4}$) the density of the
physical information $k_{ab}(\vartheta)$ given by Eq.(\ref{k EPR}).
This one is the function of the observed structural information ${\rm q^{F}}_{\!\!ab}(q_{ab})$
[which  at most can be the function of the amplitudes $q_{ab}(\vartheta)$], of the amplitudes themselves $q_{ab}(\vartheta)$,
and of
their second derivatives.
%
%

{\it The efficiency factor in the EPR-Bohm problem.}
As we mentioned in Sec.~\ref{general form of information principles},
in general, the efficiency factor $\kappa$ appears
before the density of the structural information $q$,
which in the EPR-Bohm case is equal
to
$q =  \frac{1}{4} \,q_{ab}^{2} \widetilde{{\rm q^{F}}}_{\!\!ab}(q_{ab})$.
In the EPR-Bohm case $\kappa =1$ \cite{Frieden-book};
the value $\kappa =1$ follows from the fact that except for the
information principles, no additional differential
constraints are  put upon the amplitudes $q_{ab}$.
Thus, the presented EPI model is a pure analytic
one \cite{Dziekuje informacja_2}.

Using Eq.(\ref{k EPR}) the physical information $K$, (\ref{TPI diag}),  takes the following form:
\begin{eqnarray}
\label{K analytical in EPR}
K &=& I + Q  \\
&=& \sum\limits_{ab} \int_{0}^{2 \pi} \left(- \, \frac{1}{2} \, q_{ab} {
q_{ab}^{''} }  + \frac{1}{4} \, q_{ab}^{2}(\vartheta) \, \widetilde{{\rm q^{F}}}_{\!\!ab}(q_{ab}) \right) \, d\vartheta \; . \nonumber
\end{eqnarray}
From Eq.(\ref{mikroEPR}) the expected structural information principle [see Eq.(\ref{condition from K})], for $\kappa=1$, follows:
\begin{eqnarray}
\label{expect structural IP in EPR}
I+Q=0 \; ,
\end{eqnarray}
where $I+Q$ is given by the RHS of Eq.(\ref{K analytical in EPR}).

The differential equation (\ref{mikroEPR}) is the first one from the
information principles used in the EPI method.
The second one is the variational information principle, obtained below
\cite{Frieden-book,Dziekuje informacja_1,Dziekuje informacja_2}.

\vspace{-1mm}

\subsubsection{The variational information principle}

\label{variational information principle}

\vspace{-1mm}

In order to obtain the variational information principle, we have to transform the physical information $K$, (\ref{K analytical in EPR}), into the {\it metric form}, i.e., the one quadratic in $q_{ab}^{'}$.
Therefore, after integration by parts, $K$ can be rewritten as
\begin{eqnarray}
\label{K variational in EPR}
K = I + Q = \sum\limits_{ab} \int_{0}^{2 \pi} \left( k_{ab}^{met}(\vartheta) - \frac{\textsc{c}_{ab}}{2} \,   \right) \, d\vartheta \; ,
\end{eqnarray}
where the constant $\textsc{c}_{ab}$ is equal to
\begin{eqnarray}
\label{stalaCEPR male} \textsc{c}_{ab} =
\frac{1}{{2\pi}}\left({q_{ab}\left({2\pi}\right)q_{ab}^{'}
\left({2\pi}\right)-q_{ab}\left(0\right)q_{ab}^{'}\left(0\right)}\right)
\;
\end{eqnarray}
and $k_{ab}^{met}(\vartheta)$ is the {\it metric form} of the density of the physical information:
\begin{eqnarray}
\label{k met variational in EPR}
k_{ab}^{met}(\vartheta) = \frac{1}{2} \, q_{ab}^{'2}  + \frac{1}{4} \,  q_{ab}^{2}(\vartheta) \, \widetilde{{\rm q^{F}}}_{\!\!ab}(q_{ab}) \; .
\end{eqnarray}
%
{\it The variational information principle} \cite{Frieden-book,Dziekuje informacja_1,Dziekuje informacja_2}
has the form:
\begin{eqnarray}
\label{zskalarnaEPR}
\delta_{(q_{ab})} K &\equiv&
\delta_{(q_{ab})}\left( I + Q\right) = \\
&=&
\delta_{(q_{ab})} \! \left(\,\sum_{ab} \int\limits _{0}^{2\pi}
{ ( \, k_{ab}^{met}(\vartheta) - \frac{\textsc{c}_{ab}}{2} \, ) \, d\vartheta } \right)  = 0 \; . \nonumber
\end{eqnarray}
The solution of the {\it variational} problem (\ref{zskalarnaEPR})
with respect to $q_{ab}$ is the {\it Euler-Lagrange equation}:
\begin{eqnarray}
\label{row E-L dla EPR}
\frac{d}{d\vartheta}\left(\frac{\partial
k_{ab}^{met}(\vartheta)}{\partial
q_{ab}^{'}(\vartheta)}\right)
&=&
\frac{\partial
k_{ab}^{met}(\vartheta)}{\partial q_{ab}} \;  \;\;\;\; \\
& & {\rm for
\; every }\;\; S_{ab} \in \Omega_{ab} \;  . \nonumber
\end{eqnarray}
From  this equation and for $k_{ab}^{met}(\vartheta)$ as in Eq.(\ref{k met variational in EPR}),
the following differential equation is obtained for every amplitude $q_{ab}$:
\begin{eqnarray}
\label{rozweularaEPR}
q_{ab}^{''}=\frac{1}{2}\frac{{d (\frac{1}{2} q_{ab}^{2}  \widetilde{{\rm q^{F}}}_{\!\!ab}(q_{ab}))}}{{dq_{ab}}} \;  \;\;\;\; {\rm for
\; every }\;\; S_{ab} \in \Omega_{ab} \;  .
\end{eqnarray}
As $q_{ab}^{2}(\vartheta) \widetilde{{\rm q^{F}}}_{\!\!ab}(q_{ab})$ is
explicitly the function of $q_{ab}$ only, the total
derivative has replaced the partial derivative over $q_{ab}$ present in
Eq.(\ref{row E-L dla EPR}).

The modified observed structural information principle
(\ref{mikroEPR}) and the variational information principle
(\ref{zskalarnaEPR}) (from which the Euler-Lagrange
equation (\ref{rozweularaEPR}) follows)
serve for the derivation of the equation that generates the distribution.

\vspace{-3mm}

\subsubsection{The derivation of the generating equation}

\label{derivation of the generating equation}

\vspace{-3mm}

Using the relation (\ref{rozweularaEPR}) in Eq.(\ref{mikroEPR}), we obtain
\begin{eqnarray}
\frac{1}{2}q_{ab}\frac{{d( q_{ab}^{2}
\widetilde{{\rm q^{F}}}_{\!\!ab}(q_{ab}))}}{{dq_{ab}}}
&=&  q_{ab}^{2}
\widetilde{{\rm q^{F}}}_{\!\!ab}(q_{ab}) \;  \;\;\;\; \\
& & {\rm for
\; every }\;\; S_{ab} \in \Omega_{ab} \;  . \nonumber
\end{eqnarray}
Let us rewrite the above equation in a handier form,
\begin{eqnarray}
\frac{{2dq_{ab}}}{{q_{ab}}}=\frac{{d \left(\frac{1}{2} q_{ab}^{2}
\widetilde{{\rm q^{F}}}_{\!\!ab}(q_{ab})\right)}}{{\frac{1}{2} q_{ab}^{2}
\widetilde{{\rm q^{F}}}_{\!\!ab}(q_{ab})
}}
\; ,
\end{eqnarray}
from which, after integration on both sides, we obtain
\begin{eqnarray}
\label{jEPR}
& & \!\!\!\!\!\!\!
\frac{1}{2} q_{ab}^{2}(\vartheta)
\widetilde{{\rm q^{F}}}_{\!\!ab}(q_{ab}) =
\frac{{q_{ab}^{2}(\vartheta)}}{{A_{ab}^{2}}} \nonumber \\
& & \!\!\!\!\!\!\!
{\rm hence} \;\;\;
\widetilde{{\rm q^{F}}}_{\!\!ab}(q_{ab}) = \frac{2}{A_{ab}^{2}} \;\;\;
{\rm for
\; every }\;\; S_{ab} \in \Omega_{ab}
\; ,
\end{eqnarray}
where the constants of integration $A_{ab}^{2}$ are in general
complex numbers. This result was obtained previously in
\cite{Frieden-book}, but the arrival at the structural
information principle is different in this paper and the form
of both information principles also differs slightly.

{\it The generating equation.}  By substituting Eq.(\ref{jEPR}) into
Eq.(\ref{rozweularaEPR}), we obtain the searched for differential
{\it generating equation} for the amplitudes  $q_{ab}$ \cite{Frieden-book},
\begin{eqnarray}
\label{row generujace dla amplitud w EPR}
q_{ab}^{''}(\vartheta) &=& \frac{{q_{ab}(\vartheta)}}{{A_{ab}^{2}}}\;\;\;\;\;
\\
& &
{\rm for
\; every }\;\; S_{ab} \in \Omega_{ab} \;   \;\; {\rm and} \;\; \vartheta \in V_{\vartheta} \; ,
\nonumber
\end{eqnarray}
which is the consequence of both information principles, the structural and variational ones.
%
%
%

{\it The solution of the generating equation.} As
the amplitude $q_{ab}$ is the real one, thus $A_{ab}^{2}$ has also to
be real and it can be displayed
with the aid of the other real constant $a_{ab}$ as
$A_{ab}=a_{ab}$ or $A_{ab} = i \, a_{ab}$ \cite{Frieden-book}, where here $i$ is
the imaginary unit.
Therefore, there are two classes of solutions for Eq.(\ref{row generujace dla amplitud w EPR}).
For $A_{ab} = a_{ab}$, the solution of Eq.(\ref{row generujace dla
amplitud w EPR}) is purely of an {\it exponential} character  \cite{Frieden-book}:
\begin{eqnarray}
\label{exponential}
q_{ab}(\vartheta) &=&
B_{ab}^{''}\exp\left(-\frac{\vartheta}{a_{ab}}\right) +
C_{ab}^{''}\exp\left(\frac{\vartheta}{a_{ab}}\right) \;\;\;
\\
& & {\rm for
\; every }\;\; S_{ab} \in \Omega_{ab} \; , \nonumber
\end{eqnarray}
where the $B_{ab}^{''}$ and $C_{ab}^{''}$ constants are real.
For $A_{ab}$, which is a purely imaginary number,
\begin{eqnarray}
\label{A dla trigonometric solution}
A_{ab} = i \, a_{ab} \; ,
\end{eqnarray}
we obtain the solution of Eq.(\ref{row generujace dla amplitud w EPR}) that
is purely of a {\it trigonometric} character \cite{Frieden-book},
\begin{eqnarray}
\label{qabEPR}
q_{ab}(\vartheta) &=&
B_{ab} \sin\left(\frac{\vartheta}{a_{ab}}\right) +
C_{ab} \cos\left(\frac{\vartheta}{a_{ab}}\right) \;\;\;
\\
& & {\rm for
\; every }\;\; S_{ab} \in \Omega_{ab} \; , \nonumber
\end{eqnarray}
where $a_{ab}$, $B_{ab}$, $C_{ab}$ are the real constants.

{\it The
invariance
under the rotation.}
%
The possible values of the angle $\vartheta$ between measuring devices
range from $\langle0, 2\pi)$ (see Fig.~1).
The physical periodicity is also inferred from the geometrical symmetry
of the measuring system under the rotation of the angle  $2\pi$.
Thus, the distribution $P\left(S_{ab}\left|{\vartheta}\right.\right) = \frac{1}{4} \,
q_{ab}^{2}\left(\vartheta\right)$, (\ref{inffishEPR2}), and
every amplitude $q_{ab}(\vartheta)$ are also periodic functions of $\vartheta$.
Therefore, from the solutions (\ref{exponential}) and (\ref{qabEPR}),
we choose only the one which has a
{\it trigonometric character} [in fact, we did it below Eq.(\ref{Freiden like equation with qF modyfikowane})].
Next, the functions $sin$ and $cos$ in  Eq.(\ref{qabEPR}) form
the basis
for the probability amplitudes  $q_{ab}(\vartheta)$.
%
As $q_{ab}(\vartheta)$ is determined on the parameter space
$V_{\vartheta} = \left\langle 0, 2 \pi\right)$, thus
{\it the orthogonality
condition
of the base functions},
\begin{eqnarray}
\label{ortogonalEPR}
\!\!\!\!\!\!
\int\limits_{0}^{2\pi} \! \sin \left( \frac{\vartheta}{a_{ab}} \right) \cos \left( \frac{\vartheta}{a_{ab}} \right) d\vartheta = \frac{a_{ab}}{2} \, \sin ^2\left(\frac{2}{a_{ab}}  \, \pi \right)  = 0 \; ,
\end{eqnarray}
gives
the form of the constants $a_{ab}$ \cite{Frieden-book}
\begin{eqnarray}
\label{warAEPR}
a_{ab} &=& \frac{2}{{n_{ab}}} \; ,\quad {\rm where} \; \;\; n_{ab} = \pm 1 \; {\rm or} \; \pm 2 \; {\rm or} \; ...  \;\;\; \\
& & {\rm for
\; every }\;\; S_{ab} \in \Omega_{ab} \; \nonumber
\end{eqnarray}
and in Sec.~\ref{first and second minimal} we see that only $n_{ab} = \pm 1$ (for every $S_{ab} \in \Omega_{ab}$) or $\pm 2$ (for every $S_{ab} \in \Omega_{ab}$) are permitted where the $``plus"$ or $``minus"$ signs correspond to the right-handed or left-handed polarization, respectively, of two final particles in the bipartite system.
%

{\it The requirement of the orthogonality of the amplitudes} $q_{++}(\vartheta)$ and $q_{+-}(\vartheta)$, (\ref{qabEPR}), is not invoked here.
From Sec.~\ref{Rao-Fisher metric and constants of amplitudes}, it follows that the orthogonality of
$q_{++}(\vartheta)$  and $q_{+-}(\vartheta)$ arises afterwards from
the condition of the regularity of the probability distribution.

{\it The condition of the minimal capacity $I$.} The orthogonality
condition (\ref{ortogonalEPR}) is the one that gives the possible values
of $a_{ab}$, but
its solution (\ref{warAEPR}) permits their infinite sequence.
%
%
Therefore, in \cite{Frieden-Soffer,Frieden-book}
the condition of the minimal value of
the information (kinematical) channel capacity
$I \rightarrow min$ is postulated as the one that,
first, in accordance with
the {\it additive} form (\ref{postac I koncowa w pn}) of $I>0$,
fixes the value of $N$ to $1$ and, second, also fixes
the value of $n_{ab}$ in a unique way \cite{Frieden-book}.
However, in the present paper the second consecutive value
of $I$ is also analyzed
(see Sec.~\ref{forms of the amplitude}).

[Sometimes the nonminimal
values of $I$ are also discussed as they
lead to the EPI method's models, which are
of a physical significance.
For example, the information principles of the EPI method
analyzed in the realm of classical statistical physics
\cite{Frieden-Soffer,Frieden-book}
led (for the space component of the four-momentum vector)
to the Maxwell-Boltzmann velocity law \cite{Frieden-book}, in which case
the minimal $I$ appears for $N=1$, whereas for $N>1$, nonequilibrium,
stationary solutions were obtained
(that otherwise follow from the Boltzmann transport equation) \cite{Frieden-Soffer,Frieden-book}.
For the time component of the four-momentum vector and with
$N=1$ which minimizes $I$, the Boltzmann
probability law of the equipartition of energy in the
form (\ref{exponential}) was also obtained \cite{Frieden-Soffer,Frieden-book}].

Thus, in the present paper $N=1$ and the consecutive values
of $I$ are obtained for increasing values of $|n_{ab}|$
(Sec.~\ref{forms of the amplitude}).

Now, the generating equation (\ref{row generujace dla amplitud w EPR})
allows $q_{ab}^{''}\,$ to be eliminated, thus giving the useful form
of the information channel capacity (\ref{inffishEPR}):
\begin{eqnarray}
\label{postac Iab w Aab i qab} I =
- \sum\limits_{ab}{\frac{1}{{A_{ab}^{2}}}\int\limits_{0}^{2\pi}{
q_{ab}^{2}(\vartheta) \, d\vartheta}} \; .
\end{eqnarray}
The integral in Eq.(\ref{postac Iab w Aab i qab}) is calculated as
\begin{eqnarray}
\label{piEPR}
\!\! & & \!\!\!\!
\int\limits_{0}^{2\pi} \! {q_{ab}^{2}\left(\vartheta\right) \, d\vartheta} \equiv
4 \int\limits_{0}^{2\pi} \! { P
\left({S_{ab}\left|{\vartheta} \right.} \right) \, d\vartheta }
= 4 \int\limits_{0}^{2\pi} \! { \frac{P\left(S_{ab},
\vartheta\right)}{r\left(\vartheta\right)} \, d\vartheta} = \nonumber \\
\!\!\!\!\! & & \!\!\!\!\!
=   8 \pi \int\limits_{0}^{2\pi} \! {
P\left(S_{ab}, \vartheta\right) \, d\vartheta}
=  8 \pi P \left(S_{ab}\right) = 8 \pi \, \frac{1}{4} = 2 \, \pi \; ,
\end{eqnarray}
where Eqs.(\ref{inffishEPR2}),
(\ref{P Sab warunek vartheta}), (\ref{prawdkat}), (\ref{srednia z  PSab theta}), and (\ref{laczneEPR}), respectively, have been used in the successive equalities.

Using  Eqs.(\ref{postac Iab w Aab i qab}) and (\ref{piEPR}) and  $A_{ab}
= i \, a_{ab}$, (\ref{A dla trigonometric solution}),  together with  Eq.(\ref{warAEPR}), the information channel capacity $I$ can be expressed via the constants $n_{ab}$, giving
\begin{eqnarray}
\label{I1piEPR}
I &=& - \, Q = - \sum\limits_{ab}{\frac{1}{{A_{ab}^{2}}}\int\limits _{0}^{2\pi}{ q_{ab}^{2}d\vartheta } }   \nonumber \\
&=& 2\pi\sum\limits_{ab}{\frac{1}{{a_{ab}^{2}}}} = \frac{\pi}{2}\sum\limits_{ab}{n_{ab}^{2}} \; , \\
\!\!\! & & \!\! {\rm where} \; n_{ab} = \pm 1 \; {\rm or} \; \pm 2 \; {\rm or} \; ...  \;\;\; {\rm for
\; every }\;\; S_{ab} \in \Omega_{ab} \, . \nonumber
\end{eqnarray}
Here, in the first of the above equalities, the relation $Q = - I$,
which was obtained from the expected structural information principle (\ref{expect structural IP in EPR}),
has been used.
%
%

\vspace{-1mm}

\section{The forms of the amplitude}

\label{forms of the amplitude}

\subsection{The first and second minimal $I$}

\label{first and second minimal}

\vspace{-2mm}

Let us recall only that in the entire EPI-Bohm problem analyzed in this
paper, the inner sample taken by the system alone is a $N=1$-dimensional one
(see the paragraph on {\it The condition of the minimal capacity $I$} just
above).
%
%
%

\vspace{-3mm}

\subsubsection{The spin-$\frac{1}{2}$ with $n_{ab} = \pm 1$ case of minimal $I$}

\label{first minimal}

\vspace{-2mm}

From the above condition, it follows that the minimization
condition
for $I$ will be fulfilled when \cite{Frieden-book}:
\begin{eqnarray}
\label{n_ab minimal I}
n_{ab} &=& \pm 1  ({\rm for \; every }\;\; S_{ab} \in \Omega_{ab}) \;\;  \nonumber \\
& & \Rightarrow \;\; I \rightarrow minimal \; ,
\end{eqnarray}
for arbitrary
$S_{ab}$.
According to Eq.(\ref{warAEPR}), the condition $n_{ab} = \pm 1$ corresponds to the following values of $a_{ab}$:
\begin{eqnarray}
\label{ss}
a_{ab} = \pm 2 \quad\quad {\rm for\;\; every} \;\;\; S_{ab}  \in \Omega_{ab} \; .
\end{eqnarray}
The summation in Eq.(\ref{I1piEPR}) runs over all pairs $S_{ab}$ of the spin
projections for particles 1 and 2, (\ref{4 zdarzenia EPR});
thus, for $n_{ab} = \pm 1$, we obtain the minimal value of
$I$ equal to
\begin{eqnarray}
\label{I EPR minimalna}
\!\!\!
I_{(min)}
= 2 \, \pi \; \; \;\; {\rm for} \;\; n_{ab} = \pm 1 \; .
\end{eqnarray}
Thus, the minimal value of the  information channel capacity for
the parameter $\vartheta$ is obtained.
As the amplitude  given by Eq.(\ref{qabEPR}) for $a_{ab} = \pm 2$ has the form
\begin{eqnarray}
\label{qab dla a=2} q_{ab}(\vartheta) =
B_{ab} \sin\left(\pm \frac{\vartheta}{2}\right) +
C_{ab} \cos\left(\pm \frac{\vartheta}{2}\right) \; ,
\end{eqnarray}
we notice that $q_{ab}(\vartheta + 2 \pi) = - q_{ab}(\vartheta)$
and its period  is equal to  $T_{1/2} = 4 \pi$. This means that the solution with
$n_{ab} = \pm 1$ is characteristic for the two-dimensional representation of the rotation
group operator to which the spin-$\frac{1}{2}$ particles belong.
Below we see that the EPR-Bohm problem formulas on the probabilities for the
case of the bipartite system of two spin-$\frac{1}{2}$ particles are consistent with this
finding \cite{Frieden-book}. Finally, $n_{ab} = + 1$ corresponds to the right-handed polarization of two final  electrons and similarly $n_{ab} = - 1$ corresponds to the left-handed polarization of two final electrons.

\vspace{-3mm}

\subsubsection{The spin-1 with  $n_{ab} = \pm 2$ case of second minimal $I$}

\label{second minimal}

\vspace{-1mm}

The second smallest value of $I$  is obtained for
\begin{eqnarray}
\label{n_ab second minimal I}
n_{ab} &=& \pm 2 \;\; ({\rm for \; every }\;\; S_{ab} \in \Omega_{ab})  \;\;  \nonumber \\
& & \Rightarrow \; \;\; I \rightarrow \; {\rm second} \; minimal \; ,
\end{eqnarray}
for arbitrary
$S_{ab}$. According to Eq.(\ref{warAEPR}), the
condition $n_{ab} = \pm 2$ corresponds to the following values of
$a_{ab}$:
\begin{eqnarray}
\label{ss second}
a_{ab} = \pm 1 \quad\quad {\rm for \; every }\;\; S_{ab} \in \Omega_{ab} \; .
\end{eqnarray}
The  summation in Eq.(\ref{I1piEPR}) runs over all pairs $S_{ab}$ of the spin
projections for particles 1 and 2, (\ref{4 zdarzenia EPR}), and therefore
for $n_{ab} = \pm 2$, we obtain the second minimal
$(s.min)$ value of the information channel capacity $I$  for the parameter $\vartheta$, which is
equal to
\begin{eqnarray}
\label{I EPR second minimalna}
\!\!\! \!\!\! \!\!\!
I_{(s.min)}
= 8 \pi \; \; \;\; {\rm for} \;\; n_{ab} = \pm 2 \; .
\end{eqnarray}
As the amplitude  (\ref{qabEPR}) for $a_{ab} = \pm 1$ has the form
\begin{eqnarray}
\label{qab dla a=1}
q_{ab}(\vartheta) = B_{ab} \sin\left(\pm \vartheta\right) + C_{ab} \cos\left(\pm \vartheta\right) \;
\end{eqnarray}
thus, $q_{ab}(\vartheta + 2 \pi) = q_{ab}(\vartheta)$ and
the period of the amplitude under the rotation is equal to $T_{1} = 2 \pi$.
Thus, the solution with $n_{ab} = \pm 2$ is characteristic for the bipartite system of two spin-1 particles, which belong to the three-dimensional representation of the rotation group operator.
Below, we see that the EPR-Bohm problem formulas on the probabilities for the case
of the bipartite system with spin-1 particles are also consistent with this
finding. Finally, $n_{ab} = + 2$ corresponds to the right-handed polarization of two final photons and,  similarly, $n_{ab} = - 2$ corresponds to the left-handed polarization of two final photons.

It follows from the above considerations that for $|n_{ab}| > 2$
the basic period of the amplitudes $q_{ab}$
is smaller than $2 \pi$, which would be characteristic
for bipartite systems of spin-1 massive particles
or those higher than spin-1 particles.
However, then the base space is different than $\Omega_{ab}$ given by
Eq.(\ref{4 zdarzenia EPR}) [see also
the text below Eq.(\ref{4 zdarzenia EPR})].
Therefore, we conclude that the EPI method analysis for $\Omega_{ab}$ given by
Eq.(\ref{4 zdarzenia EPR})
permits $n_{ab} = \pm 1$ or $n_{ab} = \pm 2$ only. We discuss this splitting in
Sec.~\ref{rao-Fisher metric}.
Finally, from Eqs.(\ref{qabEPR}) and (\ref{warAEPR}), we see that the amplitudes $q_{ab}$ have the form
\begin{eqnarray}
\label{qab with nab}
q_{ab}(\vartheta) &=& B_{ab} \sin\left(n_{ab}\frac{\vartheta}{2}\right) +
C_{ab} \cos\left(n_{ab}\frac{\vartheta}{2}\right) \; , \\
& &
{\rm where} \;\;  n_{ab} = \pm 1 \; {\rm or} \; \pm 2 \;\;\; \nonumber \\
& & {\rm for \; every }\;\; S_{ab} \in \Omega_{ab} \;\; {\rm and} \;\; \vartheta \in V_{\vartheta} \; . \nonumber
\end{eqnarray}

\subsection{The determination of the constants of amplitudes and the Rao-Fisher metric}

\label{Rao-Fisher metric and constants of amplitudes}

In the amplitudes  (\ref{qab with nab}) [or particularly in (\ref{qab dla a=2}) or (\ref{qab dla a=1})],
there are the constants $B_{ab}$ and $C_{ab}$, which have to be determined.

\vspace{-2mm}

\subsubsection{The restrictions from pure boundary conditions}

\label{restrictions from pure boundary conditions}

\vspace{-1mm}

In order to perform the task, we appeal at first to the
values of the joint conditional probabilities
$P\left(S_{ab}|\vartheta\right)$ for  $\vartheta=0$, which was previously determined
in Eq.(\ref{vartheta zero}) \cite{Frieden-book}.
According to Eq.(\ref{inffishEPR2}), we know that
$P\left(S_{ab}|\vartheta\right) = \frac{1}{4} \, q_{ab}^{2}$, and thus
it follows that
\begin{eqnarray}
\label{q Sab zero}
q_{ab}(\vartheta) = 0 \; \;\; {\rm if \; only} \;\;\;  P\left(S_{ab}|\vartheta\right) = 0 \; .
\end{eqnarray}
Therefore, using  Eq.(\ref{q Sab zero})
we see that for both the amplitudes given by Eq.(\ref{qab dla a=2})
for the bipartite system of the spin-$\frac{1}{2}$ particles
and for the amplitudes given by (\ref{qab dla a=1}) for the bipartite system of
the spin-1 photons, the boundary condition (\ref{vartheta zero}) in $\vartheta=0$,
$P\left(++|0\right)=P\left(--|0\right)=0$,
leads to \cite{Frieden-book}
\vspace{-1mm}
\begin{eqnarray}
\label{ce}
C_{++} = C_{--} = 0 \; .
\end{eqnarray}
Moreover, by appealing to the geometric symmetry  of the experiment,
$P\left(S_{+-}|\vartheta\right)=P\left(S_{- +}|\vartheta\right)$ and  $P\left(S_{++}|\vartheta\right)$ $=P\left(S_{--}|\vartheta\right)$, given by Eq.(\ref{symetriaEPR}), we obtain \cite{Frieden-book}
\begin{eqnarray}
\label{wspol B prim dla q w EPR}
B_{++} = B_{--}  \;\;\;  {\rm and} \;\;\; B_{+-} = B_{-+} \; ,
\end{eqnarray}
\vspace{-3mm}
and
\vspace{-2mm}
\begin{eqnarray}
\label{C+- rowne C+- EPR}
C_{+-} = C_{-+}  \; .
\end{eqnarray}
\vspace{-8mm}

\subsubsection{Restrictions from the regularity condition}

\label{restrictions from the regularity condition}

\vspace{-2mm}

After using Eqs.(\ref{ce})--(\ref{C+- rowne C+- EPR}) and
Eqs.(\ref{inffishEPR2}) and (\ref{qab with nab}), the normalization condition (\ref{normalizacja P daje wsp w qab}) reads
\begin{eqnarray}
\label{NormP z war brzeg}
& & \!\!\!\!\!\!\!\!\!\!\!\!
\sum\limits_{ab}  P\left(S_{ab}|\vartheta\right) = \frac{1}{2} \,   \left(B_{+-} \sin (\frac{n_{ab} \vartheta}{2} ) +
C_{+-} \cos (\frac{n_{ab} \vartheta}{2} )\right)^2  \nonumber \\
&+& \frac{1}{2} \, B_{++}^{2} \sin^2 (\frac{n_{ab} \vartheta}{2}) = 1 \; ,   \\
& &
{\rm where} \;  n_{ab} = \pm 1 \; {\rm or} \; \pm 2 \;\;\;
{\rm for \; every }\;\; S_{ab} \in \Omega_{ab} \; . \nonumber
\end{eqnarray}
From the above equation and the regularity condition (\ref{regularity condition}), it follows that
\begin{eqnarray}
\label{D2 NormP z war brzeg}
& & \!\!\!\!\!\!\!\!\!\!\!\!\!
\frac{\partial^{2} }{\partial\vartheta^{2}} \sum\limits_{ab}  P\left(S_{ab}|\vartheta\right)  =  \frac{1}{4} \, n_{ab}^2 \bigglb[ - 2\, B_{+-} C_{+-}  \sin (n_{ab} \vartheta) \nonumber \\
& & \!\!\!\!\!\!\!\!\! +  \left(B_{++}^{2} + B_{+-}^{2} - C_{+-}^{2} \right)
\cos (n_{ab} \vartheta)  \, \biggrb] = 0  \; ,  \;\;\;\; \vartheta \in V_{\vartheta} \; , \nonumber \\
& & \!\!\!\!\!\!\!\!\!\!\!\! {\rm where} \;\;  n_{ab} = \pm 1 \; {\rm or} \; \pm 2 \;\;\;  {\rm for \; every }\;\; S_{ab} \in \Omega_{ab}  \; .
\end{eqnarray}
This condition gives
\begin{eqnarray}
\label{rownanie dla B i C}
B_{++}^{2}
=   C_{+-}^{2}  \neq 0 \;
\end{eqnarray}
and [together with
Eq.(\ref{wspol B prim dla q w EPR})]
\begin{eqnarray}
\label{zerowanie B+-}
B_{+-} =  B_{-+} = 0 \; ,
\end{eqnarray}
as otherwise, i.e., for $C_{+-}=0$, from the conditions (\ref{ce})--(\ref{C+- rowne C+- EPR}) and
Eq.(\ref{D2 NormP z war brzeg}), we obtain the zeroing of all coefficients
$B_{ab}$ and  $C_{ab}$, which corresponds to the trivial
case of the lack of a solution to the EPR problem in the event of
not measuring any of the spin projections (i.e., no EPR-Bohm
experiment is occurring).
Thus, for the physical {\it nontrivial solution} for
the EPR-Bohm problem, we obtain
\vspace{-1mm}
\begin{eqnarray}
\label{C+- niezerowe EPR}
C \equiv C_{+-} = C_{-+} \neq 0 \; ,
\end{eqnarray}
where the equality of the coefficients follows from  Eq.(\ref{C+- rowne C+- EPR}).
Let us notice that because of Eq.(\ref{rownanie dla B i C}), the condition (\ref{C+- niezerowe EPR}), which is the condition of the existence of the nontrivial solution, means that
\begin{eqnarray}
\label{nie zerowanie B++}
B \equiv B_{++} = B_{--} \neq 0 \; ,
\end{eqnarray}
where again the condition (\ref{wspol B prim dla q w EPR}) is used in the equality.
Finally, in accordance with Eqs.\ref{rownanie dla B i C}, (\ref{C+- niezerowe EPR}) and (\ref{nie zerowanie B++}) it follows that
\vspace{-2mm}
\begin{eqnarray}
\label{B2 rowne C2}
B^{2} = C^{2} \neq 0 \; .
\end{eqnarray}
Additionally, as was mentioned in Sec.~\ref{Pojemnosc informacyjna zagadnienia EPR}, the regularity condition
that resulted in relation (\ref{D2 NormP z war brzeg}) enables the Fisher information to pass from the {\it analytical form} (\ref{postac I analytical EPR}) to the  {\it metric form} (\ref{inffishEPR-origin}).
Let us notice that the conditions (\ref{rownanie dla B i C}) and (\ref{zerowanie B+-}) also  saturate
the equality: $\frac{\partial }{\partial\vartheta} \sum\limits_{ab}
P\left(S_{ab}|\vartheta\right)  = 0$.
%
%

[{\it Frieden's quantum amplitudes.} In order to
obtain condition (\ref{zerowanie B+-}), Frieden
used another  approach \cite{Frieden-book}.
He introduced  the ``quantum amplitude''
$\psi_{ab}(\vartheta) \propto q_{ab}(\vartheta)$  of the bipartite system,
where
$q_{ab}^{2}(\vartheta) = P(S_{a}|{S_{b},\vartheta})$
(see
{\it \cite{Comment-Frieden_ampl_q}}).
Then, by the Bayes' rule, it follows that
$\psi_{ab}^{2}(\vartheta) \equiv \psi_{ab}^{2}(\vartheta|S_{ab})
\equiv  p(\vartheta|S_{ab})= \frac{P(S_{a}|S_{b}, \vartheta)
P(S_{b}) r(\vartheta)}{P(S_{a}|S_{b}) P(S_{b})}$ \cite{Frieden-book}. This
equality means that
with the appearance of the joint
configuration of spins $S_{ab}$,
the probability amplitude $\psi_{ab}(\vartheta)$ which says
that the value of the angle is equal to $\vartheta$
is proportional to the probability amplitude
$q_{ab}(\vartheta) = \sqrt{P(S_{a}|{S_{b},\vartheta})}$ \cite{Frieden-book}
of observing the spin projection $S_{a}$  of
the particle 1 under the requirement that the spin projection of
the particle 2 amounts to  $S_{b}$ and the angle is equal to $\vartheta$.
In quantum mechanics we would say that $\psi_{ab}(\vartheta)$
signifies the probability amplitude of the event that the value of
the angle is equal to  $\vartheta$ under the condition that a joint
configuration  $S_{ab}$ of spins appears. Then, Frieden
required {\it the orthogonality of quantum  amplitudes}
$\psi_{++}(\vartheta)$ and $\psi_{+-}(\vartheta)$ on $\langle 0, 2
\, \pi)$, from where the orthogonality of the
amplitudes $q_{++}$ and  $q_{+-}$ follows automatically. From this in
\cite{Frieden-book} the zeroing of $B_{+-}$
and  $B_{-+}$ that is seen in Eq.(\ref{zerowanie B+-}) follows].

\vspace{-3mm}

\subsubsection{Analysis of the Rao-Fisher metric}

\label{analysis with Rao-Fisher metric}

\vspace{-2mm}

Below we convince ourselves that the determination of the constants $B_{ab}$ and $C_{ab}$
that were obtained in Secs.~\ref{restrictions from pure boundary conditions} and
\ref{restrictions from the regularity condition} leads to the constancy of the
Rao-Fisher metric on the statistical (sub)space  ${\cal S}$.
%
%
The probability distribution of the EPR-Bohm problem (that we are looking for)
is the discrete one (\ref{spinprawdop}). It is
determined on the joint space $\Omega_{ab}$ of
the possible results
$S_{a}S_{b} \equiv S_{ab} \in
\Omega_{ab} = \{++,--,+-,-+\}$, (\ref{4 zdarzenia EPR}), and normalized to unity in
accordance with (\ref{normalizacja P daje wsp w qab}).
Let us express the double index $ab$ in a compact form, i.e.,
\begin{eqnarray}
\label{compact notation}
& & \!\!\!\!\!\!\! ab = ++,--,+-,-+ \;\;\; \nonumber \\
& & \!\!\!\!\!\!\! {\rm corresponds \; to} \;\; \; j-1 \equiv a_{b} = 0,1,2,3,
\end{eqnarray}
respectively.
The order of $ab$ can be different than $++,--,+-,-+$.

The probability amplitudes related to the distribution
(\ref{spinprawdop}) have  the following form in accordance with Eq.(\ref{inffishEPR2}):
\begin{eqnarray}
\label{rozklad na przestrzeni statystycznej EPR}
\tilde{q}_{a_{b}} \equiv q_{ab}(\vartheta) =  \pm \sqrt{ 4 \, P\left(S_{ab}|\vartheta\right)} \; .
\end{eqnarray}
%
{\it Note on the Rao-Fisher metric of the statistical (sub)space
${\cal S}$}.
%
%
For $\aleph$ results, the general relation
that determines the induced Rao-Fisher metric on the submanifold
of the $\aleph-1$-dimensional probability simplex, which is coordinatized
by a set of coordinates $(\theta^{\alpha})$, is as follows
\cite{Bengtsson_Zyczkowski} [see Section~\ref{amplitudes sphere},
Eq.(\ref{general induced g coordinatized appendix})]:
\begin{eqnarray}
\label{general induced g}
g_{\alpha\beta} =   \sum_{j=1}^{\aleph} \frac{\partial
q_{j}}{\partial \theta^{\alpha}} \frac{\partial q_{j}}{\partial
\theta^{\beta}} \; .
\end{eqnarray}
Using Eq.(\ref{qab with nab})
we see that the amplitudes $q_{ab}(\vartheta)$
have the following derivatives:
\begin{eqnarray}
\label{pochodna qab dla EPR}
\!\! \!\!\! \!\!\! \frac{\partial q_{ab}}{\partial \vartheta}
=  \frac{n_{ab}}{2}\left(B_{ab} \cos(\frac{n_{ab}\, \vartheta}{2})
- C_{ab} \sin(\frac{n_{ab}\, \vartheta}{2}) \right)  . \,
\end{eqnarray}
Thus, for
the statistical (sub)space ${\cal S}$,
(\ref{statistical space S}), coordinatized by $\vartheta$ and after making use of Eq.(\ref{rozklad
na przestrzeni statystycznej EPR}) and the above EPI result (\ref{pochodna qab dla EPR}),
the following form of the induced metric $g^{\vartheta \, \vartheta}$ on the statistical
(sub)space ${\cal S}$
is obtained (see $I_{F}(\vartheta)$, (\ref{inffishEPR-origin}), which,
in fact, defines $g^{\vartheta \, \vartheta}$ on ${\cal S}$ \cite{Amari Nagaoka book}):
\begin{eqnarray}
\label{metryka Rao-Fishera dla EPR}
& & \!\!\!\!\!\!\!\! g^{\vartheta \, \vartheta}(\vartheta)
=   \sum_{a_{b}=0}^{3} \frac{\partial \tilde{q}_{a_{b}}}{\partial \vartheta}
\frac{\partial \tilde{q}_{a_{b}}}{\partial \vartheta} =  \sum\limits_{ab} \frac{\partial
q_{ab}}{\partial \vartheta} \frac{\partial
q_{ab}}{\partial \vartheta}  \nonumber  \\
&=&   \frac{1}{4} \sum_{ab} n_{ab}^{2} \left( B_{ab}^{2} -   B_{ab} \, C_{ab} \sin(n_{ab} \, \vartheta) \right. \nonumber \\
&+& \left. \left(C_{ab}^{2} - B_{ab}^{2}\right)
\sin^{2} (\frac{n_{ab} \, \vartheta}{2})
\right)  \; ,  \\
& & \; {\rm where} \;\;  n_{ab} = \pm 1 \; {\rm or} \; \pm 2 \;\;\;
{\rm for \; every }\;\; S_{ab} \in \Omega_{ab} \; . \; \nonumber
\end{eqnarray}
After using the conditions (\ref{ce})--(\ref{C+- rowne C+- EPR}), which follow from
the boundary conditions (i.e., without using the regularity condition as
in Sec.~\ref{restrictions from the regularity condition}), the above equation reads
\begin{eqnarray}
\label{metryka Rao-Fishera dla EPR po war brzegowych}
& & \!\!\!\!\!\!\!\!
g^{\vartheta \, \vartheta}(\vartheta) = \nonumber \\
&=&  n_{ab}^{2} \left( \frac{1}{2} \, (B_{++}^{2} + B_{+-}^{2})
- \frac{1}{2} \,  (B_{+-} \,C_{+-}) \sin(n_{ab} \, \vartheta)  \right. \nonumber \\
&+& \left.  \frac{1}{2} \, (  \,C_{+-}^{2}  - B_{++}^{2} - \,B_{+-}^{2} )
\sin^{2} (\frac{n_{ab} \, \vartheta}{2}) \right)  \; ,  \\
& & \; {\rm where} \;\;  n_{ab} = \pm 1 \; {\rm or} \; \pm 2 \;\;\;
{\rm for \; every }\;\; S_{ab} \in \Omega_{ab} \; . \; \nonumber
\end{eqnarray}
Now, using the relations (\ref{rownanie dla B i C}) and (\ref{zerowanie B+-}),
which follow from the condition of regularity of the probability distribution, we obtain
\begin{eqnarray}
\label{metryka Rao-Fishera dla EPR po war brzegowych i norm}
& & \!\!\!\!\!\!\!\!
g^{\vartheta \, \vartheta}(\vartheta)
=  \frac{1}{2} \, n_{ab}^{2}  \, B^{2}  \; , \\
& & \!\!\!\!\!\!\!\! {\rm where} \;\;  n_{ab} = \pm 1 \; {\rm or} \; \pm 2 \;\;\;
{\rm for \; every }\;\; S_{ab} \in \Omega_{ab} \; , \; \nonumber
\end{eqnarray}
where the notation from Eq.(\ref{nie zerowanie B++}) is used.

\vspace{-4mm}

\subsubsection{The normalization condition to unity}

\label{normalization condition to unity}

\vspace{-3mm}

Inserting the results for the coefficients $B_{ab}$ and
$C_{ab}$ that were obtained in
Secs.~\ref{restrictions from pure boundary conditions} and \ref{restrictions from the regularity condition} into Eq.(\ref{qab with nab}),
we obtain
\begin{eqnarray}
\label{wynikqEPR}
& & \!\!\!\!\!\!\! q_{++}(\vartheta) =  B \sin\left(n_{++} \, \vartheta/2\right) , \; q_{--}(\vartheta) = B \sin\left(n_{--} \, \vartheta/2\right)  ,
\;\;\;  \nonumber \\
& &  \!\!\!\!\!\!\!  q_{-+}(\vartheta) =  C \cos\left(n_{-+} \,
\vartheta/2\right)  , \; q_{+-}(\vartheta) = C \cos\left(n_{+-} \, \vartheta/2\right)  , \nonumber
\\
& &  \!\!\!\!\!\!\!   {\rm where} \;\; n_{ab} = \pm 1 \; {\rm or} \; \pm 2 \;\;  {\rm for \; every }\;\; S_{ab} \in \Omega_{ab} \; ,
\end{eqnarray}
with the condition $B^{2} = C^{2}$ given in Eq.(\ref{B2 rowne C2}).
From Eq.(\ref{wynikqEPR}), it can be noticed that the equality of the coefficients in  relations
(\ref{C+- niezerowe EPR}) and (\ref{nie zerowanie B++}) is
a reflection of the equality of the corresponding amplitudes that
follows from the symmetry of the space reflection quantified by
Eq.(\ref{symetriaEPR}) and from using Eq.(\ref{inffishEPR2}).
Let us observe the visible orthogonality of the amplitudes $q_{++}(\vartheta)$ and $q_{+-}(\vartheta)$.
It has arisen as the result of the EPI method analysis and from the condition of the regularity of the probability distribution. It can also be noted
from the resulting property of the constancy of Rao-Fisher metric on the statistical (sub)space ${\cal S}$ of the EPR-Bohm problem discussed in Sec.~\ref{analysis with Rao-Fisher metric}.

However, we still have to determine the constants  $B$ and $C$.
From the condition of the probability  $P\left(S_{a}S_{b}|\vartheta\right)$ normalization to unity,
(\ref{normalizacja P daje wsp w qab}),  and using Eq.(\ref{inffishEPR2}), we obtain  the equation:
\begin{eqnarray}
\label{qkwa}
\frac{1}{4} \left[ q_{++}^{2}(\vartheta) + q_{--}^{2}(\vartheta) + q_{-+}^{2}(\vartheta) + q_{+-}^{2}(\vartheta) \right] = 1 \; .
\end{eqnarray}
Using Eq.(\ref{wynikqEPR}) in Eq.(\ref{qkwa}) gives:
\begin{eqnarray}
\label{rownanie norm z q2}
& & 2 \, \left( C^{2} - B^{2} \right) \cos^{2}\left(n_{ab} \, \vartheta/2 \right) + 2\,B^{2} = 4  \, ,\\
& &  {\rm where} \;\; n_{ab} = \pm 1 \; {\rm or} \; \pm 2 \;\;  {\rm for \; every }\;\; S_{ab} \in \Omega_{ab} \; . \nonumber
\end{eqnarray}
Comparing the coefficients that stand beside the appropriate functions of
$\vartheta$ on the left- and right-hand-sides of the above expression, we
obtain
\begin{eqnarray}
\label{postac B oraz C}
B^{2} = C^{2} = 2 \; ,
\end{eqnarray}
which, after putting it into Eq.(\ref{wynikqEPR}) gives the final
solution
of the EPI method for the amplitudes of the bipartite system in the EPR-Bohm problem (compare
\cite{Frieden-book}):
\begin{eqnarray}
\label{qEPR}
& & q_{++}(\vartheta) =  \pm \sqrt{2} \, \sin\left(n_{++} \, \vartheta/2 \right) \; , \;\; \nonumber \\
& & q_{--}(\vartheta) = \pm \sqrt{2} \, \sin\left(n_{--} \, \vartheta/2 \right) \; , \; \nonumber \\
& & q_{-+}(\vartheta) =  \pm \sqrt{2} \, \cos\left(n_{-+} \, \vartheta/2 \right) \; ,
\nonumber \\
& & q_{+-}(\vartheta) = \pm \sqrt{2} \, \cos\left(n_{+-} \, \vartheta/2 \right) \, ,\\
& &  {\rm where} \;\; n_{ab} = \pm 1 \; {\rm or} \; \pm 2 \;\;  {\rm for \; every }\;\; S_{ab} \in \Omega_{ab} \;\, . \nonumber
\end{eqnarray}
It is well known (see, e.g., \cite{Manoukian}) that each of the EPR-Bohm solutions (\ref{qEPR}) can be uniquely decomposed into the inner products for
either the antisymmetric or symmetric tensor product only of two one-particle states (final and initial of the detected particles), which are the electronic ones that are given by spinors or photonic ones given by vectors, respectively.
What we have obtained is that the quantization of the spin projection on
the particular direction in space, which was introduced as the boundary condition
in Sec.~\ref{determination of the spin projection}, results in the
rediscovery of the dimension of the rotation group representation
to which the particles belong, which is consistent with
the EPI solution
obtained for the bipartite system of two spin-$\frac{1}{2}$ particles
(Sec.~\ref{first minimal} for $n_{ab}=\pm 1$) or for the bipartite system of two spin-1 photons (Sec.~\ref{second minimal} for $n_{ab}=\pm 2$).
Each of the amplitudes of the EPR-Bohm problem is a point on the
3-sphere $S^{3}$ of radius 2 [see Eq.(\ref{qkwa}) and
Sec.~\ref{amplitudes sphere}].

After the transformation $q_{ab} \rightarrow q_{ab}^{S^{3}} \equiv \frac{q_{ab}}{2}$ of the  amplitudes, the sphere $S^{3}$ becomes the one of the radius~1. The isometry group of the
invariant metric on this sphere is $SO(4)$, which is isomorphic to the coset of the group product $SU(2) \times SU(2)/\mathbb{Z}_{2}$, where $\mathbb{Z}_{2} = \{{\bf 1}, \, - {\bf 1}\}$ is the cyclic group of the order 2 \cite{Bengtsson_Zyczkowski}.

Let us recall that the amplitudes (\ref{qEPR}) were obtained as the solution of the generating equation Eq.(\ref{row generujace dla amplitud w EPR}), which after using Eqs.(\ref{A dla trigonometric solution}) and (\ref{warAEPR}) has the following form:
\vspace{-4mm}
\begin{eqnarray}
\label{row generujace dla amplitud w EPR dla nab}
& & \!\!\!\!\!\!\!
q_{ab}^{''}(\vartheta) = - \,\frac{n_{ab}^{2}}{4} \, q_{ab}(\vartheta) \; , \;\;\;
{\rm where} \;\;  n_{ab} = \pm 1 \; {\rm or} \; \pm 2 \;\;\; \nonumber  \\
& &
\!\!\!\!\!\!\!
{\rm for \; every }\;\; S_{ab} \in \Omega_{ab} \;\; {\rm and} \;\; \vartheta \in V_{\vartheta} \;  .
\end{eqnarray}
Some comments on the possible connection of Eq.(\ref{row generujace dla amplitud w EPR dla nab}) with other physical equations are given in the Conclusion.

{\it The result on the probability in the EPR-Bohm experiment.}
Putting the amplitudes (\ref{qEPR}) into the relation
$P\left(S_{ab}|\vartheta\right) = \frac{1}{4} \, q_{ab}^{2}(\vartheta) \,$, (\ref{inffishEPR2}),
gives
the joint probability of getting a particular combination of
spins projections at a fixed value of the angle  $\vartheta$:
\begin{eqnarray}
\label{wynikEPR}
& &
P\left(++|\vartheta\right) = \frac{1}{2}\sin^{2}\left(n_{++} \, \vartheta/2 \right) \; , \; \;\;  \nonumber \\
& & P\left(--|\vartheta\right) = \frac{1}{2}\sin^{2}\left(n_{--} \, \vartheta/2 \right) \; , \; \nonumber \\
& &
P\left(-+|\vartheta\right) = \frac{1}{2}\cos^{2}\left(n_{-+} \, \vartheta/2 \right) \; , \; \;\;  \nonumber \\
& & P\left(+-|\vartheta\right) = \frac{1}{2}\cos^{2}\left(n_{+-} \, \vartheta/2 \right) \; ,\\
& &  {\rm where} \;\; n_{ab} = \pm 1 \; {\rm or} \; \pm 2 \;\;  {\rm for \; every }\;\; S_{ab} \in \Omega_{ab} \;\, , \nonumber
\end{eqnarray}
{\it which is also the prediction of quantum mechanics}
\cite{Manoukian}  both for the system of two spin-$\frac{1}{2}$ particles
($n_{ab} = \pm 1$) and also for the system of two spin-1 photons ($n_{ab} = \pm 2$).

\vspace{-4mm}

\subsubsection{The consistence of the Rao-Fisher metric on ${\cal S}$
and the one of the EPI method}

\label{rao-Fisher metric}

\vspace{-1mm}

By inserting Eq.(\ref{postac B oraz C}) into Eq.(\ref{metryka Rao-Fishera dla EPR po war brzegowych i norm}),
the precise form of the
Rao-Fisher metric $g^{\vartheta \vartheta}$
is obtained:
\begin{eqnarray}
\label{metryka w EPR}
g^{\vartheta \, \vartheta}(\vartheta) & = & g^{\vartheta \, \vartheta} = n_{ab}^{2} = const. \;  \;\; {\rm for} \;\;\; \vartheta \in  V_{\vartheta} \; , \\
& & \!\!
{\rm where} \;\; n_{ab} = \pm 1 \; {\rm or} \; \pm 2 \;\;  {\rm for \; every }\;\; S_{ab} \in \Omega_{ab}  \; . \nonumber
\end{eqnarray}
As the statistical (sub)space ${\cal S}$ is one-dimensional, the only $g^{\vartheta \, \vartheta}$ component is the Fisher information $I_{F}(\vartheta)$ on the parameter
$\vartheta$ \cite{Amari Nagaoka book}.
%
%
Now, the equality of $I_{F}(\vartheta)$, (\ref{inffishEPR-origin}), and
$g^{\vartheta \vartheta}$ given by
(the first line in) (\ref{metryka Rao-Fishera dla EPR}) is by no means trivial.
This equality means that the Rao-Fisher metric
$g^{\vartheta \vartheta}=I_{F}(\vartheta)$ on ${\cal S}$
was obtained dynamically by the EPI method, which uses the
information principles: the modified observed structural
one (\ref{mikroEPR}) and the variational one
(\ref{rozweularaEPR})
described in Sec.~\ref{EPR-Bohm information principles}
\cite{Dziekuje za skrypt}.
From the above description of the EPI method for the EPR-Bohm problem  \cite{Frieden-book},
we know that the Fisher information $I_{F}(\vartheta)$, (\ref{inffishEPR-origin}),
is connected with the intrinsic, $N=1$-dimensional
sampling performed by the system alone.
It enters into the information channel capacity
$I$ [compare Eq.(\ref{IF dla vartheta w EPR})],
which (i.e., primarily, its {\it analytical form}) is then  itself or {\it by means of its  density}
used in the EPI
statistical nonparametric estimation.
Thus,  let us write the final form of the
Fisher information on the parameter $\vartheta$ that is
inherent for the EPI method and calculated in accordance with
Eq.(\ref{inffishEPR})
with the amplitudes (\ref{qEPR}) that are the
solution to the EPR-Bohm problem:
\begin{eqnarray}
\label{inffishEPR-result-origin}
I_{F}(\vartheta) &=&   -  \sum\limits_{ab} q_{ab}(\vartheta) q_{ab}^{''}(\vartheta)   = n_{ab}^{2} \, , \\
& &  {\rm where} \;\; n_{ab} = \pm 1 \; {\rm or} \; \pm 2 \;\;  {\rm for \; every }\;\; S_{ab} \in \Omega_{ab} \;  . \nonumber
\end{eqnarray}
%
The result (\ref{inffishEPR-result-origin})
follows from the EPI analysis of the (inner) $N=1$-dimensional
sampling of the particles spins in the devices $a$ and $b$
{\it by the bipartite system alone}.
The obtained amplitudes $q_{ab}$ given by Eq.(\ref{qEPR})
describe the whole configuration of the system; i.e.,
$P\left(S_{ab}|\vartheta\right) = \frac{1}{4} \, q_{ab}^{2}(\vartheta)$
given by Eq.(\ref{wynikEPR}) is the true probability distribution from which,
in the further analysis described below, the data are taken by an
experimentalist.\\
Thus, Eq.(\ref{wynikEPR}) gives
the theoretical distribution from which the data are generated
during the sampling of the system {\it by the experimentalist from outside},
i.e., when the (outer) $\textsc{m}$-dimensional sample is taken,
e.g., by two of them from devices $a$ and $b$.
We  discuss the outer experiment
in Sec.~\ref{Niepewnosc wyznaczenia kata}.
%
The Rao-Fisher metric (\ref{metryka w EPR}),
which originates in
Eq.(\ref{metryka Rao-Fishera dla EPR}),
[and is the particular case of Eq.(\ref{general induced g})],
is openly related to the $\aleph=4$ outcomes that
can occur in devices $a$ and $b$ that are
observed by, this time, the outer observer.
%
%
%

{\it The equivalence of the analytical and metric models.}
Using Eq.(\ref{wynikEPR}), it can be checked that the {\it analytical form}
$I_{F}(\vartheta)$ calculated from
Eq.(\ref{postac I analytical EPR}) is equal to the {\it metric} one
calculated from  Eq.(\ref{inffishEPR-origin}) and to the {\it EPI method form}
(\ref{inffishEPR}) [given in Eq.(\ref{inffishEPR-result-origin})].
Thus, on the expected level, the analytical and metric models
are equivalent under the integral due to the regularity  condition (\ref{regularity condition}),
although it is the full
analytical model that is solved.
%
%

\vspace{-3mm}

\subsubsection{The generic property of the EPI solutions}

\label{generic property of the EPI solutions}

\vspace{-2mm}

Finally, let us notice that  in the EPR-Bohm problem
condition (\ref{metryka w EPR}) means the constancy of the Rao-Fisher
metric $g^{\vartheta \, \vartheta}(\vartheta)$,
i.e., the independence of $g^{\vartheta \, \vartheta}$  on the statistical (sub)space ${\cal S}$
on the value of the parameter $\vartheta$ (see Comment at the end of this section).
Its constancy follows, first, from the boundary conditions discussed in
Sec.~\ref{restrictions from pure boundary conditions} and, second, from the
regularity condition used in Sec.~\ref{restrictions from the regularity condition}. However, strictly speaking, for the constancy of $g^{\vartheta \, \vartheta}(\vartheta)$ on ${\cal S}$, the normalization {\it to unity} used in Eq.(\ref{qkwa}) is not necessary.
Thus, using Eq.(\ref{wynikqEPR}) in Eq.(\ref{qkwa}), but relaxing the condition of the normalization to unity on the RHS of Eq.(\ref{qkwa}) (and leaving the normalization to a finite value instead) and using Eq.(\ref{B2 rowne C2}), we obtain
\begin{eqnarray}
& &\!\!\!\!\!\! \!\!\!\!\!\!\!\!\!\!\!\!
2 \, B^{2} \sin^{2}\left(n_{ab} \, \vartheta/2\right)
+ 2 \, C^{2} \cos^{2}\left(n_{ab} \, \vartheta/2\right) = 2\,B^{2}  \, ,\\
& & \!\!\!\!\!\!\!\!\!\!\!\!\!\!\!\!\!\! {\rm where} \;\; n_{ab} = \pm 1 \; {\rm or} \; \pm 2 \;\;  {\rm for \; every }\;\; S_{ab} \in \Omega_{ab} \; , \nonumber
\end{eqnarray}
instead of Eq.(\ref{rownanie norm z q2}).  Thus, the normalization to unity in Eq.(\ref{NormP z war brzeg}) would have the form $\sum_{ab}  \tilde{P}\left(S_{ab}|\vartheta\right) = \frac{1}{2} \, B^{2}$ instead and would also produce the results (\ref{metryka Rao-Fishera dla EPR po war brzegowych i norm}) and (\ref{wynikqEPR}) although without value 2 established in Eq.(\ref{postac B oraz C}). Nevertheless, the ratio of $g^{\vartheta \, \vartheta}(\vartheta)$ given by Eq.(\ref{metryka Rao-Fishera dla EPR po war brzegowych i norm}) to $\sum\limits_{ab}  \tilde{P}\left(S_{ab}|\vartheta\right)$,
\vspace{-3mm}
\begin{eqnarray}
\label{stosunek metryki Rao-Fishera do P}
& & \frac{g^{\vartheta \, \vartheta}(\vartheta)}{\sum\limits_{ab}  \tilde{P}\left(S_{ab}|\vartheta\right)}
=  \frac{\frac{1}{2} \, n_{ab}^{2}  \, B^{2}}{\frac{1}{2} \,B^{2} }  = n_{ab}^{2}    \; , \\
& & {\rm where} \;\;  n_{ab} = \pm 1 \; {\rm or} \; \pm 2 \;\;\;
{\rm for \; every }\;\; S_{ab} \in \Omega_{ab} \; , \; \nonumber
\end{eqnarray}
is equal to the Fisher information (\ref{metryka w EPR}), that was obtained with the normalization to unity. Now, the point is that this ratio is a generic property of the solutions \cite{Pereira-Pereira}
of the pair of differential equations (\ref{mikroEPR}) and (\ref{rozweularaEPR}), which are two information principles. This means that
the ratio in Eq.(\ref{stosunek metryki Rao-Fishera do P}) is conserved for any disturbance of $B^{2} > 0$  from the value 2 (with the boundary conditions kept unchanged), where $B^{2} = C^{2}$, (\ref{B2 rowne C2}).
%

In Secs.~\ref{first minimal} and \ref{second minimal} we noticed that the EPI method analysis for $\Omega_{ab}$ given by
Eq.(\ref{4 zdarzenia EPR})
allows  $n_{ab} = \pm 1$ or $n_{ab} = \pm 2$ only for the bipartite system,
which transform according to the two- or three-dimensional representation of the rotation operator, respectively.
This means that the statistical (sub)space ${\cal S}$
splits into two
stable subspaces and each of these is the solution of the generating equation (\ref{row generujace dla amplitud w EPR dla nab}), or more precisely, the self-consistent solution of the pair of information principles,
whose particular forms are determined by
the observed structural information $\widetilde{{\rm q^{F}}}_{\!\!ab}(q_{ab})$, (\ref{jEPR}), and as a consequence by its expected value given by Eq.(\ref{I1piEPR}):
\vspace{-2mm}
\begin{eqnarray}
\label{Q EPR fo n 1}
\; Q = -  2 \pi \; \; ,  \;\; \;\;\; n_{ab} = \pm 1 \quad \;\;\;\; \quad   {\rm for}\; {\rm spin} \; \frac{1}{2} \; ,
\end{eqnarray}
\vspace{-7mm}
\begin{eqnarray}
\label{Q EPR fo n 2}
Q = -  8 \, \pi \; \;  , \;\; \;\; n_{ab} = \pm 2 \quad \;\;\;\; \quad   {\rm for}\; {\rm spin} \; 1 \; .
\end{eqnarray}
%
%
%
%
%
%
%

{\it Comment.} It is also easy to check that
the statistical model ${\cal S}$ is Amari $\alpha=0$ affine flat, where the $\alpha=0$ Amari
connection  \cite{Amari Nagaoka book} is the Riemannian one
with respect to the Rao-Fisher metric.
%



\vspace{-3mm}

\subsection{Discussion of the EPI result on the probabilities}

\label{EPI result and factorization}

\vspace{-1mm}

In quantum mechanics the relations (\ref{wynikEPR})
for  the bipartite system of two spin-$\frac{1}{2}$ particles
(here electrons) are obtained \cite{Manoukian} from the conservation
of total spin (the orbital term is assumed to be zero) and
the indistinguishability of the particles obeying the Fermi-Dirac statistics,
which are therefore described by an antisymmetric state under their exchange.
In the calculation of the amplitude for the process
[which in \cite{Manoukian} is the same as the one on Fig.~1(a)],
the opposite spins of the initial particles of the bipartite system
was
taken also into account \cite{Manoukian}. In the EPI formalism, relations
(\ref{wynikEPR}) result from the self-consistent solution of
the modified observed structural information principle (\ref{mikroEPR})
and the variational one (\ref{rozweularaEPR}).
Thus, the Fermi-Dirac statistics seems to be a reminiscence
of these information principles, that is,
of their solution, which is the generating equation
(\ref{row generujace dla amplitud w EPR dla nab}), and the precisely defined
boundary conditions.
Similarly, the relation (\ref{wynikEPR}) for two spin-1 photons
in the final state of the process [see Fig.~1(b)]
in quantum mechanics is obtained  \cite{Manoukian} due to the
indistinguishability of the particles that obey the Bose-Einstein
statistics. The opposite momenta of final photons also have to be
taken into account. Thus, in the case of the spin-1 photons,
the Bose-Einstein statistics (and not the
Maxwell-Boltzmann one) also seems to be a reminiscence of
the generating equation
(\ref{row generujace dla amplitud w EPR dla nab}) (and thus, of
the modified observed structural and variational
information principles) and the precise boundary conditions.

The  EPI result (\ref{wynikEPR}) for the EPR-Bohm problem
signifies that the effect of the dependance of the spins projections
changes strongly with the value of $\vartheta$.
A comparison of Eq.(\ref{wynikEPR}) with Eqs.(\ref{pol})
and (\ref{pol dla Sa}), in general, gives the inequality \cite{Frieden-book}
\begin{eqnarray}
\label{nierownosc P z brzegowymi dla zaleznosci od kata w EPR}
\!\!\!\!\!\!\! P\left(S_{ab}|\vartheta\right)\neq P\left(S_{a}|\vartheta\right)P\left(S_{b}|\vartheta\right)  \;\;  {\rm for \; every } \; S_{ab} \in \Omega_{ab}  \, ,
\end{eqnarray}
which
proceeds when the estimation of the probability distribution
with the EPI method that uses the differential equations (\ref{mikroEPR})
and  (\ref{rozweularaEPR}),
occurs.
That is, the inequality (\ref{nierownosc P z brzegowymi dla zaleznosci od kata w EPR})
arises in a situation in which
%
%
the averaging over  $\vartheta$ is not performed  \cite{Frieden-book}
and instead the final forms of the observed (microscopic) information
principles are self-consistently solved.
From the analysis in this paper, we see that the EPI method
coverage of quantum mechanics is due to the {\it analytical form} (\ref{postac I analytical EPR})
of the Fisher information on the parameter $\vartheta$.
%
%
%
%
%
%
%
%
%

When the averaging over $\vartheta$ is performed  instead, then the
factorization appears.
Indeed, in Eq.(\ref{laczneEPR}) the joint
probability $P\left({S_{ab}}\right)=1/4$ is determined.
On the other hand, from Eq.(\ref{Sb wycalkowane})
we see that $P(S_{b})=1/2$ [and analogously from Eq.(\ref{pol dla Sa})
$P(S_{a})=1/2$] and thus the relation follows:
\begin{eqnarray}
\label{niezalezne}
P\left({S_{ab}}\right)=P\left({S_{a}}\right)P\left({S_{b}}\right) \; \;\;  {\rm for \; every }\;\; S_{ab} \in \Omega_{ab} \;  .
\end{eqnarray}
This condition means the independence of the spin projection variables
$S_{a}$ and $S_{b}$ after the averaging over $\vartheta$
is performed \cite{Frieden-book}, i.e., when the ``third'' (nonrandom) variable,
which is $\vartheta$, is under control and therefore its effect is eliminated.

{\it The Ehrenfest analog.}
The above conclusion is in agreement with the Ehrenfest
theorem from which it follows
that in the case of averaging over angle $\vartheta$, the entangled
EPR-Bohm states again undergo the classical mechanics separation
and therefore the relation (\ref{niezalezne}) appears  \cite{Frieden-book}.

\vspace{-4mm}

\subsubsection{The hidden variables and the EPI result}

\label{Bell hidden variables}

\vspace{-2mm}

Factorization (\ref{niezalezne}) with probabilities depending
additionally on a set $\varsigma$ of random hidden variables with
values in a set $\Sigma$
is used when the Bell inequalities \cite{Manoukian} are derived.
The local hidden variables (LHV) theories are put to the test with them.

However, due to the Fisher information properties, the theories fall into two
classes because of the value
of the dimension $N$ of the sample \cite{Dziekuje informacja_1}.
For quantum mechanics, and also for classical field theory, $N$ is
finite (and the Fisher information is finite), whereas for classical mechanics
(for pointlike particles),
$N$ is infinite (and the Fisher information is infinite), which means that
quantum mechanics cannot be derived from classical mechanics
\cite{Dziekuje informacja_1}. Therefore, if hidden variables  $\varsigma$
have the classical mechanical meaning then the possibility of
deriving quantum mechanical probabilities from them is excluded
%
and the factorization rule (\ref{niezalezne})
that is used in the Bell tests
could result from classical field theory.
However, if classical theory, i.e., its equations of motion
follow from the
self-consistent
(differential) information principles, then the probabilities that are obtained
fulfill the relation
(\ref{nierownosc P z brzegowymi dla zaleznosci od kata w EPR})
instead and the Bell-like inequalities cannot be constructed.
Thus, it is possible that quantum mechanics can be covered by the EPI method modeling (see also Sec.~\ref{determination of the spin projection}).

\vspace{-2mm}

\section{The uncertainty of the estimation of the angle}

\label{Niepewnosc wyznaczenia kata}

\vspace{-2mm}

Below, the uncertainty of the estimation of the angle $\vartheta$
for the bipartite system of two spin-$\frac{1}{2}$ particles
\cite{Frieden-book} or two spin-1 photons
is analyzed. The
statistical analy\-sis,  which led to the EPR-Bohm result
(\ref{wynikEPR}), was performed in accordance with the EPI
postulates
\cite{Frieden-Soffer,Frieden-book} outlined in the Introduction.
(unless it takes into account also the influence of the measuring
devices \cite{Frieden-book}).

The basic
one
is the assumption that
the system by itself performs the ``inner'' sampling of the
configuration space and then,  in accordance with the information
principles, performs the estimation of the generation equation
whose solutions, after taking into account the boundary
conditions, are the amplitudes of the probability distributions.
The dimension of the ``inner'' sample
was chosen as equal  to $N=1$ [see Eq.(\ref{N oraz wartosci y w
funkcji wiaryg dla EPR})]. Four pairs (\ref{4 zdarzenia EPR}) of
the spin projection $S_{a}S_{b} \equiv S_{ab}  \in \Omega_{ab}$
are possible and the probability distribution $P(S_{ab}|\vartheta)$ of the random
variable $S_{ab}$ was summarized by Eq.(\ref{wynikEPR}).

\vspace{-3mm}

\subsection{The likelihood of the outer sample}

\label{likelihood of the outer sample}

\vspace{-2mm}

As mentioned earlier, there is also the ``outer'' sample of the dimension $\textsc{m}$,
which is taken by the researcher.
%
%
%
%
%
%
%
The distribution $P(S_{ab}|\vartheta)$ distinguished
by $\vartheta \in V_{\vartheta}$ forms the statistical
model ${\cal S}$, (\ref{statistical space S}), which is displayed above
by the EPI method \cite{Frieden-book}.
For the established $\vartheta$, the probabilities $\lambda_{a_b} \equiv
P(S_{ab}|\vartheta)$, $S_{ab} \in \Omega_{ab}$, can be perceived as four
parameters of the probability distribution on the space of events $\Omega_{ab}$.
[Since $\sum_{ab} P(S_{ab}|\vartheta) = 1$, three of them
are independent.] Below, we use the renaming of the double
index $ab = ++,--,+-,-+$
to $a_{b} = 0,1,2,3$, respectively
[in accordance with Eqs.(\ref{4 zdarzenia EPR}) and (\ref{compact notation})
and
the text just below Eq.(\ref{compact notation})].

Let us denote
the values of $S_{ab}$ as $\textsc{s}_{a_b}$ and let
the ``new'' random variable $S$ take the values $\textsc{s}_{a_b}$.
The lower index in $\textsc{s}_{a_{b}}$ signifies the
$a_{b}$-th value of $S$.
The probability distribution $P(\textsc{s}; \lambda)$  of this
random variable
is as follows:
\begin{eqnarray}
\label{rozklad modelu z norm}
P(\textsc{s}_{a_{b}}; \lambda) &=& \lambda_{a_{b}} \;\;\;
{\rm for} \;\;\; S=\textsc{s}=\textsc{s}_{a_{b}} \; , \\
& & {\rm where} \;\;\; 0 \leq a_{b} \leq 3 \, . \nonumber
\end{eqnarray}
Now, a statistical space ${\cal S}_{\Lambda}$ can be constructed:
\begin{eqnarray}
\label{model dla wewnetrznej proby z norm}
{\cal S}_{\Lambda} \! &=& \!
\{
P(\textsc{s}; \lambda) | \lambda \equiv
(\lambda_{a_b})_{a_{b}=0}^{3} \in \Lambda  \subset \mathbb{R}^{4}, \;   \lambda_{a_b} \geq
0 \; (\forall a_{b}) \, , \;  \nonumber \\
& &  \sum_{a_{b}=0}^{3} \! \lambda_{a_{b}} = 1 \} \, .
\end{eqnarray}
{\it Note on $g_{\Lambda}^{\vartheta \, \vartheta}$.}
The Rao-Fisher metric $g_{\Lambda}^{\vartheta \, \vartheta}$ on the
submanifold (coordinatized by $\vartheta$) of the statistical space ${\cal S}_{\Lambda}$
is equal to the Rao-Fisher metric (\ref{metryka Rao-Fishera dla EPR}) on the submanifold of the simplex of probabilities (\ref{wynikEPR}) coordinatized by $\vartheta$, which is the model ${\cal S}$ given by Eq.(\ref{statistical space S}) (see the Appendix);
that is, it has one component that is equal in accordance with Eq.(\ref{metryka w EPR}) [and Eq.(\ref{g th th on S Lambda}) in the Appendix] to $g^{\vartheta \, \vartheta} = n_{ab}^{2} = const.$

{\it The experimental distribution.} In an experiment, the
researcher obtains the $\textsc{m}$-dimensional sample with the
frequencies $\hat{\lambda}_{a_{b}}$ of
appearances of pairs of the spin projections $S_{ab}$, (\ref{4
zdarzenia EPR}). The frequencies $\hat{\lambda}_{a_{b}}$ are the
{\it unbiased} and {\it consistent} estimators of the probabilities
$\lambda_{a_{b}} \equiv P(S_{ab}|\vartheta)$ from Eq.(\ref{wynikEPR}), i.e.,
$E_{\lambda_{a_{b}}}[\hat{\lambda}_{a_{b}}] = \lambda_{a_{b}}$ and
(for all $\varepsilon >0$)  $\lim_{\textsc{m} \rightarrow \infty}
Pr_{\lambda_{a_{b}}}(|\hat{\lambda}_{a_{b}} - \lambda_{a_{b}}| >
\varepsilon ) = 0$, respectively.

In the above notation,
the characteristics of the distributions of the estimators $\hat{\lambda}_{a_{b}}$ of the parameters $\lambda_{a_{b}}$ ($a_{b}=0,1,2,3$) for the $\textsc{m}$-dimensional sample,
\begin{eqnarray}
\label{M-dimensional sample} \widetilde{S}_{(\textsc{m})}
\equiv (S^{1}, ..., S^{i}, ...
S^{\textsc{m}}) \; ,
\end{eqnarray}
are calculated with the joint probability distribution:
\begin{eqnarray}
\label{M-dimensional likelihood}
P(\,\widetilde{\textsc{s}}_{(\textsc{m})};\lambda) = \prod_{i=1}^{\textsc{m}} P(\textsc{s}^{i};\lambda) \; ,
\end{eqnarray}
where the random variables $S^{i}$, $i=1,2,...,\textsc{m}$, are independent.
The upper index in $\textsc{s}^{i}$ signifies the $i$-th data point in the sample
$\widetilde{\textsc{s}}_{(\textsc{m})} \equiv (\textsc{s}^{1}, ...,
\textsc{s}^{i}, ... \textsc{s}^{\textsc{m}})$.
The distribution $P(\textsc{s}^{i};\lambda)$ of
the variable $S^{i}$ for each single data point $i = 1,2,...,\textsc{m}$
is the same as the distribution $P(\textsc{s};\lambda)$
of $S$ that is given by Eq.(\ref{rozklad modelu z norm}).
The asymptotic local unbiasedness of four estimators
$\hat{\vartheta}$ of the parameter $\vartheta$ in the limit $\textsc{m}
\rightarrow \infty$ is proven below.
%
%
%
%
%

\vspace{-4mm}

\subsection{The estimator $\hat{\vartheta}$}

\label{estimator of vartheta}

\vspace{-2mm}

Usually in an experiment the angle $\vartheta$ is fixed.
The frequencies $\hat{\lambda}_{a_{b}}$ are measured and only then are
relations (\ref{wynikEPR}) tested (as, e.g., the signature of their
supposed quantum mechanical origin).
We saw that EPR-Bohm relations (\ref{wynikEPR}) appear as the result
of the self-consistent solution of two, in their origin classical,
statistical information principles with the conditions of regularity,
(\ref{D2 NormP z war brzeg}), and normalization to unity, (\ref{qkwa})
(see Secs.~\ref{restrictions from the regularity condition}--\ref{normalization condition to unity}).

The statistical approach to the EPR-Bohm problem enables
the implementation of all statistical techniques of the
investigation of the properties of the estimators.
Thus,  let us discuss the inverse problem, i.e., the quality of the
$\vartheta$ estimation via the frequencies $\hat{\lambda}_{a_{b}}$
of events $S_{ab} \in \Omega_{ab}$.
The  inverse of every function of the four given by
Eq.(\ref{wynikEPR}) allows the angle $\vartheta$ to be expressed as
depending on one probability $\lambda_{a_{b}}$, i.e.,
$\vartheta=\vartheta(\lambda_{a_{b}})$. We see that four
estimators $\hat{\vartheta}$ of $\vartheta \in V_{\vartheta}$ can be built;
one for each event $S_{ab} \in \Omega_{ab}$:
\begin{eqnarray}
\label{estymator theta EPR ++ -- +- -+}
& & \hat{\vartheta}(\hat{\lambda}_{a_{b}}) \! = \! \frac{2}{n_{ab}}
\arcsin\left(\sqrt{2 \,\hat{\lambda}_{a_{b}}}\;\right) \, ,  \;\;\; ab=++ \; {\rm or} \; --
\;\;\;  \nonumber \\
& & \hat{\vartheta}(\hat{\lambda}_{a_{b}}) \! = \! \frac{2}{n_{ab}}
\arccos\left(\sqrt{2 \,\hat{\lambda}_{a_{b}}}\;\right) \, ,  \;\;\; ab=+- \; {\rm or} \; -+    \nonumber \\
& &
{\rm where} \;\;  n_{ab} = \pm 1 \; {\rm or} \; \pm 2 \;\;\;
{\rm for \; every }\;\; S_{ab} \in \Omega_{ab} \, . \;
\end{eqnarray}
%
It will be proven below that each of these four estimators
$\hat{\vartheta}$ of the angle $\vartheta$ is asymptotically
locally unbiased \cite{Amari Nagaoka book};
i.e.,
\begin{eqnarray}
\label{locally unbiased estymator theta EPR} E_{\vartheta + \Delta
\vartheta}(\hat{\vartheta}) = \vartheta + \Delta \vartheta +
\tilde{o}(\Delta \vartheta) \;
\end{eqnarray}
at every $\vartheta$, where $\tilde{o}\left(\Delta\vartheta\right)$
is a small number of higher order in $\Delta \vartheta$.
(The proof is the same for every estimator.)

To determine
whether relation (\ref{locally unbiased estymator theta EPR})
really holds, let us Taylor expand the estimator $\hat{\vartheta}$,
which is treated as a function of the frequency $\hat{\lambda}_{a_{b}}$ for
each fixed $S_{ab}$
$\in  \Omega_{ab}$ around the corresponding point $\lambda_{a_{b}}$,
respectively,
\begin{eqnarray}
\label{Taylor for vartheta}
& & \!\!\!\!\!\!\! \hat{\vartheta}(\hat{\lambda}_{a_{b}})
=   \hat{\vartheta}\left(\lambda_{a_{b}}\right) + \left.\frac{\partial
\hat{\vartheta}(\hat{\lambda}_{a_{b}})}{\partial
\hat{\lambda}_{a_{b}}}\right|_{\lambda_{a_{b}}}
\!\!\!\!\!\! ( \, \hat{\lambda}_{a_{b}} - \lambda_{a_{b}} ) + o(\Delta\hat{\lambda}_{a_{b}})  \; \nonumber \\
& & \!\!\!\!\!\!\! = \vartheta - \left.\frac{\partial
\hat{\vartheta}(\hat{\lambda}_{a_{b}})}{\partial
\hat{\lambda}_{a_{b}}}\right|_{\lambda_{a_{b}}}
\!\!\!\!\!\! \lambda_{a_{b}} + \left.\frac{\partial
\hat{\vartheta}(\hat{\lambda}_{a_{b}})}{\partial
\hat{\lambda}_{a_{b}}}\right|_{\lambda_{a_{b}}}
\!\!\!\!\!\! \hat{\lambda}_{a_{b}} + o(\Delta\hat{\lambda}_{a_{b}})
\; ,
\end{eqnarray}
where $\hat{\vartheta}(\lambda_{a_{b}}) = \vartheta$ and $o(\Delta\hat{\lambda}_{a_{b}})$
consists of the higher terms of the expansion,
where $\Delta\hat{\lambda}_{a_{b}} \equiv (\hat{\lambda}_{a_{b}} - \lambda_{a_{b}})$.
Up to the first order, only the statistic $\hat{\lambda}_{a_{b}}$ is present.
Because
$\hat{\lambda}_{a_{b}}$ is an unbiased estimator of
$\lambda_{a_{b}}$, its expectation value at the parameter
${\widetilde{\vartheta} = \vartheta + \Delta \vartheta}$ is equal
to $\lambda_{a_{b}|\vartheta + \Delta \vartheta} \equiv
P(S_{ab}|\vartheta + \Delta \vartheta)$ and the Taylor expansion
of $E_{\vartheta + \Delta \vartheta}(\hat{\lambda}_{a_{b}})$
around $\vartheta$ gives
\vspace{-2mm}
\begin{eqnarray}
\label{expectation of P}
E_{\vartheta + \Delta \vartheta}(\hat{\lambda}_{a_{b}})
&=&  \lambda_{a_{b}|\vartheta + \Delta \vartheta} = \lambda_{a_{b}}
+ \left.\frac{\partial \lambda_{a_{b}|{\widetilde{\vartheta}}}}{\partial
{\widetilde{\vartheta}}}\right|_{\vartheta} \Delta \vartheta \nonumber \\
&+& o\left(\Delta\vartheta\right) \; ,
\end{eqnarray}
where $o\left(\Delta\vartheta\right)$ are the higher terms of the expansion.

Now, the lowest order term of the expectation value
$E_{\vartheta + \Delta \vartheta}(o(\Delta\hat{\lambda}_{a_{b}}))$
of the last term $o(\Delta\hat{\lambda}_{a_{b}})$ in Eq.(\ref{Taylor for vartheta})
is equal to
\begin{eqnarray}
\label{lowest order in oDeltaP}
& & \!\!\!\!\!\!\! \!\!\!\!\!\!\!
E_{\vartheta + \Delta \vartheta}\left(\frac{1}{2} \left.\frac{\partial^{2}
\hat{\vartheta}(\hat{\lambda}_{a_{b}})}{\partial
\hat{\lambda}_{a_{b}}^{2}}\right|_{\lambda_{a_{b}}} \!\!\!\!\!
(\Delta\hat{\lambda}_{a_{b}})^{2} \right)    \nonumber \\
& & \!\!\!\!\!\!\! = \frac{1}{2}  \left.\frac{\partial^{2}
\hat{\vartheta}(\hat{\lambda}_{a_{b}})}{\partial
\hat{\lambda}_{a_{b}}^{2}}\right|_{\lambda_{a_{b}}} \!\!\!\!\!
\sigma^{2}_{\vartheta + \Delta
\vartheta}\left(\hat{\lambda}_{a_{b}}\right) \stackrel{\textsc{m}
\rightarrow \infty}{\longrightarrow}  0 \; ,
\end{eqnarray}
where in the last line
the asymptotic property of the variance of the frequency $\hat{\lambda}_{a_{b}}$
(in the four-nomial distribution) is used.

Let us take the expectations on both sides of Eq.(\ref{Taylor for vartheta}):
\begin{eqnarray}
\label{expectation Taylor for vartheta}
& & \!\!\!\!\!\!\! \!\!\!\!\!\!\! \!\!\!\!\!\!\!
E_{\vartheta + \Delta
\vartheta}(\hat{\vartheta}(\hat{\lambda}_{a_{b}})) =
\vartheta - \left.\frac{\partial \hat{\vartheta}(\hat{\lambda}_{a_{b}})}{\partial
\hat{\lambda}_{a_{b}}}\right|_{\lambda_{a_{b}}} \lambda_{a_{b}} \nonumber \\
& & \!\!\!\!\!\!\! \!\!\!\!\!\!\! \!\!\!\!\!\!\!
+
\left.\frac{\partial \hat{\vartheta}(\hat{\lambda}_{a_{b}})}{\partial \hat{\lambda}_{a_{b}}}\right|_{\lambda_{a_{b}}} E_{\vartheta + \Delta \vartheta}(\hat{\lambda}_{a_{b}}) +  E_{\vartheta  + \Delta \vartheta}\left(o(\Delta\hat{\lambda}_{a_{b}})\right) \, .
\end{eqnarray}
After  using Eqs.(\ref{expectation of P}) and (\ref{lowest order in
oDeltaP}) in Eq.(\ref{expectation Taylor for vartheta}) and noticing
that $\left.\frac{\partial \lambda_{a_{b}|{\widetilde{\vartheta}}}}{\partial
{\widetilde{\vartheta}}}\right|_{\vartheta} = \frac{\partial \lambda_{a_{b}}}{\partial \vartheta}\,$
and
$\left.\frac{\partial
\hat{\vartheta}(\hat{\lambda}_{a_{b}})}{\partial
\hat{\lambda}_{a_{b}}}\right|_{\lambda_{a_{b}}} = \frac{\partial
\vartheta}{\partial \lambda_{a_{b}}}\,$, we obtain that, asymptotically,
\begin{eqnarray}
\label{expectation for esim vartheta}
& & \!\!\! \!\!\! \!\!\!  \!\!\! \!\!\! \!\!\!
E_{\vartheta + \Delta \vartheta}(\hat{\vartheta}(\hat{\lambda}_{a_{b}})) =
\vartheta + \left.\frac{\partial \hat{\vartheta}(\hat{\lambda}_{a_{b}})}{\partial \hat{\lambda}_{a_{b}}}\right|_{\lambda_{a_{b}}} \!\!\!\!
\frac{\partial \lambda_{a_{b}}}{\partial \vartheta} \, \Delta\vartheta \, \nonumber \\
& & \!\!\! \!\!\!  +  \left.\frac{\partial \hat{\vartheta}(\hat{\lambda}_{a_{b}})}{\partial \hat{\lambda}_{a_{b}}}\right|_{\lambda_{a_{b}}} \!\!\!\! o\left(\Delta\vartheta\right) =
\vartheta + \Delta\vartheta + \tilde{o}\left(\Delta\vartheta\right) \; ,
\end{eqnarray}
where $\tilde{o}\left(\Delta\vartheta\right) = \frac{\partial \hat{\vartheta}(\hat{\lambda}_{a_{b}})}{\partial \hat{\lambda}_{a_{b}}}|_{\lambda_{a_{b}}} o\left(\Delta\vartheta\right)$. Thus, the asymptotic local unbiasedness (\ref{locally unbiased
estymator theta EPR}) of each estimator $\hat{\vartheta}$,
(\ref{estymator theta EPR ++ -- +- -+}), of the
angle $\vartheta \in V_{\vartheta}$ is proven.
Therefore, the Rao-Cram{\'e}r inequality for each $\hat{\vartheta}
\in V_{\vartheta} = \langle 0, 2 \, \pi)$ can be asymptotically used.


\vspace{-3mm}

\subsection{The intrinsic error of the estimation of the parameter $\vartheta$ in the EPI method}

\label{intrinsic error of the estimation}

\vspace{-1mm}

The frequency $\hat{\lambda}_{a_{b}}$ has the normal distribution asymptotically
for some unknown (but arbitrary) value of $\vartheta$.
It follows that using one of the functions given by Eq.(\ref{estymator theta EPR ++ -- +- -+}),
a form of the asymptotic distribution of $\hat{\vartheta}$ can be obtained.
Therefore, the
confidence interval for the parameter $\vartheta$ can also be
numerically calculated. The question is: What is the
minimal error of $\vartheta$ estimation?

From the point of view of the experimentalist, the value $\textsc{s}_{a_b}$
of the random variable $S$ (Sec.~\ref{likelihood of the outer sample})
of pair of spin projections of the particles $1$ and $2$,
respectively, is observed in a single measurement.
The Rao-Cram{\'e}r inequality, which gives the bound on the accuracy of the
estimation of the parameters, which in our case is the angle
$\vartheta$,
uses the Fisher information on the parameters confined in the
$\textsc{m}$-dimensional sample (\ref{M-dimensional sample}) that is
taken by the experimentalist.
It is equal to $\textsc{m} \, g^{\vartheta \, \vartheta}(\vartheta)$
\cite{Amari Nagaoka book},
where  $ g^{\vartheta \, \vartheta}(\vartheta) = g^{\vartheta \,
\vartheta} = n_{ab}^{2}$, (\ref{metryka w EPR}), is the Fisher information
(see Note in Sec.~\ref{likelihood of the outer sample})
for the $\textsc{m} =1$-dimensional sample.
This situation has the following consequences.

The Rao-Cram{\'e}r inequality for the variance of each of the four estimators
$\hat{\vartheta}$ of the angle $\vartheta$ given by
Eq.(\ref{estymator theta EPR ++ -- +- -+})
has the form \cite{Pawitan}
\begin{eqnarray}
\label{R-C dla IF dla EPR}
\mathop{\sigma^{2}}\left(\hat{\vartheta}\right) \ge \frac{1}{{
\textsc{m}} \, g^{\vartheta \, \vartheta}(\vartheta)} \; .
\end{eqnarray}
Even if we do not know the form  of the estimator
$\hat{\vartheta}$, the Rao-Cram{\'e}r inequality (\ref{R-C dla IF dla EPR})
gives a lower Rao-Cram{\'e}r bound (LRCB) for its variance, only if
the estimator is unbiased. In our case, it is asymptotically unbiased.
Inserting the value of $g^{\vartheta \,\vartheta}(\vartheta)$
into the inequality (\ref{R-C dla IF dla EPR}), we obtain
\begin{eqnarray}
\label{Rao-Cramer w EPR}
\mathop{\sigma^{2}}\left(\hat{\vartheta}\right) \ge
\frac{1}{\textsc{m} \, n_{ab}^{2}} \; {\rm rad}^{\,2} \; , \; \;\;
n_{ab} = \pm 1, \pm 2
\;  .
\end{eqnarray}
In the case of the bipartite system of two spin-$\frac{1}{2}$ particles, $n_{ab} = \pm 1$, and from the relation (\ref{Rao-Cramer w EPR}), it follows that
\begin{eqnarray}
\label{Rao-Cramer w EPR elektron}
\mathop{\sigma^{2}}\left(\hat{\vartheta}\right) \ge \frac{1}{\textsc{m}} \, {\rm rad}^{\,2} \; , \end{eqnarray}
whereas for the bipartite system of two spin-$1$ photons, $n_{ab} = \pm 2$ and the relation (\ref{Rao-Cramer w EPR}) gives
\begin{eqnarray}
\label{Rao-Cramer w EPR foton}
\mathop{\sigma^{2}}\left(\hat{\vartheta}\right) \ge \frac{1}{4 \, \textsc{m}} \, {\rm rad}^{\,2} \; .
\end{eqnarray}
{\it Conclusion.}  The inequality (\ref{Rao-Cramer w EPR})
states that the observation of one pair of spin projections $\textsc{s}_{a_b}$
({\it in the light of the complete ignorance about the angle}
$\vartheta$) [see Eq.(\ref{prawdkat})] gives finite information on
$\vartheta$. The lowest Rao-Cram{\'e}r bound on the error of the angle $\vartheta$ estimation
$1/(\sqrt{\textsc{m}} \, |n_{ab}|)  \; {\rm rad}$, $n_{ab} = \pm 1, \pm 2\,$
is quite large for $\textsc{m} = 1$, which is connected with the flat
``ignorance'' function $r(\vartheta)$ given in Eq.(\ref{prawdkat})
\cite{Frieden-book}.  We can also notice that with an increase of
$|n_{ab}|$ from 1 to 2 and thus, with an increase of the spin of each
particle in the bipartite system, this estimation error decreases.
This means that if, e.g.,
the photons are observed, then the EPI method estimates $\vartheta$ more precisely
than in the case of the detection of electrons.
As usual, the error decreases with an increase of $\sqrt{\textsc{m}}$.


\vspace{-4mm}

\subsection{The Rao-Cram{\'e}r-Frieden inequality}

\label{zaszumienie pomiaru}

\vspace{-2mm}

In the measurement of the state of a system by an outer  observer,
we obtain the data that are influenced by the measuring apparatus.
The general EPI analysis
in the presence of the measurement and its noisy influence on  the
internal technical data (here $\vartheta$) could be the topic of the
separate study \cite{Frieden-book}.

In the analysis of the EPR-Bohm experiment \cite{Mroziakiewicz},
the measurement data of the spin projection
(let us, e.g., denote them by $\bar{S}_{a}$ for the particle 1)
are generated by the true values of the quantities $S_{a},S_{b},\vartheta$
with the noisy presence of the measuring apparatus.
%
In fact,
this noise arises in the Stern-Gerlach devices  (or polarizers)
$a$ and $b$, but
for other reasons than the (assumed as equal to zero) fluctuations of the spin projection (see Sec.~\ref{determination of the spin projection}).

{\it The Fisher information that takes into account the noise.}  The Fisher
information $I_{noise}$ on the parameter $\vartheta$ obtained from
the data, which takes into account the {\it noise} of the
measurement, fulfills the relation
\begin{eqnarray}
\label{I noise}
g^{\vartheta\vartheta}(\vartheta) \ge I_{noise}(\vartheta) \; .
\end{eqnarray}
Because $I_{noise}(\vartheta)$ enters into (\ref{R-C dla IF dla EPR})
in place of $I_{F}(\vartheta)$, this condition means the deterioration of the quality
of the estimation in comparison with the one
that follows from the Rao-Cram{\'e}r inequality (\ref{Rao-Cramer w EPR}).

{\it The information inequality.}
The information channel capacity
$I$ is the sum (\ref{pojemnoscEPR}) over the channels with
the corresponding
Fisher information $I_{F}(\vartheta)$,
(\ref{inffishEPR-result-origin}) [equal also to $g^{\vartheta \,
\vartheta}(\vartheta)$ in Eq.(\ref{metryka w EPR})]; therefore,
\begin{eqnarray}
\label{relacja pojemnosci i IF dla EPR}
& & \!\!\!\!\!\!\! \!\!\!\!\!\!\! \!
I_{(min)} =
2 \pi
> I_{F}(\vartheta) = g^{\vartheta \,
\vartheta}(\vartheta) = 1 \;\;\;\; {\rm for} \; n_{ab} = \pm 1 \; , \nonumber \\
\\
& & \!\!\!\!\!\!\! \!\!\!\!\!\!\! \!
I_{(s.min)} =
8 \pi
> I_{F}(\vartheta) = g^{\vartheta \,
\vartheta}(\vartheta) = 4 \;\;\;\; {\rm for} \; n_{ab} = \pm 2 \; ,
\;  \nonumber
\end{eqnarray}
where $I_{(min)}$ and $I_{(s.min)}$ are the first and second minimal values of $I$
given by Eq.(\ref{I EPR minimalna}) for $n_{ab} = \pm 1$ and by Eq.(\ref{I EPR second minimalna}) for $n_{ab} = \pm 2$, respectively.
From  Eqs.(\ref{relacja pojemnosci i IF dla
EPR}), (\ref{R-C dla IF dla EPR}), and (\ref{I noise}),
we  asymptotically obtain for an $\textsc{m}$-dimensional
sample
\begin{eqnarray}
\label{I oraz IF oraz var dla EPR}
\mathop{\sigma^{2}}\left(\hat{\vartheta}\right) \geq
\frac{1}{\textsc{m} \, I_{noise}(\vartheta)}
& \geq &
\frac{1}{\textsc{m} \, g^{\vartheta \, \vartheta}(\vartheta)} =
\frac{1}{\textsc{m} \, I_{F}(\vartheta)}  \nonumber \\
&>& \frac{1}{\textsc{m} \,
I} \; ,
\end{eqnarray}
where $I$ is equal to $I_{(min)}$ or $I_{(s.min)}$ for $n_{ab} = \pm 1$ or $n_{ab} = \pm 2$,
respectively.

[In the lack of noise, it is $1/[\textsc{m} g^{\vartheta \,
\vartheta}(\vartheta)]$ and not $1/(\textsc{m} I)$, which is
the proper lower bound on the variance of the
estimator $\hat{\vartheta}$ of $\vartheta$].

Without the anticipating contribution of the system  [described
here by $I_{F}(\vartheta)$ of the EPI method], the measurement
would be absolutely impossible.
In this sense the information
$I_{noise}(\vartheta)$ is generated by the
Fisher information $I_{F}(\vartheta)$,
(\ref{postac I analytical EPR}),
which as it is obtained by {\it the EPI estimation procedure
is the part of the information that
manifests itself in the measurement}.
Thus, (with a lack of noise) this composition of the Rao-Cram{\'e}r inequality with the
result of the EPI method asymptotically leads to the inequality
\begin{eqnarray}
\label{C-R EPR-Bohm inequality}
\mathop{\sigma^{2}}\left(\hat{\vartheta}\right) \,
I_{F}(\vartheta) \geq \frac{1}{\textsc{m}} \; .
\end{eqnarray}


\vspace{-5mm}

\subsection{The Frieden approach to $\vartheta$}

\label{Frieden approach to vartheta}

\vspace{-2mm}

In the original Frieden analysis, the parameter $\vartheta$ in
four joint conditional probabilities $P\left(S_{ab}|\vartheta
\right)$, (\ref{spinprawdop}), is, in fact, the expected one, i.e.,
$E_{\vartheta}(\hat{\vartheta}_{N=1}) = \vartheta$. Here, the
estimator $\hat{\vartheta}_{N=1}$ is an additional random
variable that characterizes the inner property of the system on
the same rights as the projections $S_{a}$ and $S_{b}$ of its
component particles 1 and 2,  and the index $N=1$ signifies
that the analysis concerns the inner sample that is taken by the system
alone. Only under this condition can the Rao-Cram{\'e}r inequality
be applied directly to the result of the EPI method on $I_{F}(\vartheta)$,
giving $\mathop{\sigma^{2}}\left(\hat{\vartheta}_{N=1}\right) \geq
1/I_{F}(\vartheta)$. It can be rewritten as follows
\cite{Frieden-book}:
\begin{eqnarray}
\label{Frieden Heisenberg-like inequality}
\mathop{\sigma^{2}}\left(\hat{\vartheta}_{N=1}\right) \,
I_{F}(\vartheta) \geq 1 \; .
\end{eqnarray}
Frieden also analyzed this kind of inequality for the
position-momentum representations and obtained a kind of
Heisenberg-like inequality \cite{Frieden-book}; therefore, it can be
called the Rao-Cram{\'e}r-Frieden  inequality. The difference
between   Eqs.(\ref{Frieden Heisenberg-like inequality}) and
(\ref{C-R EPR-Bohm inequality}) is such that
Eq.(\ref{Frieden Heisenberg-like inequality})
characterizes the inner accuracy of estimating the angle
$\vartheta$ as a result of the EPI estimation and by the system
alone, whereas Eq.(\ref{C-R EPR-Bohm inequality})
characterizes the accuracy of estimating
$\vartheta$ in the outer experiment in which data are generated by
the theoretical distribution (\ref{wynikEPR}) obtained as a
result of the EPI estimation.

\vspace{-1mm}

\section{Conclusion}

\label{The conclusions}

\vspace{-2mm}

One of the main intellectual indicators, which seemed to support the
indispensability of quantum mechanics, is the well-known EPR-Bohm relations
\cite{Manoukian} that are obtained for both spin-$\frac{1}{2}$ particles
(electrons in this paper) and for massless spin-1 photons.
In the quantum mechanical approach, they are obtained under
the indistinguishability property of particles and Fermi-Dirac (for electrons)
and the Bose-Einstein (for photons) statistics (for other conditions, see \cite{Manoukian},
which were also mentioned in Sec.~\ref{EPI result and factorization}).
In the
EPI method, the statistical formalism invented
by Frieden and Soffer,
%
%
the EPR-Bohm result (\ref{wynikEPR}) for the probability
distribution for the bipartite system of
particles, was originally derived in \cite{Frieden-book}; however,
the one presented in
this paper \cite{Dziekuje za skrypt} differs in a few points.
First, the boundary conditions (Sec.~\ref{Warunki brzegowe})
were formulated in a way that is easier to understand \cite{Mroziakiewicz}.
Second, the observed physical information
used directly in the structural information
principle
was consistently obtained from the analyticity condition
of the log-likelihood function without any jump from its
{\it analytic} to its {\it metric form}.
The generating equation (\ref{row generujace dla amplitud w EPR}), which is the
central output of the EPI method, was obtained from the observed structural and variational
information principles (\ref{mikroEPR}) and (\ref{zskalarnaEPR}), respectively.
Third, in \cite{Frieden-book} the appeal to the orthogonality property of quantum
mechanical amplitude (introduced there) was used,
whose condition as the introductory one we succeeded in avoiding.
Instead,
the regularity condition of the probability distribution
(\ref{regularity condition}) (or as a consequence, the condition of the
independence of the Rao-Fisher metric (\ref{metryka w EPR})
on the statistical (sub)space
${\cal S}$
on the value of the parameter $\vartheta$) was
used.
This (in addition to the boundary conditions) enabled
the integration constants of the general solution of the
generating equation (\ref{row generujace dla amplitud w EPR}) to be determined,
which led to the final form  (\ref{qEPR}) of EPR-Bohm amplitudes.
Then,  the well-known solution (\ref{wynikEPR}) for the probability
distribution in the EPR-Bohm problem \cite{Frieden-book} was obtained.

{\it Frieden's coverage of quantum mechanics.}
In the Frieden approach,
instead of using
Eq.(\ref{D2 NormP z war brzeg}),
the quantum mechanical amplitudes
$\psi_{ab}(\vartheta) \propto q_{ab}(\vartheta)$
are constructed \cite{Frieden-book}
(see
the paragraph {\it Frieden's quantum amplitudes}
in Sec.~\ref{restrictions from the regularity condition}
and \cite{Comment-Frieden_ampl_q}).
Together with the approach used in this paper (which does not
use the quantum mechanics notions in the EPI method estimation),
the Frieden's construction of the quantum mechanical amplitudes means
that when (inversely) defining quantum amplitudes via the classical ones,
the classical statistics coverage of the quantum mechanical approach
(even without any mention of the hidden variables)
is obtained. We call this the Frieden's ``coverage of quantum mechanics''.

Fourth, the analysis
was made more homogeneous
in the sense that the amplitude of the bipartite system
exhibits either the periodicity that is
characteristic for spin-$\frac{1}{2}$ particles or for the spin-1
massless photons [possessing the same mathematical form
(\ref{wynikEPR})]. Thus, it is slightly easier to notice that
the EPI method of the EPR-Bohm problem determines the wave function
representations than in \cite{Frieden-book}.
Fifth, this approach inevitably connects the EPR-Bohm result (\ref{wynikEPR}) with
the frequencies of the events that are registered by the experimentalist
(Sec.~\ref{Niepewnosc wyznaczenia kata}). Finally (in agreement with the previous point), the estimation of
the angle $\vartheta$ is pinned to the outer $\textsc{m}$-dimensional sample taken by the
experimentalist. In addition to these, the differences between the
Frieden approach and the one presented in this paper are pointed out in the text.

Let us note that the generating equation [written, e.g., in the form
given by Eq.(\ref{row generujace dla amplitud w EPR dla nab})]
is the stationary one, indifferently
whether two measuring
devices are close to each other or infinitely far apart.
Interestingly, Eq.(\ref{row generujace dla amplitud w EPR dla nab})
can be postulated, e.g., from the stationary telegraphic equation \cite{Ohtaka-Trigg}
for the field that is rotating with an infinite velocity in the $\vartheta$ direction
(which is formally equivalent to the Klein-Gordon equation \cite{Ohtaka-Trigg} for the
tachyon propagating in the $\vartheta$ direction).
Thus, the generating equation
describes the instantaneous twist around the propagation axis of the field
of a bipartite system at the moment of detection.
The spatial part of the field of a bipartite system is one of the amplitudes
(\ref{qEPR}) of the EPR-Bohm problem.
The lack of a finite time component is seen in the boundary conditions
and is in agreement with present-day experiments \cite{Photons-Never-Coexisted-Entanglement}.
However, this could ``gra\-vitate'' to the conclusion that the information principles (\ref{mikroEPR}) and (\ref{rozweularaEPR}), which result in the generating equation (\ref{row generujace dla amplitud w EPR dla nab}), describe the physics of the collapse of the wave function of a
bipartite system in the EPR-Bohm problem. That would provide us
with a richer theoretical structure than the quantum mechanical one, which is
based merely on the definition of the transition amplitude for the problem \cite{Manoukian}.
When this richer structure is disclosed, it
might appear that the bipartite system is an extended, composed  object, which,
when detected, knows about this event in the whole medium of its entity by the kind
of interaction that propagates inside it.
Could it be a gravitational one whose speed of propagation is experimentally unknown until the present day  \cite{speen of gravity} or an electromagnetic one whose velocity is experimentally measured only locally
and obviously outside of such compact systems as discussed in the EPR-Bohm problem?
If the medium is not known, the velocity could, in fact, even be infinite. This is in agreement with the EPR-Bohm-type experiments \cite{Photons-Never-Coexisted-Entanglement} (unless one believes in entanglement without any interaction).
%
%
%

Next, the EPI method provides
the general formalism for the
description of entangled states \cite{Frieden-Soffer,Frieden-book,Dziekuje za channel}.
This is particularly true in the case of the EPR-Bohm problem for which pairs $S_{ab}$
of spin projections
of particles 1 and 2 of the bipartite system are detected in the experiment
with observed frequencies and the entanglement concerns these frequencies.
Indeed, the outer data, which form the $\textsc{m}$-dimensional sample, are generated
from the (supposed) probability distribution (\ref{wynikEPR}), which this time is obtained
theoretically.
They are reflected in the Rao-Fisher metric
%
$g^{\vartheta\vartheta}$ of the statistical (sub)space ${\cal S}_{\Lambda}$
and the statistical model ${\cal S}$ (see
the Appendix) and
therefore in the Fisher information $I_{F}(\vartheta)$ and the information
channel capacity $I$ of the EPI method
(see Sec.~\ref{rao-Fisher metric}).

[The Rao-Fisher
metric $g^{\vartheta\vartheta}$ is equal to the Fisher
information $I_{F}(\vartheta)$, (\ref{inffishEPR-origin}),
which enters into the information channel capacity $I$, (\ref{pojemnoscEPR}),
via Eq.(\ref{IF dla vartheta w EPR}) (see Sec.~\ref{rao-Fisher metric})].

Therefore (in the case of the inseparability of $I$ into the sum of the proper subsystems'
information channel capacities and $Q$ into the corresponding
structural information terms),
the possibility of describing the entangled states follows from
the fact that the structural information principle $I = -  \, Q$,
(\ref{condition from K}), describes
the relation of the outer data of the $\textsc{m}$-dimensional sample (connected with $I$)
{\it with the unobserved configuration of the system},
which is described by structural information~$Q$.
%

In this paper, the unobserved configuration of the system depends on the angle $\vartheta$ between the particles spin projections in two measuring devices.
%
%
%
%
%
%
%
Therefore, in spite of the lack of any direct insight into
the {\it inner} properties of the system, we can
(in accordance with Sec.~\ref{Niepewnosc wyznaczenia kata}) make an inference about
the ({\it inner}) angle $\vartheta$ by simply observing the frequencies of the pairs
of the spin projections. This can be done with the finite accuracy that is also discussed
in Sec.~\ref{Niepewnosc wyznaczenia kata}.
The mere fact that the possibility of such an inference exists
determines the situation that is called the {\it entanglement of the states}
of the particles of a bipartite system that is perceived in the dependence
of the frequencies of both particle spin projections
on the inner angle $\vartheta$ of the
system \cite{Comment-Frieden-EPR}.
This is the reason why the experimentalist is able
to estimate the value of $\vartheta$. However, in a more complete
approach to the EPI-method estimation, its maximal accuracy also
has to be discussed  \cite{Dziekuje za channel}.

Recently, increasing problems with the
experimental validation of the uncertainty relation (UR)
\cite{Bang-Berger,Mu-Wu-Yang} in its quantum mechanical Heisenberg
form have been reported.~[Unless one forgets that the intrinsic quantum mechanical uncertainties of
the complementary variables on the quantum state are,
for the purpose of the verification of any
Heisenberg uncertainty principle (as it takes place for
any form of UR), always estimated experimentally.]
%
One of them is connected with an experiment of
the successive projective measurements of two noncommuting neutron
spin components
\cite{Erhart-Sponar-Sulyok-Badurek-Ozawa-Hasegawa}.
The other group is connected with
the diffraction-interferometric experiments for a photon, where both
UR and the meaning of the half-widths of a pair of functions
(time and frequency), which are related by the Fourier transform, are examined
\cite{Roychoudhuri}.
Thus, the quantum mechanical state-dependent formulation of the
simultaneous measurability of the observables (that also uses
the notion of their closer unknown noise operators)
was proposed, which led to a deep theoretical
reformulation of UR in the form of the measurement disturbance relationship
(MDR) \cite{Ozawa,Fujikawa-Umetsu,Branciard}.
However, the proper implementation of the Rao-Cram{\'e}r
inequality can also be considered \cite{Amari Nagaoka book,Dziekuje za channel}
and the call
to abandon UR on behalf of the more information oriented inequality has
already been sounded. It was discussed in
Secs.~\ref{intrinsic error of the estimation}--\ref{Frieden approach to vartheta}
for the EPR-Bohm problem.
%
%
%
%

Next, the irrelevance of the hidden variable idea for
the construction of Bell-like inequalities in the EPI method
was discussed in Sec.~\ref{Bell hidden variables} and, therefore,
we conclude that the Bell-like tests do not put quantum
theory on a pedestal at all. Conversely, the EPI method suggests that
although quantum mechanics is the experimentally reliable one,
the foundation of its underlying method could be of the statistical information
theory background. The introductory steps for the construction of
both the Maxwell and Dirac equations using the EPI method
and the quantization of the helicity of the free electromagnetic field
and spin for the Dirac field were mentioned
in the Introduction and
Sec.~\ref{determination of the spin projection}. In this paper,
using the EPI method, which follows the previous findings,
it was shown (Secs.~\ref{first minimal}--\ref{second minimal})
that  as a solution of the generating
equation with the constants of integration $n_{ab}$ equal to $\pm 1$ or $\pm 2$
for every $S_{ab} \in \Omega_{ab}$, the bipartite EPR-Bohm
amplitudes for the rotation group representation of the spin-$\frac{1}{2}$
or spin-$1$  particles, respectively, appear.
That is, the Fermi-Dirac statistics used in the case of electrons
for a quantum mechanical description of the EPR-Bohm problem
seems to be a reminiscence of the statistical information principles.
Thus, the Pauli exclusion principle may
also have a statistical information theory background.
The same statistical information background is also suggested for
the Bose-Einstein statistics that is used for a description of
the EPR-Bohm-like problem for photons.

Finally, the general mathematical thought
that is behind
the results of this paper is the self-consistency of the solution
of a proper set of partial differential equations.
%
%
After they are solved self-consistently, all degrees of freedom
are removed and what remains is one particular state of the system.
In this paper, this is a particular solution of the bipartite amplitude
of the EPR-Bohm problem that is obtained by the self-consistent
solution of the pair of (differential) information principles.
In the case of an electronic field fluctuation coupled to
its electromagnetic self-field, the self-consistent treatment
of the Dirac equation and classical Maxwell equations resulted
in the (obtained iteratively)  solution for the Lamb shift \cite{relativ Lamb}.
In the case of the self-consistent model of classical field interactions
of the electroweak model that is solved in the presence of
nonzero extended fermionic charge density fluctuations, the solution
obtained in \cite{Dziekuje-za-self-consistent-state} was a spin zero
electrically uncharged droplet, interpreted as the state of mass equal
to $\sim 126.5$ GeV, which was observed recently in an LHC experiment.
In \cite{LORD-Biesiada-Syska-Manka,Dziekuje za neutron,Dziekuje za redshift}
the dynamical compactification of a six-dimensional model of the space-time
to the four-dimensional, locally Minkowskian space-time, which resulted
from the self-consistent solution of the coupled Einstein
and Klein-Gordon equations, was presented.
%

Thus, in this paper it can also be seen that the necessity to introduce
the quantum (mechanical or field theory) approach
is the result of the
previous neglect of the self-consistency of a proper set of partial
differential equations, which seems to be the primary property of physical structures.
Finally, each of the solutions (\ref{qEPR}) of the EPR-Bohm problem describes the compound,
extended in space, state of a bipartite system.

\vspace{-4mm}

\begin{acknowledgments}
\vspace{-3mm}
This work has been supported by L.J.CH..\\
This work has been also supported by the Department of Field
Theory and Particle Physics, Institute of Physics, University of
Silesia and by the Modelling Research Institute, 40-059 Katowice,
Drzyma{\l}y 7/5, Poland.
\end{acknowledgments}

\section*{Appendix: The Rao-Fisher metrics on the probability simplex and its submanifolds}

\label{amplitudes sphere}

Let in the result of an experiment the finite number $\aleph$ of possible outcomes
$\textsc{s}_{j}$, $j=1,2,...,\aleph$ of a random variable $S$ be obtained.
They span a base space of events $\Omega$. The probabilities of the outcomes $\textsc{s}_{j}$
are $P_{j} \equiv P(\textsc{s}_{j})$ and
\vspace{-3mm}
\begin{eqnarray}
\label{rozklad prawdopod}
P_{j} \equiv P(\textsc{s}_{j})  \ge 0 \;\;\;\;  {\rm and} \;\;\;\;  \sum \limits_{j=1}^{\aleph} P_{j} = 1 \; .
\vspace{-2mm}
\end{eqnarray}
Let us choose
$\lambda_{j}:= P_{j}$ as the $j$-th component of the set $\lambda \equiv (\lambda_{j})_{j=1}^{\aleph}$
of parameters. Thus, the probability distribution $P$ is parametrized by  $\lambda$, i.e., $P_{j} \equiv P(\textsc{s}_{j}; \lambda)$.
Now, let us construct the statistical space:
\begin{eqnarray}
\label{model dla wewnetrznej proby z norm appendix}
{\cal S}_{\Lambda} &=& \{
P(\textsc{s}; \lambda) | \lambda \equiv
(\lambda_{j})_{j=1}^{\aleph} \in \Lambda \subset \mathbb{R}^{\aleph}, \; \;  \lambda_{j} \geq
0 \;\; (\forall j) \;\;  \nonumber \\
& & {\rm and} \;\;   \sum_{j=1}^{\aleph} \lambda_{j} = 1 \} \; . \nonumber
\end{eqnarray}
In the case of the EPR-Bohm problem the number of events is equal to $\aleph=4$ and the probability distribution $P(\textsc{s}_{j}; \lambda)$, (\ref{rozklad modelu z norm}), is determined by Eq.(\ref{wynikEPR}), where in Eq.(\ref{compact notation}) the following  denotation of the index $j$ was introduced: The value $j-1 \equiv a_{b} = 0,1,2,3$  corresponds to $ab = ++,--,+-,-+$, respectively.

The ($\aleph-1$){\it-simplex of the probability distributions} $P$, is defined as follows:
\begin{eqnarray}
\label{simplex N-1}
\Delta^{\aleph-1} &=& \{(P_{1}, P_{2}, ..., P_{\aleph}) \in \mathbb{R}^{\aleph} ; P_{j} \ge 0, \; j=1,..., \aleph, \; \nonumber \\
& & \;\; {\rm for} \; \sum_{j=1}^{\aleph}{P_{j}} =  1 \} \; .
\end{eqnarray}
It is the convex set, i.e. each probability vector $\vec{P} = (P_{j})_{j=1}^{\aleph} \in \Delta^{\aleph-1}$ can be expressed as $\vec{P} = \sum_{j=1}^{\aleph} \varepsilon_{j} \, P_{j}$, where $\sum_{j=1}^{\aleph} \varepsilon_{j} = 1$ \cite{Bengtsson_Zyczkowski}.
%
%

The squared  infinitesimal distance on the probability simplex $\Delta^{\aleph-1}$ is equal to
\begin{eqnarray}
\label{infinitezimal square distans}
ds^2 &=&  \sum_{j=1}^{\aleph} g^{jj} d P_{j} \, d P_{j}  = \sum_{j=1}^{\aleph} \frac{1}{P_{j}} d P_{j} \, d P_{j}  \; ,
\end{eqnarray}
where the diagonality of the Rao-Fisher metric,
\begin{eqnarray}
\label{g on simplex}
(g^{ij}) = {\rm diag}\left( \frac{\delta^{ij}}{P_{j}} \right) \; ,
\end{eqnarray}
on the $\aleph-1$ simplex $\Delta^{\aleph-1}$ \cite{Bengtsson_Zyczkowski} was
used.

Let us perform the transformation $P_{j} \rightarrow q_{j}$ given by $4 \, P_{j} = q_{j}^2$, $j=1,...,\aleph$, from probabilities to amplitudes
[compare Eq.(\ref{rozklad na przestrzeni statystycznej EPR})].
%
Thus, on the coordinatized by a set of $d$ parameters $\Theta=(\theta^{\alpha})_{\alpha=1}^{d}$
statistical (sub)space of the ($\aleph-1$)-dimensional probability simplex $\Delta^{\aleph-1}$,
the metric $g^{ij}$ induces the Rao-Fisher metric $g_{\alpha\beta}$, which has the form \cite{Bengtsson_Zyczkowski}
\begin{eqnarray}
\label{general induced g coordinatized appendix}
g_{\alpha\beta} &=&  \sum_{j=1}^{\aleph} P_{j} \frac{\partial \ln P_{j}}{\partial \theta^{\alpha}} \frac{\partial \ln P_{j}}{\partial \theta^{\beta}} =  \sum_{j=1}^{\aleph} \frac{1}{P_{j}} \frac{\partial
P_{j}}{\partial \theta^{\alpha}} \frac{\partial P_{j}}{\partial
\theta^{\beta}}  \nonumber \\
& \equiv &
\sum_{i,j=1}^{\aleph} g^{ij} \frac{\partial
P_{i}}{\partial \theta^{\alpha}} \frac{\partial P_{j}}{\partial
\theta^{\beta}}  = \sum_{j=1}^{\aleph} \frac{\partial
q_{j}}{\partial \theta^{\alpha}} \frac{\partial q_{j}}{\partial
\theta^{\beta}}   \; ,
\end{eqnarray}
where in the second line the diagonality, (\ref{g on simplex}), of the Rao-Fisher metric $g^{ij}$ on the $\aleph-1$ simplex, was emphasized.
%

Due to
the normalization $\sum_{j=1}^{\aleph} P_{j} = 1$, it follows that
[compare Eqs.(\ref{normalizacja P daje wsp w qab}) and (\ref{qkwa})]
\begin{eqnarray}
\label{sfera q o prom 2}
\sum_{j=1}^{\aleph} q_{j} q_{j} = 4 \; .
\end{eqnarray}
Thus, we notice that the amplitudes' sphere is of the radius~2.
%

{\it For example}, for the EPR-Bohm problem, in this paper the statistical (sub)space ${\cal S}$, (\ref{statistical space S}), with $d=1$ parameter $\vartheta$ is investigated:
\begin{eqnarray}
\label{statistical space S appendix}
{\cal S} =
\{
P\left(S_{ab}|\vartheta\right)|\;\;
\vartheta \in \langle 0 \; 2 \,\pi) \equiv V_{\vartheta} \subset
\mathbb{R}^{1} \} \; . \nonumber
\end{eqnarray}
The model ${\cal S}$ is coordinatized by parameter $\vartheta$, one-dimensional submanifold of
$\aleph-1=3$-dimensional probability simplex $\Delta^{3}$.
Thus, the Rao-Fisher metric (\ref{general induced g coordinatized appendix}) on ${\cal S}$ of the EPR-Bohm problem, is equal to:
\begin{eqnarray}
\label{metryka Rao-Fishera dla EPR appendix}
g^{\vartheta \, \vartheta}(\vartheta) = \sum_{j=1}^{\aleph=4} \frac{\partial \tilde{q}_{j}}{\partial \vartheta} \frac{\partial \tilde{q}_{j}}{\partial \vartheta} \; , \; \nonumber
\end{eqnarray}
where $\tilde{q}_{j} \equiv q_{ab}(\vartheta)$ are defined in (\ref{rozklad na przestrzeni statystycznej EPR}).
After the calculations presented in Sec.~\ref{Rao-Fisher metric and constants of amplitudes},
the Rao-Fisher metric (\ref{metryka w EPR}) on ${\cal S}$ appeared to be constant, i.e.  $g^{\vartheta \, \vartheta} = n_{ab}^{2} = const.$, where $n_{ab} = \pm 1$ or $\pm 2\,$ for the bipartite state of electrons or photons, respectively.

Now, let us determine the Rao-Fisher metric $g_{\Lambda}^{kl}$ on the statistical (sub)space ${\cal S}_{\Lambda}$, (\ref{model dla wewnetrznej proby z norm appendix}), of the ($\aleph-1$)-dimensional probability simplex. Knowing, in accordance with Eq.(\ref{general induced g coordinatized appendix}), that it is coordinatized by the set of parameters $\lambda \equiv (\lambda_{k})_{k=1}^{d=\aleph}$,
we obtain
\begin{eqnarray}
\label{general induced g on S Lambda appendix}
g^{kl}_{\Lambda} &=&  \sum_{i,j=1}^{\aleph} g^{ij} \frac{\partial P_{i}}{\partial \lambda_{k}} \frac{\partial P_{j}}{\partial \lambda_{l}} = \sum_{i,j=1}^{\aleph} g^{ij} \frac{\partial P_{i}}{\partial P_{k}} \frac{\partial P_{j}}{\partial P_{l}}  \nonumber \\
&=& \sum_{i,j=1}^{\aleph} g^{ij} \delta_{i}^{k} \delta_{j}^{l}  = g^{kl} \; .
\end{eqnarray}
We see that the components $g^{kl}_{\Lambda}$, $k,l = 1,2,..., \aleph$, of the searched for metric are equal to the corresponding components $g^{kl}$ of the Rao-Fisher metric on
the probability simplex $\Delta^{\aleph-1}$, i.e., in accord with Eq.(\ref{g on simplex}), $g^{ij}_{\Lambda} = \frac{\delta^{ij}}{P_{j}}$. Thus, we can finally compute the Rao-Fisher metric $g_{\Lambda}^{\vartheta\vartheta}$ on the submanifold coordinatized by one parameter $\vartheta$, which is induced by the metric $g^{kl}_{\Lambda}$ on the
statistical space ${\cal S}_{\Lambda}$:
\begin{eqnarray}
\label{induced g on S Lambda appendix}
g^{\vartheta\vartheta}_{\Lambda} &=&  \sum_{k,l=1}^{\aleph} g^{kl}_{\Lambda} \frac{\partial P_{k}}{\partial \vartheta} \frac{\partial P_{l}}{\partial \vartheta} = \sum_{l=1}^{\aleph} \frac{1}{P_{l}} \frac{\partial P_{l}}{\partial \vartheta} \frac{\partial P_{l}}{\partial \vartheta} \nonumber \\
&=& \sum_{l=1}^{\aleph} \frac{\partial q_{l}}{\partial \vartheta} \frac{\partial q_{l}}{\partial \vartheta} = g^{\vartheta\vartheta}  \; .
\end{eqnarray}
In conclusion, the Rao-Fisher metric $g^{\vartheta\vartheta}_{\Lambda}$ on the submanifold (coordinatized by $\vartheta$) of the statistical space ${\cal S}_{\Lambda}$ is equal to the Rao-Fisher metric  $g^{\vartheta \, \vartheta}$ on the submanifold (coordinatized by $\vartheta$) of the simplex of probabilities $\Delta^{\aleph-1}$. Thus, due to the Rao-Fisher metric $g^{\vartheta\vartheta}$ derived on ${\cal S}$ for the EPR-Bohm problem, (\ref{metryka w EPR}), we finally obtain
\begin{eqnarray}
\label{g th th on S Lambda}
g_{\Lambda}^{\vartheta \, \vartheta} &=& g^{\vartheta \, \vartheta} = n_{ab}^{2} = const. \;\; , \\
& & {\rm where} \;\;  n_{ab} = \pm 1 \; {\rm or} \; \pm 2 \;\;\;
{\rm for \; every }\;\; S_{ab} \in \Omega_{ab} \; . \; \nonumber
\end{eqnarray}

\end{document}